\documentclass[11pt, a4paper]{article}
\usepackage[margin=2.0cm]{geometry} % Márgenes más razonables
\usepackage{authblk} % Paquete esencial para autores y afiliaciones múltiples

\emergencystretch=\maxdimen
\usepackage[utf8]{inputenc} % Recomendado para caracteres especiales
\usepackage{lineno}
\usepackage{hyperref} 
\hypersetup{
    colorlinks=true,             % Activa el color en el texto (quita los recuadros)
    linkcolor={blue!80!black},   % Color para Figuras, Ecuaciones y Secciones
    citecolor={blue!80!black},   % Color para las citas bibliográficas [1]
    urlcolor={blue!80!black}     % Color para enlaces web
}

\usepackage{amsmath}
\usepackage{amsfonts}
\usepackage{amssymb}
\usepackage{graphicx}
\usepackage{mathrsfs}
\usepackage{times}
\usepackage{color}
\usepackage{psfrag}
\usepackage[numbers, sort&compress]{natbib} % Configuración estándar de natbib
\usepackage{float}
\usepackage{booktabs}
\usepackage{multirow}
\usepackage{xcolor}
\usepackage{listings}
\usepackage{xurl}
\usepackage{setspace}
\usepackage{subfig}
\usepackage{enumitem}
\usepackage[font={small}]{caption}
\setlist[enumerate]{label*=\arabic*.}
\usepackage[normalem]{ulem}

\usepackage[acronym]{glossaries}
\glsdisablehyper
\newglossary[nlg]{nomencl}{nli}{nlo}{List of Symbols}
\makeglossaries 
\newacronym{AFBC}{AFBC}{Arbitrary Flow Boundary Condition}
\newacronym{CFD}{CFD}{Computational Fluid Dynamics}
\newacronym{CY}{C-Y}{Carreau-Yasuda Model}
\newacronym{DPD}{DPD}{Dissipative Particle Dynamics}
\newacronym{FENE}{FENE}{Finitely Extensible Nonlinear Elastic}
\newacronym{GMM}{GMM}{Generalized Maxwell Model}
\newacronym{HMM}{HMM}{Heterogeneous Multiscale Method}
\newacronym{IK}{IK}{Irving-Kirkwood}
\newacronym{ILT}{ILT}{Inverse Laplace Transform}
\newacronym{LHMM}{LHMM}{Lagrangian Heterogeneous Multiscale Method}
\newacronym{MD}{MD}{Molecular Dynamics}
\newacronym{PAC}{PAC}{Periodic Array of Cylinders}
\newacronym{RPF}{RPF}{Reverse Poiseuille Flow}
\newacronym{RPM}{RPM}{Random Porous Medium}
\newacronym{SPH}{SPH}{Smoothed Particle Hydrodynamics}
\newacronym{Wi}{Wi}{Weissenberg number}
\newacronym{Re}{Re}{Reynolds number}
\newacronym{2D}{2D}{Two-dimensional}
\newacronym{DEM}{DEM}{Discrete Element Method}
\newacronym{BCF}{BCF}{Brownian Configuration Fields}

\newglossaryentry{Hfene}{
  name=$H_{\text{FENE}}$,
  description={Spring constant, FENE parameter},
  symbol={\ensuremath{H_{\text{FENE}}}},
  type=nomencl
}
\newglossaryentry{R0}{
  name=$R_0$,
  description={The maximum bond extension, FENE parameter},
  symbol={\ensuremath{R_0}},
  type=nomencl
}
\newglossaryentry{aij}{
  name=$a_{ij}$,
  description={conservative force coefficient, DPD parameter},
  symbol={\ensuremath{a_{ij}}},
  type=nomencl
}
\newglossaryentry{rc}{
  name=$r_c$,
  description={Cutoff radius, DPD parameter},
  symbol={\ensuremath{r_c}},
  type=nomencl
}
\newglossaryentry{dx}{
  name=$dx$,
  description={The average initial distance between beads, DPD parameter},
  symbol={\ensuremath{dx}},
  type=nomencl
}
\newglossaryentry{sigmaDPD}{
  name=$\sigma^R$,
  description={Random force amplitude, DPD parameter},
  symbol={\ensuremath{\sigma^R}},
  type=nomencl
}
\newglossaryentry{gammaDPD}{
  name=$\gamma^{D}$,
  description={Dissipation coefficient, DPD parameter},
  symbol={\ensuremath{\gamma}},
  type=nomencl
}
\newglossaryentry{T}{
  name=$T$,
  description={Temperature},
  symbol={\ensuremath{T}},
  type=nomencl
}
\newglossaryentry{kB}{
  name=$k_B$,
  description={Boltzmann constant},
  symbol={\ensuremath{k_B}},
  type=nomencl
}
\newglossaryentry{mi}{
  name=$m_i$,
  description={Mass of a DPD particle i},
  symbol={\ensuremath{m_i}},
  type=nomencl
}
\newglossaryentry{eta0}{
  name=$\bar{\eta}^0_p$,
  description={Zero-shear-rate viscosity from the macroscopic},
  symbol={\ensuremath{\bar{\eta}^0_p}},
  type=nomencl
}
\newglossaryentry{eta0prime}{
  name=$\eta'^0_p$,
  description={Zero-shear-rate viscosity prescribed by the microscale system},
  symbol={\ensuremath{\eta'^0_p}},
  type=nomencl
}
\newglossaryentry{eta_tot}{
  name=$\eta^0_{\rm tot}$,
  description={Total effective viscosity defined as $\eta^0_{\rm tot} = \bar{\eta}^0_p + \eta_s =\alpha\eta^0_p + \eta_s$},
  symbol={\ensuremath{\eta^0_{\rm tot}}},
  type=nomencl
}
\newglossaryentry{Uo}{
  name=$U_o$,
  description={Characteristic velocity, average velocity along time in the entire macro-domain},
  symbol={\ensuremath{U_o}},
  type=nomencl
}
\newglossaryentry{lambda}{
  name=$\lambda$,
  description={Characteristic relaxation time},
  symbol={\ensuremath{\lambda}},
  type=nomencl
}
\newglossaryentry{psi1}{
  name=$\psi_1$,
  description={First normal stress coefficient, $\psi_1= N_1/\dot{\gamma}_{xy}^2$},
  symbol={\ensuremath{\psi_1}},
  type=nomencl
}
\newglossaryentry{N1}{
  name=$N_1$,
  description={First normal stress difference},
  symbol={\ensuremath{N_1}},
  type=nomencl
}
\newglossaryentry{Nb}{
  name={$N_{\rm d}$},
  description={Number of beads per polymer chain},
  type=nomencl
}
\newglossaryentry{Gt}{
  name={$G(t)$},
  description={Relaxation modulus},
  type=nomencl
}
\newglossaryentry{Gprime}{
  name={$G'$},
  description={Storage modulus},
  type=nomencl
}
\newglossaryentry{Gprime2}{
  name={$G''$},
  description={Loss modulus},
  type=nomencl
}
\newglossaryentry{Ht}{
  name={$H(\tau)$},
  description={Relaxation spectrum},
  type=nomencl
}
\newglossaryentry{eta_complex}{
  name={$\eta'(\omega)$},
  description={Complex viscosity},
  type=nomencl
}
\newglossaryentry{dh}{
  name={$\Delta h$},
  description={Average inter-particle distance, SPH parameter},
  type=nomencl
}
\newglossaryentry{h}{
  name={$h$},
  description={The compact support of the bell-shaped kernel, SPH parameter},
  type=nomencl
}
\newglossaryentry{rho}{
  name=$\rho$,
  description={Fluid density},
  type=nomencl
}
\newglossaryentry{pi}{
  name={$\boldsymbol{\pi}$},
  description={Viscous stress tensor},
  type=nomencl
}
\newglossaryentry{tau}{
  name={$\boldsymbol{\tau}$},
  description={Total stress tensor},
  type=nomencl
}
\newglossaryentry{d}{
  name={$\mathbf{d}$},
  description={Macroscopic strain-rate tensor computed from continuum velocity fields},
  type=nomencl
}
\newglossaryentry{Omega}{
  name={$\Omega'$},
  description={Microscopic control volume},
  type=nomencl
}
\newglossaryentry{alpha}{
  name=$\alpha$,
  description={Viscosity ratio between the macroscopic and microscopic scale, $\alpha = \frac{\bar{\eta}^0_p}{\eta'^0_p}$},
  type=nomencl
}
\newglossaryentry{lambdaprime}{
  name=$\lambda^*$,
  description={Time-scale ratio between the macroscopic and microscopic scale, $\lambda^* = \frac{\bar{\lambda}}{\lambda'}$},
  type=nomencl
}
\newglossaryentry{Omega_core}{
  name=$\Omega'_{\rm core}$,
  description={Microscopic core region, AFBC parameter},
  type=nomencl
}
\newglossaryentry{Omega_bc}{
  name=$\Omega'_{\rm bc}$,
  description={Microscopic boundary-condition region, AFBC parameter},
  type=nomencl
}
\newglossaryentry{Omega_0}{
  name=$\Omega'_{0}$,
  description={Buffer region, AFBC parameter},
  type=nomencl
}
\newglossaryentry{l_core}{
  name=$l_{\text{core}}$,
  description={Length of the core region, AFBC parameter},
  type=nomencl
}
\newglossaryentry{l_bc}{
  name=$l_{\text{bc}}$,
  description={Length of the boundary condition region, AFBC parameter},
  type=nomencl
}
\newglossaryentry{l_buffer}{
  name=$l_{\text{buffer}}$,
  description={Length of the buffer region, AFBC parameter},
  type=nomencl
}
\newglossaryentry{l_total}{
  name=$l_{\text{total}}$,
  description={Length of the total domain, AFBC parameter},
  type=nomencl
}
\newglossaryentry{vi_prime}{
  name=$\mathbf{v}_i'$,
  description={Imposed velocity field for a microscopic particle},
  type=nomencl
}
\newglossaryentry{beta}{
  name=$\beta$,
  description={Polymeric viscosity fraction, $\beta=\eta_p/\eta_{\rm tot}$},
  type=nomencl
}
\newglossaryentry{p}{
  name=$p$,
  description={Pressure},
  type=nomencl
}
\newglossaryentry{f}{
  name={$\mathbf{f}$},
  description={External body force vector applied in the flow domain},
  type=nomencl
}
\newglossaryentry{Fbody}{
  name=$F$,
  description={Magnitude of the applied body force per unit mass},
  symbol={\ensuremath{F}},
  type=nomencl
}
\newglossaryentry{eta_s}{
  name={$\eta_{s}$},
  description={Solvent viscosity (shear viscosity of the fluid)},
  type=nomencl
}
\newglossaryentry{varepsilon}{
  name=$\varepsilon$,
  description={The polymer microstructure of characteristic scale variable},
  symbol={\ensuremath{\varepsilon}},
  type=nomencl
}
\newglossaryentry{eta_p}{
  name=$\bar{\eta}_p$,
  description={Effective microscopic viscosity arising from polymer},
  symbol={\ensuremath{\bar{\eta}_p(\mathbf{x},\varepsilon)}},
  type=nomencl
}
\newglossaryentry{eta_ps}{
  name=$\langle\eta'_{p,s}\rangle$,
  description={Shear viscosity arising from polymer},
  type=nomencl
}
\newglossaryentry{eta_pe}{
  name=$\bar{\eta}_{p,\varepsilon}$,
  description={Extensional viscosity arising from polymer},
  type=nomencl
}
\newglossaryentry{eta_eff}{
  name=$\eta_{\rm tot}$,
  description={The total effective viscosity combining solvent and polymer contributions, $\eta_{\rm tot} = \eta_{\rm s} + \bar{\eta}_p$},
  symbol={\ensuremath{\eta_{\rm tot}}},
  type=nomencl
}
\newglossaryentry{dbar}{
  name=$\bar{\mathbf{d}}$,
  description={Macroscopic strain-rate tensor prescribed or reconstructed from microscopic variables},
  symbol={\ensuremath{\bar{\mathbf{d}}}},
  type=nomencl
}
\newglossaryentry{pibar}{
  name=$\bar{\boldsymbol{\pi}}$,
  description={Macroscopic stress tensor},
  symbol={\ensuremath{\bar{\boldsymbol{\pi}}}},
  type=nomencl
}
\newglossaryentry{pis}{
  name=$\bar{\boldsymbol{\pi}}_{\rm s}$,
  description={Solvent stress tensor, representing the ideal Newtonian contribution},
  symbol={\ensuremath{\bar{\boldsymbol{\pi}}_{\rm s}}},
  type=nomencl
}
\newglossaryentry{pip}{
  name=$\langle\boldsymbol{\pi}'_{p}\rangle$,
  description={Microscopic polymeric stress tensor, including microstructural non-ideal contributions},
  symbol={\ensuremath{\langle\boldsymbol{\pi}'_{p}\rangle}},
  type=nomencl
}
\newglossaryentry{dgamma}{
  name=$\dot{\gamma}$,
  description={Characteristic or reference shear rate, used for non-dimensionalization or defining flow regimes},
  symbol={\ensuremath{\dot{\gamma}}},
  type=nomencl
}
\newglossaryentry{Delta_t_SPH}{
  name=$\Delta \bar{t}_{\text{SPH}}$,
  description={SPH macroscopic time step},
  type=nomencl
}
\newglossaryentry{Delta_t_DPD}{
  name=$\Delta t'_{\text{DPD}}$,
  description={DPD microscopic time step},
  type=nomencl
}
\newglossaryentry{delta_t}{
  name=$\delta t$,
  description={Exchange time step between macro- and micro-scales},
  type=nomencl
}
\newglossaryentry{M}{
  name=$\bar{M}$,
  description={Positive integer for the SPH time step scaling, $\delta t = \bar{M} \cdot \Delta \bar{t}_{\text{SPH}}$},
  type=nomencl
}
\newglossaryentry{K}{
  name=$k'$,
  description={Positive integer for the DPD time step scaling, $\delta t = k' \cdot \Delta t'_{\text{DPD}}$},
  type=nomencl
}
\newglossaryentry{ri_prime}{
  name=$\mathbf{r}_i'$,
  description={Position of a microscopic particle relative to its micro-domain center},
  type=nomencl
}
\newglossaryentry{grad_vI}{
  name=$\nabla \mathbf{v}_I$,
  description={Macroscopic velocity gradient at SPH particle I},
  type=nomencl
}
\newglossaryentry{c}{
  name=$c$,
  description={Speed of sound, SPH parameter},
  type=nomencl
}
\newglossaryentry{zeta}{
  name=$\zeta_s$,
  description={Bulk viscosity, SPH parameter},
  type=nomencl
}
\newglossaryentry{rho0}{
  name=$\rho_0$,
  description={Reference density, SPH parameter},
  type=nomencl
}
\newglossaryentry{mI}{
  name=$m_I$,
  description={Mass at SPH particle I},
  type=nomencl
}
\newglossaryentry{mDPD}{
  name=$m_{\rm DPD}$,
  description={Mass of a DPD particles},
  type=nomencl
}
\newglossaryentry{nDPD}{
  name=$n_{\rm DPD}$,
  description={DPD number density},
  type=nomencl
} 
\newglossaryentry{pI}{
  name=$p_I$,
  description={The pressure at SPH particle I},
  type=nomencl
}
\newglossaryentry{rhoI}{
  name=$\rho_I$,
  description={The density at SPH particle I},
  type=nomencl
}
\newglossaryentry{eta0cy}{
  name=$\eta_{\text{cy}}^0$,
  description={Zero-shear viscosity from Carreau–Yasuda fit},
  symbol={\ensuremath{\eta_{\text{cy}}^0}},
  type=nomencl
}

\newglossaryentry{etainf}{
  name=$\eta_{\text{cy}}^\infty$,
  description={Infinite-shear viscosity from Carreau–Yasuda fit},
  symbol={\ensuremath{\eta_{\text{cy}}^\infty}},
  type=nomencl
}

\newglossaryentry{acy}{
  name=$a_{\text{cy}}$,
  description={Transition sharpness parameter from Carreau–Yasuda fit},
  symbol={\ensuremath{a_{\text{cy}}}},
  type=nomencl
}

\newglossaryentry{ncy}{
  name=$n_{\text{cy}}$,
  description={Power-law index from Carreau–Yasuda fit},
  symbol={\ensuremath{n_{\text{cy}}}},
  type=nomencl
}
\newglossaryentry{gammadotxy}{
  name=$\dot{\gamma}_{xy}$,
  description={Shear rate associated with simple (pure) shear in the $xy$ plane},
  symbol={\ensuremath{\dot{\gamma}_{xy}}},
  type=nomencl
}
\newglossaryentry{taumaxRPF}{
  name={$\tau_{\rm max}^{\rm RPF}$},
  description={Maximum theoretical shear stress from the RPF analytical solution, given by $\tau_{\rm max}^{\rm RPF} = F \rho L_y / 4$},
  type=nomencl
}
\newglossaryentry{tauStarPAC}{
  name={$\tau_{\rm PAC}^{*}$},
  description={Dimensionless Stress parameter to PAC cases},
  type=nomencl
}
\newglossaryentry{tauStarPore}{
  name={$\tau_{\rm obstracle}^{*}$},
  description={Dimensionless Stress parameter to PAC cases},
  type=nomencl
}
\newglossaryentry{vxStar}{
    name={\(v^*_x\)},
    description={the dimensionless magnitude of the x-component of velocity, defined as \(v^*_x={\|v_x\|}/{U_o}\)},
    type=nomencl
}
\newglossaryentry{vyStar}{
    name={\(v^*_y\)},
    description={the dimensionless magnitude of the y-component of velocity, defined as \(v^*_y={\|v_y\|}/{U_o}\)},
    type=nomencl
}
\newglossaryentry{vStar}{
    name={\(v^*\)},
    description={the dimensionless magnitude of the velocity, defined as \(v^*={\|\textbf{v}\|}/{U_o}\)},
    type=nomencl
}
\newglossaryentry{N1Star}{
    name={\(\bar{N}^*_{1}\)},
    description={the dimensionless total normal stress, defined as \(\bar{N}^*_{1}={\|N_{1}\|}/{\tau_{\text{PAC}}^{*}}\)},
    type=nomencl
}
\newglossaryentry{Wilocal}{
  name={$\text{Wi}^{\rm local}$},
  description={local Weissenberg number},
  type=nomencl
}
\newglossaryentry{gammadotTensor}{
  name={$\underline{\underline{\dot{\boldsymbol{\gamma}}}}$},
  description={Rate-of-strain tensor},
  type=nomencl
}
\newglossaryentry{piStarP}{
  name={$\bar{\pi}_p^*$},
  description={Dimensionless polymeric stress magnitude},
  type=nomencl
}
\newglossaryentry{piStar}{
  name={$\bar{\pi}^{*}$},
  description={Dimensionless total stress magnitude},
  type=nomencl
}
\newglossaryentry{piStarS}{
  name={$\bar{\pi}^{*}_{\text{s}}$},
  description={Dimensionless solvent stress},
  type=nomencl
}
\newglossaryentry{piStarPxy}{
  name={$\bar{\pi}_{p,xy}^*$},
  description={Dimensionless microscopic polymeric shear stress},
  type=nomencl
}
\newglossaryentry{piStarXY}{
  name={$\bar{\pi}^{*}_{xy}$},
  description={Dimensionless total shear stress},
  type=nomencl
}
\newglossaryentry{piXY}{
  name={$\bar{\pi}_{xy}$},
  description={Total shear stress denotes for the $xy$-component of the $\bar{\boldsymbol{\pi}}$},
  type=nomencl
}
\newglossaryentry{piXX}{
  name={$\bar{\pi}_{xx}$},
  description={The $xx$-component stress of the $\bar{\boldsymbol{\pi}}$},
  type=nomencl
}
\newglossaryentry{piYY}{
  name={$\bar{\pi}_{yy}$},
  description={The $yy$-component stress of the $\bar{\boldsymbol{\pi}}$},
  type=nomencl
}
\newglossaryentry{piStarSXY}{
  name={$\bar{\pi}^{*}_{\text{s},xy}$},
  description={Dimensionless solvent shear stress},
  type=nomencl
}
\newglossaryentry{Lx}{
  name=$L_x$,
  description={Domain length in the $x$-direction},
  type=nomencl
}

\newglossaryentry{Ly}{
  name=$L_y$,
  description={Domain length in the $y$-direction},
  type=nomencl
}
\newglossaryentry{pip_xy}{
  name={$\langle\pi'_{p,xy}\rangle$},
  description={Average polymeric shear stress, $xy$-component of the polymer stress tensor ($\boldsymbol{\pi}'_p$)},
  type=nomencl
}
\newglossaryentry{pis_xy}{
  name={$\bar{\pi}_{\text{s},xy}$},
  description={Average solvent shear stress, $xy$-component of the polymer stress tensor ($\bar{\boldsymbol{\pi}}_{\rm s}$)},
  type=nomencl
}
\newglossaryentry{Rcy}{
  name={$R_{cl}$},
  description={Radius of cylinder},
  type=nomencl
}
\newglossaryentry{Rpore}{
  name={$r_{\rm obstacle}$},
  description={Radius of pore},
  type=nomencl
}
\newglossaryentry{Rav}{
  name={$R_{\rm obstacles}$},
  description={The average radius of the pores},
  type=nomencl
}

\definecolor{mygreen}{rgb}{0,0.6,0}
\definecolor{myred}{rgb}{1,0,0}
\definecolor{mygray}{rgb}{0.5,0.5,0.5}
\definecolor{mymauve}{rgb}{0.88,0,0.82}
\definecolor{violet}{rgb}{0.5,0,0.5}
\definecolor{blue2}{rgb}{0.5,0,1}

\begin{document}

% --- TÍTULO Y AUTORES (Formato Genérico) ---
\title{Lagrangian Heterogeneous Multiscale Method (LHMM) for Simulating Polymer Solutions/Melts Behavior under Complex Flows using DPD-SPH}

% Autor 1
\author[1]{Edgar A. Pati\~no-Nari\~no\thanks{\texttt{epatino@bcamath.org}}}
% Autor 2
\author[1]{Nicolas Moreno\thanks{\texttt{nmoreno@bcamath.org}}}
% Autor 3
\author[1,2,3]{Marco Ellero\thanks{\texttt{mellero@bcamath.org}}}

% Afiliaciones
\affil[1]{\small Basque Center for Applied Mathematics (BCAM), Alameda de Mazarredo, 14, 48400 Bilbao, Spain.}
\affil[2]{\small IKERBASQUE, Basque Foundation for Science, Calle de María Díaz de Haro 3, 48013 Bilbao, Spain.}
\affil[3]{\small Complex Fluids Research Group, Department of Chemical Engineering, Faculty of Science and Engineering, Swansea University, Swansea SA1 8EN, UK.}

\date{} % O \date{\today} para quitar la fecha

\maketitle
\vspace{-1.0cm} % 

\begin{abstract}
We present a Lagrangian Heterogeneous Multiscale Method (LHMM) for simulating the non-Newtonian rheology of polymer melts in complex two-dimensional flows. The method couples Dissipative Particle Dynamics (DPD) at the microscale with a GENERIC-compliant Smoothed Particle Hydrodynamics (SPH) at the macroscale, in a concurrent framework, overcoming the limitations of traditional Eulerian-based methods in capturing long-memory and history-dependent effects. At the microscale, DPD serves as a virtual rheometer, employing FENE (Finitely Extensible Nonlinear Elastic) bead-spring polymer chains. This approach provides key rheological properties, including shear-thinning and zero-shear-rate viscosities, relaxation times, and viscoelastic dynamics, which are quantified via Carreau-Yasuda fitting and spectral analysis. The LHMM couples SPH-derived strain rates with microscopic stress responses using the Irving-Kirkwood formalism. This approach enables a concurrent interaction between macroscopic strain rates and microscopic stress tensors, ensuring a consistent viscoelastic response across scales. The method is validated against benchmark flows, including Reverse Poiseuille Flow and flow through a Periodic Array of Cylinders, across Weissenberg numbers $0.5 < \text{Wi} < 30$ and low Reynolds numbers ($\text{Re} < 1$). A final demonstration of flow in a 2D porous medium highlights LHMM’s capability to handle highly heterogeneous geometries. The LHMM is implemented in LAMMPS, making it suitable for integrating multiple models to describe microscales. In contrast, large-scale simulations efficiently utilize GPU and CPU resources, managing multiple coupling and time-scaling levels to maintain numerical stability and accuracy. The framework offers a predictive, constitutive-free tool that links microscopic polymer dynamics to macroscopic flow behavior, making it suitable for multiscale applications.
\end{abstract}

\vspace{0.5cm}
\section{Introduction} \label{sec:intro}
Polymer melts exhibit complex rheological behaviors, including pronounced viscoelasticity and memory effects, particularly under nontrivial flow conditions. Accurately simulating these dynamics remains challenging due to the intrinsically nonlinear coupling between microscopic and macroscopic scales \cite{Ge2019,Vassaux2020,Hughes1998}.  Over the past few decades, numerous multiscale simulation strategies have emerged to bridge the gap between \gls{MD} and continuum mechanics \citep{Gooneie2017,Schmid2022,Smith2023}. These strategies include adaptive resolution schemes that facilitate a seamless transition between atomistic and coarse-grained models \cite{Zavadlav2017,Zavadlav2016,Delgado-Buscalioni2008}, often utilizing the MARTINI force field for efficient sampling \cite{Marrink2007,Souza2021,Zavadlav2017,Zavadlav2016}. This approach has effectively linked molecular structure to macroscopic rheological behavior \citep{Boyko2025,Xie2023,Svaneborg2016,Borg2018,Borg2014}. However, despite these advancements, a computational framework that accurately and efficiently captures the non-Newtonian and history-dependent behavior of polymer melts under arbitrary flow conditions remains cumbersome.

In the context of multiscale modelling, Laso and \"Ottinger \cite{Laso1993} introduced the so-called CONNFFESSIT method, combining Eulerian finite-element solvers (fixed-grid formulation) with stochastic polymer dynamics \cite{Feigl1995}, establishing a foundation for Eulerian–Lagrangian multiscale formulations \cite{Sherck2021,Khajeh2024,De2013}. These efforts have shown relevant applicability in the modelling of polymer flow characterized by simple deformation histories \cite{Boyko2024,Seryo2020,Laso1999}, in both dilute and semi-dilute systems \cite{Sato2019,Schieber2020,Murashima2013}. This approach establishes a direct connection between the macroscopic and microscopic models via the stress and strain rate tensor fields \cite{Sato2017,Sato2025a,Zhao2021}. Subsequent advancements have included the integration of Brownian Configuration Field (BCF) methods \cite{Ottinger1997a,Hulsen1997,Borgbjerg1994,Khajeh2024a}. In BCF, for instance, macroscopic domains are represented using Eulerian descriptions, while the microscopic state of the system is typically characterized at each grid point. The microscale features are included through a configuration tensor or ensemble of dumbbell/bead–spring replicas denoted by  $\mathbf{Q}$, which evolves according to an Eulerian Langevin-type equation: $d\mathbf{Q} = \mathcal{A}_{\text{flow}}(\mathbf{Q})\,dt 
    \;+\; \mathcal{R}_{\text{relax}}(\mathbf{Q})\,dt
    \;+\; \mathcal{N}_{\text{thermal}}\,,$
 where the first term in the equation accounts for both advection and deformation resulting from the macroscopic flow, the second term represents the elastic relaxation toward equilibrium, and the third term captures stochastic fluctuations. Thus, $\mathbf{Q}(t)$ serves as a configuration descriptor, effectively capturing local stretching and orientation~\cite{De2006,De2013,Liu2019}. While this scheme yields manageable models~\cite{Yasuda2010,Yasuda2011}, the description of the microscopic state is limited to Langevin-type models (e.g., BD or slip-link) that simplify polymer dynamics~\cite{Xu2023a,Vassaux2020}.

 Weinan and Engquist introduced the \gls{HMM}~\cite{Weinan2007,E2009} as a comprehensive framework for coupling macro- and microscale models. This coupling allows to deal with more complex macroscopic domains and provide detailed microscopic representations of systems, such as Synchronized Molecular Dynamics (SMD) \cite{Yasuda2009,Yasuda2010,Yasuda2011,Wu2021} and non-equilibrium molecular dynamics (NEMD) \cite{Tedeschi2021,Mortazavi2015}. HMM strategies \cite{E2009,Lockerby2012,Ren2005} have evolved into robust schemes (see also \cite{Weinan2007}) that integrate numerical algorithms that independently resolve macro- and microscales. Where microscopically derived quantities (using a variety of available algorithms, \gls{MD}, NEMD, \gls{DPD}, \gls{DEM}, to name a few) inform the macroscopic momentum equations \citep{Ren2005,Stalter2018,Vassaux2020}. Following the CONNFFESSIT spirit, Eulerian HMM-based schemes approximate the microscopic stress at a grid point \(i\) as: $\boldsymbol{\sigma}_i \approx \left\langle\mathbf{F}(\mathbf{Q}_j, \dot{\gamma}_j) \cdot \mathbf{r}_{ij}\right\rangle_{\Omega_i}$, where $\mathbf{F}$ denotes the microscopic force derived from configuration descriptors $\mathbf{Q}_j$, $\dot{\gamma}_j$ represents the local shear rate, and the operator $\langle\cdot\rangle_{\Omega_i}$ indicates statistical sampling over the microscopic configurations within the sampling volume $\Omega_i$. These methods offer a flexible route to estimate microscopic stresses. However, its accuracy is critically dependent on rapid relaxation and timescale separation ($\lambda \ll \Delta t$)~ \cite{Vassaux2019,Morii2021}, conditions that are often not met in polymeric flows.

Alternatively, there are more general HMM-based microscopic coupling strategies available \cite{Ren2005,E2009,Vassaux2019}. These include embedded microscopic simulations at certain macroscopic grid points, where stresses are obtained using \gls{IK} averaging \cite{Irving1950,Lockerby2013,Stalter2018}. Variants such as the Seamless \gls{HMM} \cite{E2009,Lockerby2013} effectively address the issue of repeated reinitialization in microscopic simulations. Nevertheless, they still rely on the assumption of a large timescale separation between macro and micro descriptions. While there have been \gls{HMM} schemes aiming to bridge different temporal scales for polymer mechanics~\citep{Vassaux2019,Vassaux2020,Leiter2023}, these Eulerian-based descriptions of the macroscopic scales often struggle to capture the history-dependent nature of polymer melts, limiting their applicability in realistic flow scenarios~\citep{E2009,Vassaux2019,Vassaux2020,Moreno2023}.

In polymer melts, the viscoelastic stress depends on the complete strain history of fluid elements \cite{Bird1987a,Bird1995,Ottinger1996}. This memory dependence can be represented as $M(\mathbf{x},t) = M\!\left[ \dot{\gamma}\!\left(X(t'),t'\right);\, t' \leq t \right]$, where $X(t')$ denotes the Lagrangian trajectory. While Eulerian approaches require challenging backward-tracking or interpolation \citep{Vassaux2019,Vassaux2020}, recent developments in fully Lagrangian (HMM) frameworks inherently integrates this memory, modelling both macro and microscales using Lagrangian descriptions (LL)\cite{Weinan2007,Vassaux2020,Schmid2023,Moreno2023}. LL alternatives have spurred the development of multiscale schemes that adopt the \gls{SPH} method to discretize the macroscales, which inherently track material trajectories and effectively maintain viscoelastic stress buildup \cite{Vazquez-Quesada2009,Vazquez-Quesada2012,Ye2019,LeTouze2025a}. Whereas at the micro scales, they can adopt a variety of models —ranging from Brownian dynamics \citep{Xu2016b,Xu2023a,King2021} and slip-link models \citep{Feng2016a,Sato2017,Sato2019} to coarse-grained particle methods that merge \gls{MD} and \gls{DPD} \cite{Tang2015,Oda2025,NietoSimavilla2022}, as well as data-driven approaches \cite{Seryo2020,Sato2025a}. Current LL implementations often limit microscopic feedback to polymer stress, treating the solvent or matrix at a continuum level \citep{Feng2016a, Smith2023, Tang2015}. Moreover, establishing consistent boundary conditions at the microscale presents a considerable challenge, particularly in complex or transient flow situations \cite{Schmid2023,Vassaux2020}. As a result, many LL schemes tend to concentrate on idealized conditions, such as simple shear, or on mean-field polymer models, which restricts their applicability to a wider range of non-Newtonian systems. This underscores the necessity for a more adaptive, bidirectional multiscale coupling strategy.
 
Herein, we adopt a GENERIC (General Equation for Non-Equilibrium Reversible-Irreversible Coupling) compliant \gls{LHMM}~\cite{Moreno2023} framework that enables bidirectional coupling between micro- and macroscales while accurately preserving the full deformation history within a fully Lagrangian framework. In this context, our implementation employs a thermodynamics structure-preserving, GENERIC-consistent \gls{SPH} discretization at the macroscopic level, and a \gls{DPD}/\gls{FENE} coarse-grained model at the microscopic level \cite{Fedosov2010, Litvinov2014}. This formulation allows the polymeric fluid to be represented as a combination of solvent particles and spring-chain polymers of varying lengths, with their configurations evolving dynamically. Following Moreno \& Ellero~\cite{Moreno2023}, the \textit{top-down coupling} is executed via the \gls{AFBC} protocol \cite{Moreno2021}. Within this approach, the macroscopic velocity field $\mathbf{u}_{\text{macro}}(\mathbf{x},t)$ derived from SPH serves as the kinematic input for imposing the local velocity gradient at the microscale: $\nabla \mathbf{u}_{\text{macro}}$. This coupling strategy guarantees that the microscale shear rate adheres to the relation $\boldsymbol{\dot{\gamma}_{\text{micro}}} = \nabla\mathbf{u}_{\text{macro}}$. On the other hand, \textit{bottom-up} information is transmitted through the \gls{IK} formalism \cite{Yang2012}, which calculates the microscopic stress tensor, enabling the evolving microstructural configuration to influence the macroscopic momentum balance on-the-fly. This formulation effectively maintains the complete deformation history across scales while accurately capturing nonlinear viscoelastic responses derived from detailed DPD simulations—without the need for the assumptions commonly associated with generalized Langevin models~\cite{Moreno2023}. As a result, fundamental rheological phenomena—such as viscoelastic relaxation, shear thinning, and normal stress differences—arise naturally at the microscale, \textit{without the need for a prescribed evolution equation like $\mathbf{Q}$}~\cite{Schieber2020,Datta2022,Winters2024}. Overall, the \gls{LHMM} naturally integrates microstructural memory, making it especially well-suited for complex viscoelastic systems that require an in-depth understanding of molecular behavior.

We validate the \gls{LHMM} using benchmark flows and complex porous-media configurations \cite{Kumar2023a,Haward2021c,Mokhtari2024,Browne2020}. The implementation is accelerated using GPUs within the LAMMPS environment \cite{Tang2015,Thompson2022,Nguyen2015}, enabling near-linear scaling in massively parallel computation and stable concurrent time stepping between the \gls{SPH} and \gls{DPD} solvers. The \gls{LHMM} thus provides a robust and efficient multiscale avenue for polymer solutions and melts, with potential applications to a broader class of fluid–structure interaction problems dominated by history effects and complex kinematics \cite{Matous2017}.

The efficacy and broad applicability of the proposed methodology are demonstrated through benchmark cases such as Reverse Poiseuille Flow and flow through a periodic array of cylinders, exhibiting excellent numerical convergence and accurate recovery of both transient and steady-state viscoelastic behaviors. Finally, we investigate flow through \gls{2D} porous media with heterogeneous permeability, demonstrating the \gls{LHMM} 's ability to capture localized stress concentrations and spatially varying deformation paths. Owing to GPU acceleration, simulations with more than 200 million particles were achieved while maintaining stability across concurrent time stepping between the \gls{SPH} and \gls{DPD} solvers. The structure of this paper is organized as follows. Sections~\ref{sec:LHMM} and \ref{sec:algorithm} introduce the LHMM methodology and algorithmic formulation, including the coupling scheme and numerical implementation. Section~\ref{sec:micro_results} presents the microscale rheological characterization using \gls{DPD} simulations. Section~\ref{sec:macro_results} addresses benchmark validations and explores complex flow geometries. Finally, Section~\ref{sec:conclusions} summarizes the main findings and discusses future directions.

\section{Lagrangian Heterogeneous Multiscale Method (LHMM)} \label{sec:LHMM}
The \gls{LHMM} provides a theoretical framework for coupling continuum descriptions across macroscopic and microscopic scales. At the macroscopic level, the dynamics of a simple incompressible fluid are governed by the Navier–Stokes equations, which can be written as:

\begin{equation}
    \nabla \cdot \textbf{v}=0, \quad 
    \rho\frac{d\textbf{v}}{dt}=\nabla\cdot\boldsymbol{\tau}+\textbf{f},
    \label{eq:NS_01} 
\end{equation}
where the total stress tensor is given by \gls{tau}$= -p\mathbf{I} + \boldsymbol{\pi}$, with \gls{p} denoting the pressure and \gls{pi} the extra stress, and \gls{f} the external body forces. For incompressible Newtonian fluids, the stress ($\boldsymbol{\pi}$) is purely viscous and is linearly related to the strain-rate tensor (\gls{d}), $\boldsymbol{\pi} = 2\eta_{s}\mathbf{d}$ with $\mathbf{d}=\frac{1}{2}\left(\nabla\textbf{v} + \nabla\textbf{v}^{\rm T}\right)$, where \gls{eta_s} is the solvent viscosity. This linear constitutive relation allows the flow field to be fully described by Eq.~\eqref{eq:NS_01}. However, for non-Newtonian fluids, such as colloidal suspensions or polymer melts, this linear relationship no longer holds, and the stress must instead be obtained from either constitutive equations \cite{Bird1987a} or an explicit microscale description of the extra stress \gls{pi}, as is customary in HMMs~\cite{Weinan2007,Abdulle2012}. 

To maintain clarity in notation, we use an overbar to denote macroscopic quantities and variables computed from microscopic estimates (e.g., the macroscopic stress tensor containing microscopic information is denoted as \gls{pibar}). While purely microscopic quantities or local microscopic control domains (viewed from the macroscopic level) are indicated with a prime (e.g., \gls{Omega} denotes a microscopic control volume)~\cite{Moreno2023}. Consequently, using an ensemble-averaged of the microscopic stresses measured over \gls{Omega}, yields the macroscopic counterpart given by 
\begin{align}
    \bar{\boldsymbol{\pi}}(\mathbf{x})=\frac{1}{\Omega'}\int_{\Omega'}\boldsymbol{\pi}'{\rm d}\Omega' = \langle\boldsymbol{\pi}'\rangle, 
\end{align}
where \( \langle \cdot \rangle \) denotes spatial averaging over the microscopic domain. In general, the macroscopic stress tensor can be decomposed as \gls{pibar}$(\mathbf{x},t) = \bar{\boldsymbol{\pi}}_h(\mathbf{x},t) + \bar{\boldsymbol{\pi}}_m(\mathbf{x},t) + \bar{\boldsymbol{\pi}}_k(\mathbf{x},t),$ where the subscripts $h$, $m$, and $k$ correspond to the hydrodynamic, microstructural non-hydrodynamic (e.g  polymer chains, colloidal particles, or interfaces), and kinetic contributions, respectively. Which leads to $\bar{\boldsymbol{\pi}}(\mathbf{x})=\langle\boldsymbol{\pi}'_h\rangle+\langle\boldsymbol{\pi}'_m\rangle+\langle\boldsymbol{\pi}'_k\rangle$. Notice that the term \( \boldsymbol{\pi}_h'\) accounts for the hydrodynamic contributions at the microscale~\cite{Feng2016a,Vazquez-Quesada2009a}. However, as discussed by Moreno and Ellero \cite{Moreno2023}, in LHMM the ideal hydrodynamic contributions can be conveniently accounted for either from macro, microscale, or a combination of both scales. 

Herein, for polymeric systems, the solvent stress contribution (\gls{pis}) manifests at the macroscopic level as an ideal Newtonian component, and for a spatially homogeneous solvent this leads to \(\langle \boldsymbol{\pi}'_h \rangle =\) \gls{pis} \(= \boldsymbol{\pi}_s = 2 \eta_{\rm s} \mathbf{d} \). Thus, we consider that hydrodynamic contributions are fully captured by a Newtonian description of the fluid at macroscales. In contrast, the microstructural non-hydrodynamic contributions \( \bar{\boldsymbol{\pi}}_m(\mathbf{x}) = \langle\boldsymbol{\pi}'_m\rangle\) completely arise from the polymer dynamics modelled at the microscale, such that \(\langle\boldsymbol{\pi}'_m\rangle = \bar{\boldsymbol{\pi}}_{p}(\bar{\mathbf{d}},\varepsilon)=\langle\boldsymbol{\pi}'_{p}\rangle(\mathbf{d}',\varepsilon)\). Where, \(\varepsilon\) denotes a characteristic microscopic length-scale associated with polymer structural features (e.g the end-to-end distance, monomer size, or radius of gyration) depending on the level of coarse-graining, which governs the local non-ideal (non-Newtonian) response and depends on the macroscopic strain-rate field \gls{dbar}. Accordingly, the total macroscopic stress tensor can be expressed as the sum of solvent and polymeric contributions:
\begin{subequations} \label{eq:Hydro_hybrid}
\begin{align}
\bar{\boldsymbol{\pi}} (\mathbf{x})
    &={\bar{\boldsymbol{\pi}}_{s}} +  \langle\boldsymbol{\pi}'_{p}\rangle(\mathbf{d}',\varepsilon), \\
    \bar{\boldsymbol{\pi}} (\mathbf{x}) 
    &= \underbrace{2\eta_{\rm s} \mathbf{d}}_{\text{macroscopic (solvent)}}
    + \underbrace{\bar{\boldsymbol{\pi}}_{p}(\bar{\mathbf{d}},\varepsilon)}_{\text{micro (polymer-microstructure)}}.
\end{align}
\end{subequations}

The macroscopic strain-rate tensor \gls{dbar} is determined by the macroscopic velocity gradient field. The ensemble-averaged polymeric stress \gls{pip} defines also an effective polymeric viscosity \gls{eta_p}$(\mathbf{x},\bar{\mathbf{d}},\varepsilon)$. Thus, the total effective viscosity (\gls{eta_eff}) characterizes the complex fluid and is defined as:
\begin{equation}\label{eq:eta_eff}
 \eta_{\rm tot}(\bar{\mathbf{d}},\varepsilon) = \eta_{\rm s} + \bar{\eta}_p(\bar{\mathbf{d}},\varepsilon),
\end{equation}
where \gls{eta_p} denotes the polymeric contribution resulting from the evolving microstructure. In accordance with the work of Moreno and Ellero~\cite{Moreno2023}, the macroscopic scales are discretized using the well-established \gls{SPH} method~\cite{LeTouze2025a,Pozorski2024,Patino-Narino2025}. At the microscopic level, the polymeric systems are modeled using the \gls{DPD} method, which has been extensively utilized in simulating complex polymeric materials~\cite{Fedosov2008a,Moreno2020,Lei2017a}.

\subsection{Viscoelastic coupling and nondimensionalization}
The polymeric stress tensor, \(\langle\boldsymbol{\pi}'_{p}\rangle(\mathbf{d}',\varepsilon)\), is evaluated through Irving–Kirkwood averaging over the microscopic domains (see Section~\ref{sub_micro_IK}), contributing to the total macroscopic stress tensor (Eq.~\eqref{eq:Hydro_hybrid}). To ensure consistent coupling between the SPH and DPD discretizations, reproducing the proper viscoelastic response of the polymeric system,  we require a coherent transfer of microscopic stresses to the macroscopic flow field. To this end, we introduce a set of  characteristic viscosity and times scales for macro and micro descriptions, that define characteristic stresses $\bar{\tau}^0$ and ${\tau'}^0$. These characteristic stresses serve as reference to conduct the communication between scales via non-dimensional values. This scaling parameter is crucial when numerical methods differ across scales (such as \gls{SPH} and \gls{DPD}), making the scaling approximation essential for achieving proper coupling between varying quantities and dimensions.

Under steady shear conditions, the polymer melt exhibits a zero-shear viscosity ($\eta_p^0$) and a characteristic relaxation time ($\lambda$), where the critical shear rate corresponding to the crossover of the storage and loss moduli satisfies $\dot{\gamma}^{c}_{xy}\approx 1/\lambda$~\cite{Santelli2024,Becerra2020,Katzarova2015}. These parameters define the viscoelastic transition between the Newtonian plateau and the shear-thinning regime, characteristic of polymer melts ~\cite{Fedosov2010,Litvinov2014}, and within the range determined by the microstructure configuration. Accordingly, a convenient characteristic stress (associated with the zero-shear viscosity and relaxation time) can be defined as $\bar{\tau}^0=\bar{\eta}^0_p/\bar{\lambda}$ and $\tau'^0=\eta'^0_p/\lambda'$ for the macroscale and microscale, respectively. Thus, if we require that the dimensionless polymeric stress tensors remain consistent across scales, then $\bar{\boldsymbol{\pi}}_p/\bar{\tau}^0=\langle\boldsymbol{\pi}'_{p}\rangle/\tau'^0$. Consequently, the polymeric stress contribution at the macroscopic level can be readily expressed as:
\begin{equation}\label{eq:pip}
\bar{\boldsymbol{\pi}}_p 
= \frac{\bar{\tau}^0}{\tau'^0}\langle\boldsymbol{\pi}'_{p}\rangle
= \frac{(\bar{\eta}^0_p/\eta'^0_p)}{(\bar{\lambda}/\lambda')}\langle\boldsymbol{\pi}'_{p}\rangle
\dot{=} \frac{\alpha}{\lambda^*}\langle\boldsymbol{\pi}'_{p}\rangle,
\end{equation}
where the introduce
\begin{equation}\label{eq:ratios}
\alpha \dot{=} \frac{\bar{\eta}^0_p}{\eta'^0_p} 
\qquad \text{and} \qquad 
\lambda^* \dot{=} \frac{\bar{\lambda}}{\lambda'},
\end{equation}
represent the viscosity and time-scale ratios between the macroscopic and microscopic descriptions, respectively.

At the macroscopic level, from Eq.~\eqref{eq:eta_eff}, the total effective viscosity in the zero-shear rate limit is given by \gls{eta_tot}$=$ \gls{eta0} $+$ \gls{eta_s}. It should be noted that \gls{eta_tot} acts as a reference quantity under zero-shear conditions; however, within the LHMM framework, the effective viscosity evolves dynamically with the local deformation and the underlying microstructure. In the context of viscoelastic polymeric systems, it is customary to define the viscosity ratio \gls{beta} as the fraction of the polymeric contribution to the total viscosity. The total viscosity is often related to material properties such as polymer concentration or solvent quality. This relationship can be expressed as \gls{beta}$=$\gls{eta0}/\gls{eta_tot}, where the polymeric viscosity employed at the macroscopic SPH level is defined, according to Eq.~\eqref{eq:ratios}, as \gls{eta0}$=$\gls{alpha}\gls{eta0prime}. Hence:
\begin{equation} \label{eq:beta}
    \beta = \frac{\bar{\eta}^0_p}{\eta^0_{\rm tot}}=\frac{\alpha \eta'^0_p}{\alpha\eta'^0_p + \eta_s},
\end{equation}   
where \gls{eta0prime} denotes the zero-shear viscosity prescribed by the microscale system, extracted from the microscopic \gls{DPD} characterization  (Section~\ref{subsec:steady_shear}). For consistency, the shear-rate tensor \(\bar{\mathbf{d}}\) at the macroscopic level must be rescaled by \gls{lambdaprime} to ensure synchronized deformation timescales between both scales, such that \(\mathbf{d}' = \lambda^*\bar{\mathbf{d}}\), where \(\mathbf{d}'\) is the actual shear-rate tensor imposed at the microscale. The scaling factor \(\alpha\) is varied to represent different polymer melt configurations, accounting for changes in the polymeric fraction or molecular structure at the macroscopic level. The characteristic time ratio \gls{lambdaprime}$= \bar{\lambda}/\lambda'$ is set to unity, reflecting synchronous description of relaxation times across scales; this is generally the case when there is no scale separation between micro-structured timescale and the macroscopic one (as in long-memory behavior fluids), and can be handled by the same time stepping schemes. On the other hand, $\lambda^{*}\neq1$ should be chosen when dealing with short-memory fluids; its role in temporal coupling is discussed in Section~\ref{sec:Time-DepentLHMM}.

As shown in Eq.~\eqref{eq:Hydro_hybrid}, the first term represents the ideal Newtonian response of the solvent, evaluated at the \gls{SPH} level. In contrast, the second term incorporates the microstructural contribution from coarse-grained \gls{DPD} simulations and is introduced through Eq.~\eqref{eq:pip}. This hybrid approach consistently integrates multiscale physics, enabling the capture of time-dependent phenomena such as shear thinning, thickening, and strain hardening without relying on empirical constitutive models. Consequently, the total stress \gls{tau} is formulated within a Lagrangian framework, incorporating the stress decomposition from Eq.~\eqref{eq:Hydro_hybrid} into the momentum balance (Eq.~\eqref{eq:NS_01}), leading to:
\begin{equation}\label{eq:macro} 
    \rho\frac{d\mathbf{v}}{dt}=-\nabla p + \nabla\cdot\left(2\eta_{\rm s} \mathbf{d}\right) + \frac{\alpha}{\lambda^*}\nabla\cdot\langle\boldsymbol{\pi}'_{p}\rangle+ \mathbf{f}.
\end{equation}
Here, \( d/dt \) denotes the material derivative, which represents the change in velocity of a fluid parcel as it moves through space and time.

\subsection{Macroscale: Smoothed Particle Hydrodynamics}
At the macroscale, the fluid is discretized into $I$ particles with their positions $\mathbf{r}_I$ and velocities $\mathbf{v}_I$. For clarity, we use uppercase indices $I$ and $J$ to denote the particles in the macroscopic domain. A thermodynamically consistent SPH formulation of the continuity and momentum equations, following Español, Revenga \& Ellero ~\cite{Espanol2003,Ellero2018} and consistent with Eq.~\eqref{eq:macro}, is given by:
\begin{equation} \label{eq:sph_continuity}
    m_I\frac{d\mathbf{v}_I}{dt} = \sum_J \left[\frac{p_I}{d_I^2} + \frac{p_J}{d_J^2}\right] F_{IJ}\mathbf{r}_{IJ} 
- \left[a\mathbf{v}_{IJ} + b(\mathbf{v}_{IJ}\cdot\mathbf{e}_{IJ})\mathbf{e}_{IJ}\right]\frac{F_{IJ}}{d_I d_J}
- \underbrace{\overbrace{\frac{\alpha}{\lambda^*}\langle\boldsymbol{\pi}'_{p}\rangle_{IJ}}^{\bar{\boldsymbol{\pi}}_{p,IJ}}F_{IJ}\mathbf{r}_{IJ}}_{\text{Micro-to-Macro Coupling}}.
\end{equation}
Here, \(\mathbf{r}_{IJ} = \mathbf{r}_I - \mathbf{r}_J\), \(\mathbf{v}_{IJ} = \mathbf{v}_I - \mathbf{v}_J\), and \(\mathbf{e}_{IJ} = \frac{\mathbf{r}_{IJ}}{\|\mathbf{r}_{IJ}\|}\). The mass of particle \(I\) is denoted as \gls{mI}, and the friction coefficients are given by \(a = 2\eta_s - \zeta_s\) and \(b = 4\zeta_s + 2\eta_s\), which are related to the solvent shear (\gls{eta_s}) and bulk (\gls{zeta}) viscosities. The particle number density is expressed as \(d_I = \sum_{J=1}^N W(\|\mathbf{r}_{IJ}\|, h)\), leading to the definition \gls{rhoI}\(= m_Id_I\) for the explicit summation density conservation~\cite{Espanol2003}. The kernel gradient is calculated as \(F_{IJ} = -\nabla W_{IJ}/\|\mathbf{r}_{IJ}\|\). A bell-shaped kernel with compact support, \gls{h}$=4$\gls{dh}, is used where \gls{dh} is the mean SPH particle spacing, specifically adopting the Lucy kernel (\ref{AppSPH}). 

The last term on the right-hand side of Eq.~\eqref{eq:sph_continuity} corresponds to an inter-particle average approximation that incorporates the polymeric stresses, $\langle\boldsymbol{\pi}'_{p}\rangle_{IJ}=\left({\langle\boldsymbol{\pi}'_{p}}\rangle_{I}/{d_I^2} + {\langle\boldsymbol{\pi}'_{p}\rangle_{J}}/{d_J^2}\right)$, where the stresses of particles $I$ and $J$ are obtained from the microscale through the \gls{IK} formulation, thereby facilitating the micro-to-macro stress coupling. The pressure of particle $I$ (\gls{pI}) is governed by the Tait equation of state, expressed as:

\begin{equation} \label{eq:tait}
p_I = \frac{c^2 \rho_0}{7} \left[ \left(\frac{\rho_I}{\rho_0}\right)^7 - 1 \right] + p_b.
\end{equation}
In this context, \gls{c} represents the speed of sound, while \gls{rho0} denotes the reference density. The background pressure is defined as \(p_b = \theta c^2 \rho_0 / 7\) (with \(\theta < 1\)), which ensures that \(p_I > 0\). For numerical stability, it is essential that the Mach number \({\rm Ma} = \bar{v}/c \leq 0.1\) (where \(\bar{v}\) is the average characteristic velocity). This condition maintains the fluid's weak compressibility and enables feasible time steps (refer to \ref{SoundAn}).

\subsection{Microscale: Polymer Melt Model — DPD and FENE} \label{subsec:micro}
The \gls{DPD} method, initially proposed by Hoogerbrugge and Koelman (1992)~\cite{Hoogerbrugge2007}, is employed here as a mesoscopic simulation technique in which clusters of molecules are represented as Lagrangian particles. It enables simulations over time and length scales often inaccessible to atomistic \gls{MD}~\cite{Groot1997b, Fedosov2010, Moreno2014b, Moreno2015a}. Within the \gls{LHMM} framework, polymer melts at the microscale are modeled using \gls{DPD} particles representing coarse-grained polymer chains or their constituent beads. To maintain clarity in our multiscale notation, lowercase indices $i$ and $j$ are used for particles in the microscopic domain. Chain connectivity and finite extensibility are ensured through the \gls{FENE} potential between DPD polymer beads~\cite{Litvinov2014,Lei2017a}. This combined \gls{DPD}/\gls{FENE} approach captures viscous fluid behavior via dissipative and random forces, conserves momentum, maintains constant temperature, and provides the microstructural information required for bottom-up stress calculations.

Each \gls{DPD} particle $i$, with mass $m_i$, evolves according to Newton’s equations of motion:
\begin{equation}
m_i \frac{d\mathbf{v}'_i}{dt} = \mathbf{F}_i^{\text{DPD}} + \sum_{k \in \text{bonds to } i} \mathbf{F}_{i,k}^{\text{FENE}},
\label{eq:motion}
\end{equation}
where $\mathbf{F}_i^{\text{DPD}}$ is the sum of all non-bonded interactions with other particles $j$:
\begin{equation}
\mathbf{F}_i^{\text{DPD}} = \sum_{j \neq i} \left( \mathbf{F}_{ij}^C + \mathbf{F}_{ij}^D + \mathbf{F}_{ij}^R \right).
\label{eq:F_DPD}
\end{equation}
The pairwise forces are defined as follows. 
The conservative force:
\begin{equation}
\mathbf{F}_{ij}^C = a_{ij} \left(1 - \frac{\|\mathbf{r}'_{ij}\|}{r_c}\right) \mathbf{e}'_{ij}, 
\quad \|\mathbf{r}'_{ij}\| < r_c,
\label{eq:conservative}
\end{equation}
where \gls{aij} is the conservative force amplitude and \gls{rc} is the cutoff radius. 
The dissipative force:
\begin{equation}
\mathbf{F}_{ij}^D = -\gamma^D \, w^D(\|\mathbf{r}'_{ij}\|) \, (\mathbf{e}'_{ij}\cdot\mathbf{v}'_{ij}) \, \mathbf{e}'_{ij},
\label{eq:dissipative}
\end{equation}
with \gls{gammaDPD} the dissipation coefficient, $w^D(\|\mathbf{r}'_{ij}\|)$ its weight function, \(\mathbf{r}'_{ij} = \mathbf{r}'_i - \mathbf{r}'_j\), \(\mathbf{v}'_{ij} = \mathbf{v}'_i - \mathbf{v}'_j\), and \(\mathbf{e}'_{ij} = \frac{\mathbf{r}'_{ij}}{\|\mathbf{r}'_{ij}\|}\). 
The random force:
\begin{equation}
\mathbf{F}_{ij}^R = \sigma^R \, w^R(\|\mathbf{r}'_{ij}\|) \, \theta_{ij} \, \mathbf{e}'_{ij},
\label{eq:random}
\end{equation}
where \gls{sigmaDPD} is the random force amplitude, $w^R(\|\mathbf{r}'_{ij}\|)$ its weight function, and $\theta_{ij}$ a Gaussian random variable with zero mean and unit variance. 
All pairwise forces vanish beyond the cutoff distance \gls{rc}. To ensure momentum conservation and maintain a constant temperature, the fluctuation–dissipation theorem \cite{Espanol1995} stipulates that \( w^D(\|\mathbf{r}'_{ij}\|) = \left[ w^R(\|\mathbf{r}'_{ij}\|) \right]^2 \) and that \(\sigma^D = (2 \gamma^D k_B T)^{1/2¨} \). In these equations, \gls{T} denotes the temperature, while \gls{kB} represents the Boltzmann constant. Consequently, polymer chains are formed by connecting beads with \gls{FENE} springs. The bonded term in Eq.~\eqref{eq:motion} is expressed as follows:
\begin{equation}
\mathbf{F}_{ij}^{\text{FENE}}(\mathbf{r}'_{ij}) 
% = -\nabla U(\mathbf{r}_{ij}) 
= \frac{H_{\text{FENE}} \mathbf{r}'_{ij}}{1 - (\|\mathbf{r}'_{ij}\|/R_0)^2},
\label{eq:fene_force}
\end{equation}
where \gls{Hfene} is the spring constant and \gls{R0} the maximum bond extension. This potential captures the finite extensibility of polymer chains, enabling accurate predictions of shear-dependent and extensional viscosities. The DPD simulations utilize a reduced (dimensionless) unit system defined by the bead mass ($m_i$), cutoff radius (\gls{rc}), and thermal energy (\gls{kB}\gls{T}), with further details provided in Section~\ref{subsec:polymer_details}.

\subsection{Time-Dependent Dynamics and Relaxation}~\label{sec:Time-DepentLHMM}
Polymer melts are viscoelastic fluids characterized by a flow resistance (viscosity) that is strongly influenced by shear rate, temperature, and molecular structure~\cite{Schieber2020,Morii2021}. Coarse-grained \gls{DPD} models can be calibrated to accurately reproduce the microstructural dynamics of a polymer melt and predict its specific shear-thinning behavior~\cite{Litvinov2014,Fedosov2010}. This non-Newtonian response arises from the microscopic dynamics of polymer chains, making the \gls{Wi} a key dimensionless parameter in the study of viscoelastic fluid dynamics~\cite{NietoSimavilla2022,Sun2024}. It quantifies the ratio between elastic and viscous forces and is defined as:
\begin{equation} \label{eq:WI}
{\rm Wi} = \dot{\gamma} \lambda,
\end{equation}
where \gls{dgamma} is the characteristic shear rate, and \gls{lambda} is the material's characteristic relaxation time. The relaxation time, influenced by the microscopic configuration and reflected in the polymeric stress tensor (\gls{pip} in Eq.~\eqref{eq:Hydro_hybrid}), governs the transient response of the melt. It is determined through \gls{DPD} simulations using virtual rheometry, as described in Section~\ref{subsec:steady_shear}, for \gls{Wi} values ranging from 0.5 to 30.

To ensure consistency in timescaling within the \gls{LHMM} framework established in Eq.~\eqref{eq:pip}, the characteristic relaxation time at the macroscale (\(\bar{\lambda}\)) is related to the microscale measurement (\(\lambda'\)) by the time-scale ratio \(\lambda^* = \bar{\lambda}/\lambda'\) (Eq.~\eqref{eq:ratios}). For this study, we assume that the relaxation effects arising from the microstructural system of the polymeric melt, as simulated with \gls{DPD}, and the temporal coupling are configured to satisfy the condition \(\bar{\lambda} = \lambda'\), leading to \(\lambda^* = 1\). This assumption therefore reflects the long microstructured relaxation effects present in polymeric systems and facilitates temporal synchronization between the microscopic and macroscopic solvers, resulting in a cohesive viscoelastic response across the coupled system and enabling the characterization of a global rheological parameter. As both levels of description use identical relaxation dynamics at the microscale, we will refer to the relaxation time as \(\lambda\) from this point forward.

\subsubsection{Temporal Coupling Scheme and Invariant micro-macro Weissenberg Number (\gls{Wi})}~\label{sec:Time-Coupling}
In the \gls{LHMM}, maintaining consistent temporal resolution is critical when coupling \gls{SPH} for macroscopic scales and \gls{DPD} for microscopic scales. Thus, given the macroscopic Eq.~\eqref{eq:sph_continuity} and microscopic Eq.~\eqref{eq:motion} evolution equations, the information exchange occurs at every exchange time step (\gls{delta_t}). The \gls{SPH} time step (\gls{Delta_t_SPH}) governs the macroscopic flow evolution, while the \gls{DPD} and \gls{FENE} time step (\gls{Delta_t_DPD}) resolve the dynamics of the finer microscopic structures. A fundamental requirement for accurately capturing microscopic relaxation processes is that the \gls{DPD} time step must be significantly smaller than the characteristic relaxation time, i.e., $\Delta t'_{\text{DPD}} \ll \lambda$.

Our proposed Continuous Intermittent (CI) method~\cite{Lockerby2013} effectively manages this temporal consistency. The exchange time step (\gls{delta_t}) defines the fixed interval at which information is transferred between the macroscopic and microscopic models~\cite{Moreno2023}. This interval is expressed as \(\delta t = \bar{M} \cdot \Delta \bar{t}_{\text{SPH}} = k' \cdot \Delta t'_{\text{DPD}}\), where \gls{M} and \gls{K} are positive integers. To accurately map transient behaviors between scales and ensure proper capture of relaxation effects, a hierarchical relationship must be satisfied: $\delta t \ll \lambda$. This setup allows the microscopic \gls{DPD} model to evolve continuously, capturing relaxation effects, while \gls{SPH} resolves macroscopic variations.

Due to the ongoing execution of the microscale DPD simulation and its intermittent sampling by the macroscale SPH at defined time intervals (\(\delta t\)), the relaxation time (\(\lambda\)) derived from microscopic dynamics is consistently aligned across both scales and is systematically mapped to the physical time step of the SPH simulation through the integers \gls{K} and \gls{M}. Both parameters must satisfy the condition \(\delta t = \bar{M} \cdot \Delta \bar{t}_{\text{SPH}} = k' \cdot \Delta t'_{\text{DPD}}\). This direct scaling ensures that the relaxation time determined by DPD is implicitly reflected in macroscopic behavior, fulfilling the assumption of \(\lambda^* = 1\). Consequently, a significant advantage of this methodology is the invariance of the Weissenberg number (Eq.~\eqref{eq:WI}) across scales: \(\text{Wi}_{\text{SPH}} = \text{Wi}_{\text{DPD}}\). 

The consistency of the characteristic time scale, combined with the definition of the macroscopic-to-microscopic rate-of-strain tensor \gls{dbar}$=\lambda^*\mathbf{d}'$ as a locally averaged variable coupling the scales, ensures that \gls{Wi} remains effectively identical across them. A detailed analysis of the influence of different \gls{delta_t} values and of the $k'\cdot\Delta t_{\text{DPD}}=\bar{M}\cdot\Delta t_{\text{SPH}}$ relationships is presented in Section~\hyperref[SM]{SM5} for various test cases in the \hyperref[SM]{Supplementary Material}.

\subsection{Macro-to-Micro Coupling}~\label{sub_micro_bc}
The coupling between macroscopic and microscopic scales (top-down) occurs at discrete time intervals defined as $\delta t = \bar{M}\Delta \bar{t}_{\rm SPH}$. At each interval $\delta t$, the macroscopic velocity gradient, denoted as \gls{grad_vI}, is computed at the position of \gls{SPH} particle $I$ and subsequently assigned to the corresponding microscopic domain \gls{Omega}. This procedure, proposed and validated by Moreno \& Ellero (2023) in the context of \gls{LHMM}~\cite{Moreno2023}, determines \gls{grad_vI} through the SPH kernel approximation:
\begin{equation} \label{eq:macro_micro}
\nabla \mathbf{v}_I = \sum_J F_{IJ} \mathbf{r}_{IJ} \mathbf{v}_{IJ}.
\end{equation}

For consistency, the \gls{grad_vI} defining the boundary condition of the microscopic simulations must be rescaled by the time-scale ratio $\lambda^*$ to ensure synchronization between levels, i.e., $\nabla\mathbf{v}'_I = \lambda^* \nabla\mathbf{v}_I$. The microscopic domain (\gls{Omega}) applies the \gls{AFBC} methodology adapted from Moreno \& Ellero~\cite{Moreno2021, Moreno2023} for \gls{DPD}/\gls{FENE} implementations. The \gls{AFBC} divides \gls{Omega} into three regions: the core region (\gls{Omega_core}), the boundary-condition (BC) region (\gls{Omega_bc}), and the buffer region (\gls{Omega_0}). For each microscopic DPD particle $i$ within \gls{Omega_bc}, the reconstructed microscopic velocity field, denoted $\mathbf{v}_i'$, is obtained using the macroscopic velocity gradient from Eq.~\eqref{eq:macro_micro}: 
\begin{equation} \label{eq:micro_bc}
\mathbf{v}_i' = \mathbf{r}_i' \cdot \lambda^*\nabla \mathbf{v}_I, \quad \forall i \in \Omega'_{\rm bc},
\end{equation}
where \gls{ri_prime} denotes the position of microscopic particle \( i \) relative to the center of mass of its micro-domain. The velocity gradient \gls{grad_vI} corresponds to that derived from the macroscopic particle \( I \) (as defined in Eq.~\eqref{eq:macro_micro}) at each coupling interval \(\delta t = \bar{M}\Delta t'_{\rm SPH}\). The application of \gls{vi_prime} to \gls{Omega_bc} establishes the macro-to-micro (top-down) coupling. In particular, the macroscopic rate-of-strain tensor \gls{dbar}, defined in Eq.~\eqref{eq:Hydro_hybrid}, is effectively imposed on the microscopic fluid via the corresponding \gls{grad_vI}. Further details on the coupling protocol are provided in \ref{subsec:coupling_protocol} and in~\cite{Moreno2021,Moreno2023}.

\subsection{Micro-to-macro Coupling}~\label{sub_micro_IK}
The Irving–Kirkwood (\gls{IK}) formalism serves as the foundational framework for transferring information from microscale DPD/FENE simulations to macroscale SPH particles \cite{Moreno2021,Moreno2023}. Central to the bottom-up approach, the \gls{IK} formalism allows for the direct computation of the macroscopic polymeric stress tensor (\gls{pip}), which functions as the micro-to-macro coupling term in the SPH momentum equations as approximated in Eq.~\eqref{eq:sph_continuity}. This process is integrated over the temporal interval \gls{delta_t}, capturing the microstructural response of the polymer melt. The formalism maps statistical averages of microscopic quantities—such as particle positions, velocities, and interparticle forces—within the microscale core region (\gls{Omega_core}) to a representative contribution in the macroscopic stress tensor (\gls{pibar}).

For a given macroparticle SPH \(I\), its associated polymeric stress tensor \gls{pip}\(_{I}\) is obtained by microscopically averaging the \gls{DPD}/\gls{FENE} interactions within its corresponding micro-domain. That is, \gls{pip}\(_{I}=\langle\boldsymbol{\pi}'^{\rm IK}_{p}\rangle_I\). This local stress tensor within \gls{Omega_core}, denoted as:
\begin{equation} \label{eq:IKtotal}
    \langle\boldsymbol{\pi}'^{\rm IK}_{p}\rangle_I(\mathbf{x}',t) = \langle\boldsymbol{\pi}'^{\rm kin}_{p}\rangle_I(\mathbf{x}',t) + \langle\boldsymbol{\pi}'^{\rm pot}_{p}\rangle_I(\mathbf{x}',t).
\end{equation}
The kinetic stress (\(\langle\boldsymbol{\pi}'^{\rm kin}_{p}\rangle_I\)) accounts for momentum transport resulting from the thermal motion and flow-induced fluctuations of individual \gls{DPD} beads. It captures the energetic and diffusive aspects of the fluid's response to deformation, and it is computed as:
\begin{equation} \label{eq:kinetic_stress}
\langle\boldsymbol{\pi}'^{\rm kin}_{p}\rangle_I = -\frac{1}{N'_t} \sum_{n=1}^{N'_t} \left[
\sum_i m_i\, w_{\text{IK}}(\mathbf{r}'_i(n)-\mathbf{x}')\, \Delta\mathbf{v}'_i(n) \otimes \Delta\mathbf{v}'_i(n)
\right]
\end{equation}
Where \(N'_t \) is the number of microscopic time steps used in the interval \(\delta t' = k' \Delta t_{\rm DPD}\). The quantity \(\Delta\mathbf{v}'_i(n)\) is the velocity of particle \(i\) relative to the local streaming velocity at time step \(n\), and \(w_{\text{IK}}(\mathbf{r}'_i(n)-\mathbf{x}')\) is a spatial weighting function that projects microscopic quantities onto the macroscopic point \(\mathbf{x}\).

The potential stress (\(\langle\boldsymbol{\pi}'^{\rm pot}_{p}\rangle_I\)) arises from interparticle forces. This includes the \gls{FENE} potential, which governs the elasticity and finite extensibility of polymer bonds. It also accounts for non-bonded stochastic hydrodynamic interactions such as \gls{DPD} conservative, dissipative, and random forces. These contributions reflect the stored elastic energy and short-range interactions that characterize the melt's viscoelastic behavior. Assuming a central force decomposition where \(\mathbf{f}_{ij} = \varphi_{ij} \mathbf{e}_{ij}\), the potential stress is given by:
\begin{equation} \label{eq:potential_stress}
\langle\boldsymbol{\pi}'^{\rm pot}_{p}\rangle_I = \frac{1}{2N'_t} \sum_{n=1}^{N'_t} \left[
\sum_{\substack{i,j \\ i \neq j}} \mathbf{f}_{ij}(n) \otimes \mathbf{r}'_{ij}(n)\, \mathcal{B}(\mathbf{x}; \mathbf{r}'_i(n), \mathbf{r}'_j(n))
\right]
\end{equation}
The bond function \(\mathcal{B}(\mathbf{x}'; \mathbf{u}', \mathbf{v}')\) integrates the weighting function \(w_{\text{IK}}\) along the path connecting the two particles:
\begin{equation}
\mathcal{B}(\mathbf{x}'; \mathbf{u}', \mathbf{v}') = \int_0^1 w_{\text{IK}}\left((1-s)\mathbf{u}' + s\mathbf{v}' - \mathbf{x}'\right) ds
\end{equation}
This formulation yields the Hardy stress. Different choices of the weighting function \(w_{\text{IK}}\) recover other well-known microscopic stress definitions. For more details, we refer the reader to the work of Moreno \& Ellero (2021)~\cite{Moreno2021}.

\section{LHMM Methodology Algorithm} \label{sec:algorithm}
The \gls{LHMM} proposed in this work operates as a two-way coupling methodology. The macroscale kinematics informs the microscale model at a specific, constant frequency defined by the coupling time step \gls{delta_t} as a top-down coupling. Conversely, it is bottom-up since the stress computed via \gls{IK} formalism at the microscale closes the macroscopic conservation equation by the same coupling time step \gls{delta_t}. The coupling between models at different scales is achieved through data exchange \cite{Smith2023,Matous2017}, specifically stress tensors (\gls{pip}) from the microscale and velocity gradients (\gls{grad_vI}) from the macroscale. These top-down and bottom-up approach contrasts with methods that link scales directly through their solution fields \cite{Larson2015,King2021}.

The \gls{LHMM} framework necessitates careful synchronization between the time integration schemes of the macroscopic (\gls{SPH}) and microscopic (\gls{DPD}/\gls{FENE}) solvers, adhering to principles established in earlier studies \cite{Moreno2023, Moreno2021}. Given the inherent differences in time scales—where \gls{DPD}/\gls{FENE} captures the dynamics of polymers at the microstructural level while \gls{SPH} addresses macroscopic flow—a time-stepping strategy is employed to ensure both numerical stability and physical consistency. This strategy is governed by the relationship \(\delta t = \bar{M} \cdot \Delta \bar{t}_{\text{SPH}} = k' \cdot \Delta t'_{\text{DPD}}\). It builds upon concepts discussed in previous Sections~\ref{sec:Time-DepentLHMM}.

\subsection{Implementation Framework and Workflow} \label{subsec:flowchart_overview}
The \gls{LHMM} multiscale coupling scheme employs a C++/MPI architecture that coordinates LAMMPS simulations \cite{Xu2024,Hue2025}. The overall workflow and data exchange between scales are summarized in Figure~\ref{fig:LHMM_Flowchart} and detailed in \ref{sec:Flowchart}, while the step-by-step coupling protocol is provided in \ref{subsec:coupling_protocol}. Together, these appendices complement the description below by offering both a schematic overview and the full operational procedure.

The \gls{LHMM} architecture achieves strong scaling performance through three main features: (1) dedicated MPI sub-communicators connecting macro- and micro-domains (Fig.~\ref{fig:LHMM_Flowchart}), (2) suitable GPU-accelerated for the integration of multiple microscopic computations in parallel distribution \cite{Nguyen2015,Thompson2022}, and (3) direct tensor transfer via the LAMMPS C++ interface \cite{Oda2025,Xie2023}. The full implementation is available at \url{https://github.com/BCAM-CFD/LHMM/tree/main/LHMM-PolymerDPD}. The computational performance and scalability of the coupling scheme were evaluated across various hardware architectures, encompassing both CPU and GPU computational resources. This approach involved managing multiple coupling and time-scaling levels to ensure numerical stability and accuracy. Our implementation adeptly accommodates a particle range from \(1 \cdot 10^7\) to \(1 \cdot 10^8\), integrating all particles from each microsystem with the macro particles within the framework of High-Performance Computing (HPC) nodes and workstations. Detailed benchmarks—including CPU/GPU comparisons and single-GPU performance—are detailed in Section~\hyperref[SM]{SM10} in the \hyperref[SM]{Supplementary Material}.

The GPU-enabled \texttt{DPD/GPU} module (\ref{sec:Flowchart}) further boosts microscale performance by enabling the parallel execution of independent microdomains. Since each micro simulation proceeds in a decoupled manner, the domains can be distributed across multiple NVIDIA GPUs with minimal communication overhead. This strategy noticeably outperforms single-GPU setups and provides a scalable pathway toward multi-GPU implementations for large-scale simulations.

\subsection{Microscale Domain and Boundary Conditions} \label{sec:boundary_regions}
\begin{figure}[H]
    \centering
    \includegraphics[scale=0.45]{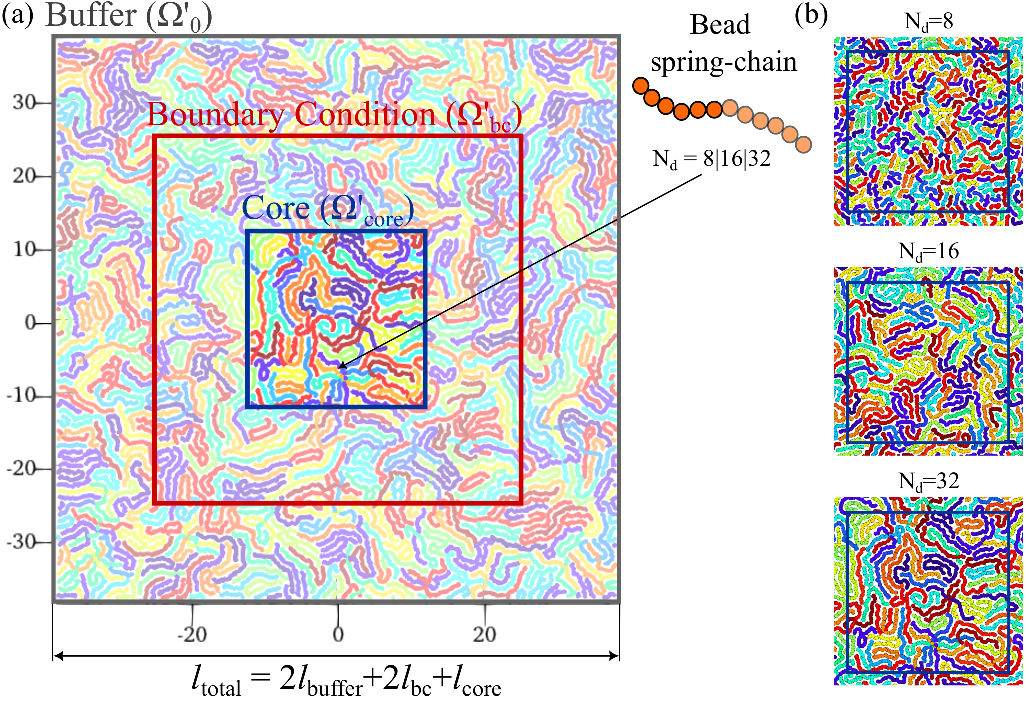}
    \caption{(a) Sketch of the Arbitrary Flow Boundary Conditions and their definition of buffer (\gls{Omega_0}), Boundary Condition (\gls{Omega_bc}), and Core (\gls{Omega_core}) regions for the \gls{DPD}/\gls{FENE} simulation. This diagram also illustrates how each micro-domain is delimited within the \gls{LHMM} framework for monodisperse bead-spring chains (polymer melts). Here, \gls{l_total} is the total length of the arbitrary flow boundary setup, \gls{l_core} is the length of the core region, \gls{l_bc} is the length of the boundary condition region, and \gls{l_buffer} is the length of the buffer region. (b) Details of the initial configuration within the core regions for polymer melts with different chain lengths, \gls{Nb}$\in \{8, 16, 32\}$.}
    \label{Fig:1}
\end{figure}

For all cases, the polymer melt system considered here consists solely of flexible, monodisperse bead-spring chains with $N_{\rm d} \in \{8, 16, 32\}$ beads connected by \gls{FENE} springs. As polymer melts, these systems contain no solvent particles, and chains are distinguished solely by their bead numbers. The computational domain for these microscale \gls{DPD} simulations, defining the individual micro-domains within the \gls{LHMM} framework, is configured using \gls{DPD} units as illustrated in Fig.~\ref{Fig:1}(a). This domain implements an \gls{AFBC} to facilitate the coupling between the macroscopic and microscopic scales \cite{Moreno2021,Moreno2023}. Table \ref{Tab:2} outlines the dimensions of the core, buffer, and boundary condition regions specific to this setup.

\begin{table}[H]
    \centering
    \caption{Dimensions of the \gls{DPD} simulation regions for Arbitrary Flow Boundary Conditions (\gls{AFBC}) used with monodisperse bead-spring chains. \gls{l_core} is the length of the core region, \gls{l_bc} is the length of the boundary condition region, \gls{l_buffer} is the length of the buffer region, and \gls{l_total} is the total domain length. For more details, see Fig.~\ref{Fig:1}(a).}
    \begin{tabular}{l*{4}{c}}
    \hline 
    $l_{\text{core}}$ & $l_{\text{bc}}$ & $l_{\text{buffer}}$ & $l_{\text{total}}$ \\
    \hline 
    24  & 13  & 13  & 76 \\
    \hline 
    \end{tabular}
    \label{Tab:2}
\end{table}
  
As illustrated in Fig.~\ref{Fig:1}(a), this computational domain employs three specialized regions to model complex flow configurations and enable the micro-macro coupling:
\begin{itemize}
    \item The \textit{Core region} (\gls{Omega_core}) represents the active simulation area where the fluid's physical behavior is simulated and where the polymer melt's viscometric properties are measured. Specifically, the average polymer stress \gls{pip} is computed through the \gls{IK} formalism (Eq.~\eqref{eq:IKtotal}). As shown in Fig.~\ref{Fig:1}(b), this panel displays the initial configuration within the \gls{Omega_core} for polymer melts with different chain lengths, $N_{\rm d} \in \{8, 16, 32\}$. 
    \item The \textit{Boundary-Condition region} (\gls{Omega_bc}) is crucial for the top-down coupling. It imposes target velocity fields through interpolation functions, specifically utilizing Eq.~\eqref{eq:micro_bc} (Section~\ref{sub_micro_bc}) for the transfer of discrete macroscopic velocity gradients (\gls{grad_vI}) from the associated macro-particle. This region ensures smooth velocity transitions and allows for the specification of arbitrary flow profiles, including both open and closed boundary configurations.
    \item The \textit{Buffer region} (\gls{Omega_0}) is designed to stabilize mass and momentum fluxes between periodic boundaries and the core domain. This region effectively acts as a thermal reservoir, with its capacity primarily controlled by adjusting its size (\gls{l_buffer}) or modulating its temperature \cite{Moreno2021}. While larger buffer regions ensure greater stability at increased computational cost, temperature modulation offers an alternative for maintaining equilibrium without requiring an excessive number of particles. This design is crucial for handling strong velocity gradients while preventing disruption of the core region dynamics.
\end{itemize}

\subsection{Polymer Melt Model Details} \label{subsec:polymer_details}
Table \ref{Tab:1} presents the set of \gls{DPD} and \gls{FENE} simulation parameters (referencing Eqs.~\eqref{eq:motion}--\eqref{eq:fene_force}) adopted for all microsystems investigated in this work. These parameters are consistent with those successfully employed in previous studies for polymer melts \cite{Litvinov2014,Fedosov2010} and were applied uniformly across both the microscale characterization and the \gls{LHMM} simulations for all presented cases.

\begin{table}[H]
    \centering
    \caption{\gls{DPD} and \gls{FENE} simulation parameters for monodisperse bead-spring chains. \gls{kB} is the Boltzmann constant, \gls{T} is the temperature, \gls{aij} is the conservative force coefficient, \gls{gammaDPD} is the amplitude of the dissipative force, \gls{mDPD} is mass of a DPD particles, \gls{nDPD} is the DPD number density, and \gls{dx} is the average initial distance between beads. \gls{R0} is the maximum extension of \gls{FENE} springs, \gls{rc} is the \gls{DPD} cutoff radius, and \gls{Hfene} is the \gls{FENE} spring constant. \gls{Delta_t_DPD} is the \gls{DPD} time step.}
    \begin{tabular}{l*{9}{c}}
    \hline 
    $k_{b}T$ & $a_{ij}$ & $\gamma^D$ & $m_{\rm DPD}$ & $n_{\rm DPD}$ & $dx=n^{-1/2}$ & $R_{0}$ & $r_{c}$ & $H_{\text{FENE}}$ & $\Delta t_{\rm DPD}$\\
    \hline 
    1.0 & 25 & 4.5 & 1 & 4 & 0.5 & 1.5 & 1.0 & 50 & 0.005 \\
    \hline 
    \end{tabular}
    \label{Tab:1}
\end{table}

The \gls{DPD} method makes use of reduced units to ensure the generality of simulation results. In this framework, physical quantities are made dimensionless by normalising them with characteristic scales derived from a set of base units~\cite{Serrano1999,Groot1997}. 
For this work, we define the particle mass, cutoff radius, and thermal energy as base units: \(m_{\rm DPD} = 1\), \(r_c = 1\), and \(k_B T = 1\) (Table~\ref{Tab:1}). From these, we derive the characteristic scales for time, \(\tau_{\rm DPD} = r_c \sqrt{m_{\rm DPD}/(k_B T)}\), and stress, \(\pi_{\rm DPD} = k_B T / r_c^3\). Since the base units are unitary, these scales also become one, simplifying the interpretation of results. All reported quantities are thus implicitly dimensionless. For example, the integration step \(\Delta t_{\rm DPD}=0.005\) in Table~\ref{Tab:1} is equivalent to \(0.005\,\tau_{\rm DPD}\), and all stress tensor components are normalized by \(\pi_{\rm DPD}\). This systematic non-dimensionalisation is essential for the consistent reporting and subsequent use of the \gls{DPD} data in our multiscale coupling scheme.

\section{Rheological characterization} \label{sec:micro_results}
The local rheological characterization is conducted at the microscopic level using \gls{DPD}/\gls{FENE} simulations, specifically designed to show the complex rheological behavior of our sample method. These preliminary simulations are conducted within a single micro-domain, whose configuration is illustrated in Fig.~\ref{Fig:1} and detailed in Table~\ref{Tab:2}. Within this specific \gls{AFBC} micro-domain, which functions as a rheometer, the prescribed pure shear and extensional deformation rates in the \gls{Omega_bc} dictate the flow behavior. Subsequently, the rheological response is measured using the \gls{IK} stress tensor within the \gls{Omega_core}. This methodology is based on the approach developed by Moreno \& Ellero (2021)~\cite{Moreno2021}, with detailed implementation specifics for the current work provided in Section~\hyperref[SM]{SM1} in the \hyperref[SM]{Supplementary Material}. For further insights into the definition and measurement of planar extensional viscosity, please refer to Section~\hyperref[SM]{SM2} in the \hyperref[SM]{Supplementary Material}.

\subsection{Rheological analysis melt: steady-state shear flow} \label{subsec:steady_shear}
This section details the rheological characterization of the proposed \gls{DPD}/\gls{FENE} model for polymer melts under steady shear flow. We investigated melts with chain lengths (\gls{Nb}) of 4, 8, 16, and 32 beads. The rheometer with \gls{AFBC} in the micro-domain determines the polymeric shear viscosity (\gls{eta_ps}) using Eq.~(\hyperref[SM]{S4}) (refer to the \hyperref[SM]{Supplementary Material}) by varying the shear rate \gls{gammadotxy} and the resulting polymeric \gls{pip} obtained from \gls{IK} (see Section~\hyperref[SM]{SM1} in the \hyperref[SM]{Supplementary Material})~\cite{Moreno2021}.

\begin{figure}[!h]
\centering
\includegraphics[scale=0.8]{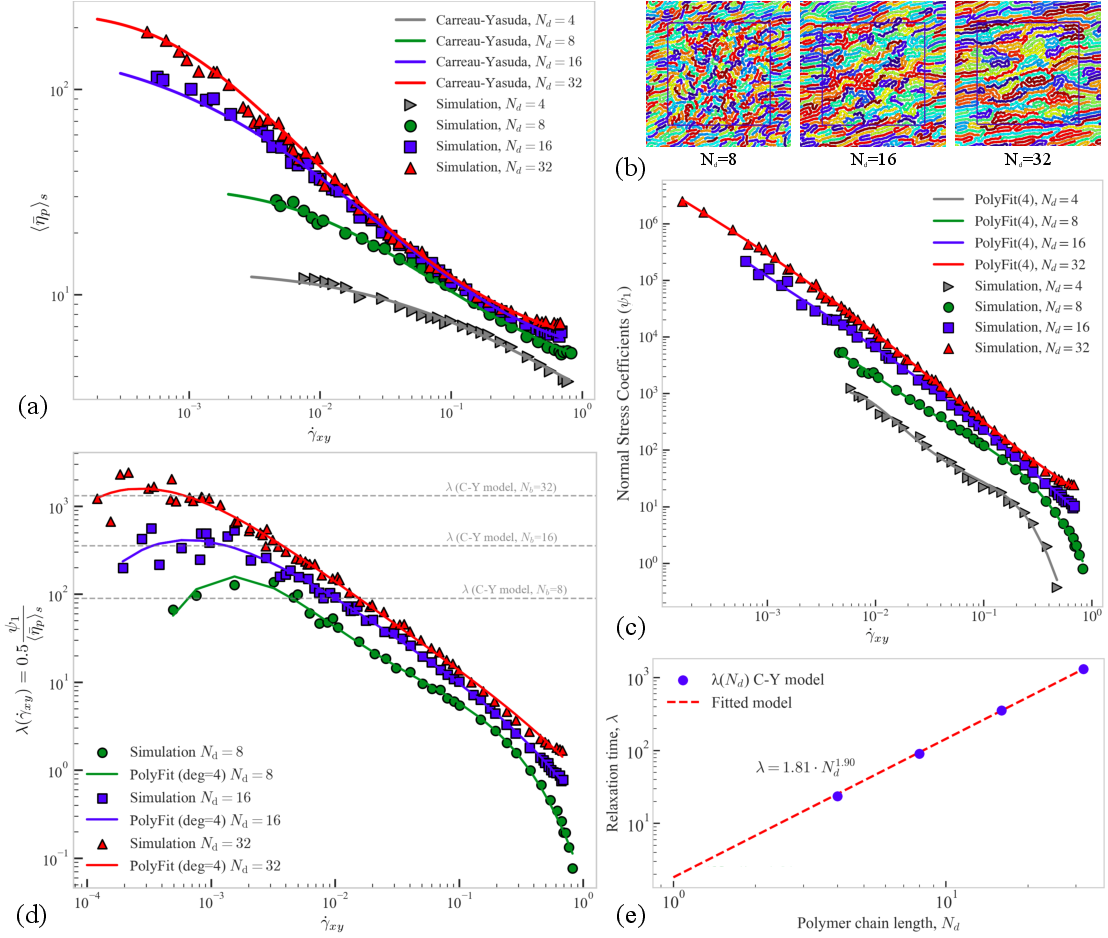}
\caption{(a) Carreau-Yasuda model validation for polymer melts under steady shear flow. Lines represent fitted viscosity using the Carreau-Yasuda model, and symbols correspond to \gls{DPD} simulation data for various polymer chain lengths (\(N_{\rm d}\)). (b) Details of the distribution of bead-chains within the \gls{Omega_core} regions for polymer melts with different $N_{\rm d}=[8;16;32]$ after reaching the shear rate steady-state of $\dot{\gamma}_{xy}=0.02$. (c) Normal stress coefficients (\(\psi_1\)) for polymers under steady shear flow. Lines represent fitted values based on the Carreau-Yasuda model, and symbols indicate \gls{DPD} simulation data for different polymer chain lengths (\(N_{\rm d}\)). (d) Validation of relaxation time using constant shear rate for \gls{Nb}\(= 8\) to \gls{Nb}\(= 32\). The asymptotic values of \(\lambda\) were determined as the shear rate (\(\dot{\gamma}_{xy} \to 0\)) approaches zero. (e) Relaxation Time (\(\lambda\)) as a function of polymer chain length (\gls{Nb}). The fitted curve shows a power-law relationship \(\lambda \sim N_{\rm d}^{1.90}\), consistent with the Rouse model scaling \cite{Goto2021,Murashima2011}.}
\label{Fig:2_VirtualRheo}
\end{figure}

The non-Newtonian viscous behaviour of these melts is well described by fitting the \gls{CY} model to our simulation data. As illustrated in Fig.~\ref{Fig:2_VirtualRheo}(a), \gls{DPD} simulation data for viscosity at various shear rates are fitted to this model, with optimized parameters for each polymer chain length. The Carreau-Yasuda (\gls{CY}) equation (Eq.~\eqref{eq:Yasuda}) describes this behavior, and its fitted parameters are summarised in Table~\ref{tab:yasuda_params}. Further details on the \gls{CY} model and its parameters are provided in \ref{app:CY_details}. The case with \gls{Nb}$=4$ is included as a reference point in our analysis. The \gls{DPD} and \gls{FENE} parameters used for these microsystems are detailed in Table~\ref{Tab:1}.

As depicted in Figure~\ref{Fig:2_VirtualRheo}(b), the polymer chains within the \gls{Omega_core} exhibit significant deformation and alignment under a steady-state shear rate of \gls{gammadotxy}=0.02. This shear-induced anisotropy, particularly evident for longer chains (\gls{Nb}=32), contrasts sharply with the initial isotropic configuration (Fig.~\ref{Fig:1}(b)) and highlights the pronounced impact of the flow on the polymer microstructure.

\begin{table}[ht]
\centering
\caption{Fitted Carreau–Yasuda parameters for different polymer chain lengths (\gls{Nb}). Here, \gls{eta0cy} denotes the zero-shear viscosity, \gls{etainf} the infinite-shear viscosity, \gls{lambda} the mean relaxation time, \gls{acy} the transition sharpness parameter, and \gls{ncy} the power-law index.}
\begin{tabular}{@{}cccccc@{}}
\toprule
\gls{Nb} & \(\eta_{\text{cy}}^0\) & \(\eta_{\text{cy}}^\infty\) & \(\lambda\) & \(a_{\text{cy}}\)  & \(n_{\text{cy}}\) \\ \midrule
4   & 13.20     & 0             & 23.50     & 0.71 & 0.60 \\
8   & 34.29     & 3.15          & 89.75     & 0.95 & 0.37 \\
16  & 173.18    & 4.45          & 353.48    & 0.55 & 0.18 \\
32  & 242.83    & 4.81          & 1307.88   & 1.23 & 0.29 \\ 
\bottomrule
\end{tabular}
\label{tab:yasuda_params}
\end{table}

The viscometric characterization of the DPD/FENE simulations provides a clear indication of the complex non-Newtonian response exhibited by the microscopic polymer sample, which directly informs the macroscopic model. The \gls{CY} model offers an excellent quantitative description of the simulation data, accurately capturing both the shear-thinning behavior and the Newtonian plateaus across all polymer chain lengths investigated (Fig.~\ref{Fig:2_VirtualRheo}(a)).

This microscale analysis is fundamental to the coupled LHMM framework, as it supplies the macroscopic solver with key rheological parameters. The zero-shear viscosity, \gls{eta0cy}, obtained from the CY fit (Table~\ref{tab:yasuda_params}), serves as the reference viscosity for defining the viscosity ratio \gls{beta} (Eq.~\eqref{eq:beta}), where \gls{eta0prime}$=$\gls{eta0cy}. Additionally, this analysis determines the characteristic relaxation time, \gls{lambda}, which plays a central role in defining the Weissenberg number (\gls{Wi}) and, consequently, in characterizing the viscoelastic response of the fluid across different flow regimes. The same relaxation time also establishes the scaling condition $\lambda^* = 1$, ensuring consistent coupling between micro- and macro-level dynamics.

\subsubsection{Normal Stress Coefficients and Relaxation Time}
The first normal stress coefficient (\gls{psi1}) is a key parameter for characterizing the elastic properties of viscoelastic fluids, such as polymer melts \cite{Fedosov2010, Litvinov2014}. Its definition, the first normal stress by the square of the applied shear rate, is detailed in \ref{app:NormalStressRelaxation}.

Figure~\ref{Fig:2_VirtualRheo}(c) illustrates the normal stress coefficients (\gls{psi1}) for polymers under steady shear flow. The symbols correspond to the raw simulation data for various polymer chain lengths (\gls{Nb}), and the lines represent a fourth-degree polynomial interpolation. These results exhibit trends similar to those reported in previous work~\cite{Fedosov2010, Padding2003}. To further validate the relaxation time obtained from the \gls{CY} model, we examined the asymptotic values of the \gls{lambda} as the \gls{gammadotxy} approaches zero. The calculation of this asymptotic relaxation time, as discussed by \cite{Fedosov2010}, is provided in~\ref{app:NormalStressRelaxation}.

Figure~\ref{Fig:2_VirtualRheo}(d) illustrates the \gls{lambda}, calculated according to Eq.~\eqref{eq:tau0_def_appendix} as \(\lambda = 0.5 \psi_1 / \eta'^0_p\). This figure highlights the asymptotic behavior for polymer chain lengths ranging from \gls{Nb}$=8$ to \gls{Nb}$= 32$. The results exhibit a strong correlation between the relaxation time \gls{lambda} predicted by the \gls{CY} model and \(\lambda(\dot{\gamma}_{xy} \to 0)\), thereby reinforcing the consistency and reliability of our simulation data in accurately depicting the viscoelastic behavior of polymer melts, especially within the zero-shear plateau region.

These findings underscore the robustness of the \gls{IK} approach for capturing not only shear stress but also normal stresses, demonstrating its reliability in describing the elastic properties of polymer melts, with results consistent with those in \cite{Fedosov2010, Padding2003}. This behavior is consistent with the predictions of the Carreau-Yasuda model. 

Following the \gls{CY} model analysis, we characterized the system's dynamics by examining how the characteristic relaxation time, \gls{lambda}, scales with the polymer chain length, $N_{\rm d}$. These $\lambda$ values were obtained from the \gls{CY} fits of our \gls{DPD} simulation data (Fig.~\ref{Fig:2_VirtualRheo}(a)) for \gls{Nb}$=[4;8;16;32]$. As shown in Fig.~\ref{Fig:2_VirtualRheo}(e), a power-law fit reveals that the relaxation time increases with chain length following \(\lambda \sim N_{\rm d}^b\). The fitted exponent \(b = 1.90\) is in close agreement with the theoretical prediction of the Rouse model \cite{Rouse1953}, which predicts $\lambda \sim N_{\rm d}^2$ for unentangled chains \cite{Kremer1990,Padding2002,Cohen-Addad1983}. This scaling confirms that our polymer chains behave as unentangled within the simulated micro-domains.

The slight deviation of the fitted exponent ($b=1.90$) from the expected Rouse prediction of $2.0$ can be attributed to several factors. One significant contributor is the finite chain length effects, as our simulations are limited to relatively short chains (\gls{Nb} ranging from 4 to 32 beads), which may display minor deviations from ideal asymptotic scaling \cite{Kalathi2015}. Additionally, the use of soft \gls{DPD} potentials, while inherently accounting for some hydrodynamic interactions, may induce slight correlations that could sightly alter the ideal Rouse scaling \cite{Groot1997b,Goto2021}. Nevertheless, despite these minor discrepancies, the observed scaling behavior clearly remains consistent with the Rouse regime.

\subsection{Relaxation Behavior Characterization} \label{subsec:relaxation}
To complement the steady-state rheological analysis, we conducted recovery tests to characterize the viscoelastic behavior of the polymer melts and to validate the characteristic relaxation time (\gls{lambda}) obtained from the \gls{CY} model. These tests involved initially subjecting the system to a constant shear rate ($\dot{\gamma}_{xy} \approx 1/\lambda$) until it reached a steady state. Subsequently, the external shear was abruptly removed, allowing the system to relax towards equilibrium. The decay of the shear stress, \gls{pip_xy} (the $xy$-component of \gls{pip}), was monitored over time through the relaxation modulus \gls{Gt}, defined as $G(t) = \langle\pi'_{p,xy}(t)\rangle/\gamma_{xy}$, where \gls{pip_xy}$(t)$ is the shear stress at time $t$ after the cessation of shear and $\gamma_{xy}$ is the total applied strain. Further details on the recovery test methodology and the evolution of \gls{Gt} are provided in Section~\hyperref[SM]{SM3} in the \hyperref[SM]{Supplementary Material}, including visual illustrations for different chain lengths.

The recovery tests provide qualitative validation with sample's properties calibrated from the \gls{CY} model's predictions, demonstrating an evident decay in shear stress after the external force is removed. This behavior is consistent with the higher intrinsic viscosity and longer characteristic relaxation time expected for longer polymer chains.

Further quantitative analysis of the linear viscoelastic behavior was performed using spectral analysis, specifically the \gls{ILT} and the \gls{GMM}. The methodologies for obtaining the storage modulus (\gls{Gprime}) and loss modulus (\gls{Gprime2}) from these methods are detailed in Section~\hyperref[SM]{SM4} in the \hyperref[SM]{Supplementary Material}.

\subsection{Comparative Analysis of Viscoelastic Properties} \label{subsubsec:comparative_analysis}

\begin{figure}[h]
    \centering
    \includegraphics[width=0.80\textwidth]{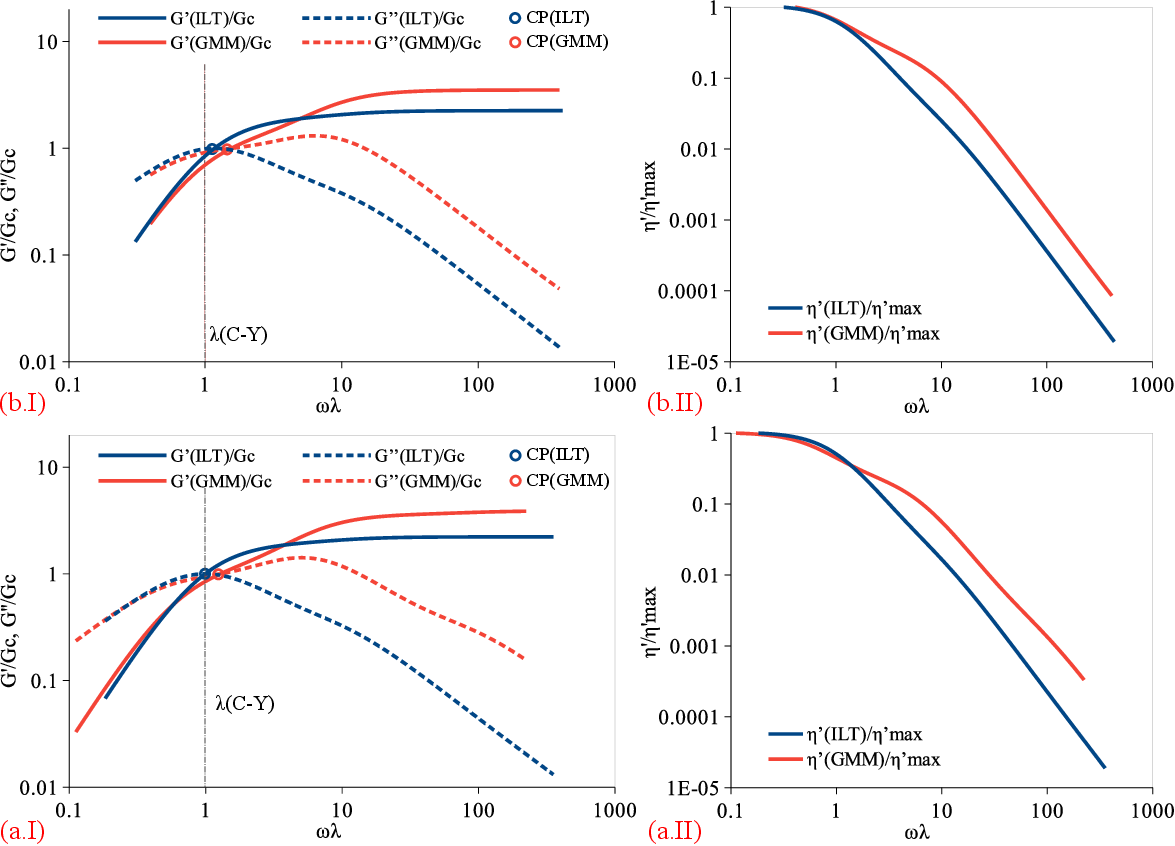}
    \caption{Comparison of viscoelastic properties for \gls{Nb}$=16$ (a) and \gls{Nb}$=32$ (b). (I) Storage modulus ($G'$) and loss modulus ($G''$) derived from \gls{GMM} and \gls{ILT}, normalized by their crossover point value $\text{G}_\text{c}$ (\gls{Gprime}=\gls{Gprime2}). (II) Complex viscosity \gls{eta_complex} from \gls{GMM} and \gls{ILT} (See Section~\hyperref[SM]{SM4} in the \hyperref[SM]{Supplementary Material}), normalized by the maximum complex viscosity $\eta'_{\text{max}}$. $1/\lambda$ is the inverse of the characteristic relaxation time from the \gls{CY} model, represented in Panels (I) as a vertical line, and CP is the crossover point.}
    \label{fig:complex_viscosity}
\end{figure}

To provide a comprehensive characterization of the material's viscoelastic properties, we perform a detailed comparative analysis of the results obtained from three distinct approaches: \gls{GMM}, \gls{ILT} (unsteady), and the \gls{CY} model fit (steady). The key findings from these methodologies are summarized in Table~\ref{tab:rheology_results_en} and visually represented in Figure~\ref{fig:complex_viscosity}. The table and figure include parameters from the \gls{CY} viscosity fit, as implemented and discussed in Section~\ref{subsec:steady_shear}.
\begin{table}[h!]
    \centering
    \caption{Virtual Rheology Results for \gls{Nb}$=16$ and \gls{Nb}$=32$, showing parameters from \gls{GMM}, \gls{ILT}, and \gls{CY} fitting. $\tau_i^{\text{GMM}}$ represents the $i$-th relaxation time constant. $\tau_j^{\text{Peak}}$ denotes the $j$-th peak relaxation time from the \gls{ILT} relaxation spectrum, \gls{Ht} (See Section~\hyperref[SM]{SM4} in the \hyperref[SM]{Supplementary Material}). Crossover Freq. indicates the frequency where the storage and loss moduli are equal (\gls{Gprime}=\gls{Gprime2}). $\lambda$ is the characteristic relaxation time from the \gls{CY} model.}
    \label{tab:rheology_results_en}
    \begin{tabular}{lccccccccccc}
        \toprule
        \multirow{2}{*}{$N_{\rm d}$} & \multicolumn{3}{c}{GMM} & \multicolumn{3}{c}{ILT} & \multicolumn{2}{c}{C-Y Fit} \\
        \cmidrule(lr){2-4} \cmidrule(lr){5-7} \cmidrule(lr){8-9}
        & $\tau_2^{\text{GMM}}$ & $\tau_3^{\text{GMM}}$ & Crossover Freq. & $\tau_1^{\text{Peak}}$ & $\tau_2^{\text{Peak}}$ & Crossover Freq. & $\lambda$ & $1/\lambda$ \\
        \midrule
        16 & 63.9 & 547.0 & 0.00364 & 40.3 & 379.7 & 0.00271 & 353.48 & 0.002829 \\
        32 & 168.4 & 1331.4 & 0.00126 & 130.3 & 1139.0 & 0.000902 & 1307.88 & 0.000765 \\
        \bottomrule
    \end{tabular}
\end{table}

Table~\ref{tab:rheology_results_en} and Figure~\ref{fig:complex_viscosity} highlight a strong compatibility between the \gls{GMM} and \gls{ILT} characterizations. For both samples, the relaxation times derived from \gls{GMM} ($\tau_i^{\text{GMM}}$) and \gls{ILT} ($\tau_j^{\text{Peak}}$) show a consistent shift towards slower processes as chain length increases. While absolute values may differ, their relative magnitudes and trends are remarkably consistent, a fact that is made more explicit in Figure~\ref{fig:complex_viscosity}. Similarly, the crossover frequencies (where $G'=G''$) consistently decrease for larger \gls{Nb} values, reflecting the longer relaxation times.

A key finding is also the excellent agreement between the inverse of the characteristic relaxation time from the \gls{CY} fit ($1/\lambda$) and the crossover frequency from the \gls{ILT} analysis. This alignment reinforces the physical relevance of the \gls{ILT}-derived crossover frequency as a characteristic relaxation rate of the material, consistent with established empirical models \cite{Yao2020}.

Furthermore, both \gls{GMM} and \gls{ILT} accurately describe the complex viscosity curves, particularly the identification of an apparent zero-shear viscosity plateau at low frequencies and the subsequent shear-thinning behavior at higher frequencies, as shown in Figure \ref{fig:complex_viscosity} (a.II)-(b.II). The consistency observed across these three distinct analysis (discrete \gls{GMM}, continuous \gls{ILT}, and \gls{CY} model fit) strengthens the confidence in our virtual rheological characterization, confirming that our simulated systems exhibit well-defined viscoelastic properties consistent with established theoretical frameworks.

% % 5. Numerical results: macroscopic flow
\section{Numerical Results: Macroscopic Flow} \label{sec:macro_results}
This section presents the numerical results of the macroscopic flow simulations, which validate the large-scale response and explore the influence of microstructures within the \gls{LHMM} framework. Three benchmark cases are considered: \gls{RPF}, a \gls{PAC}, and flow through a \gls{RPM}. These cases are used to demonstrate the long-term memory effects of polymer relaxation on stresses and flow dynamics, as well as to assess how microstructural distribution and spacing affect both steady and transient regimes. For \gls{RPF} and \gls{PAC}, we employ a $L_x \times L_y$ (\gls{Lx}$=$\gls{Ly}$=$4) box with periodic boundary conditions in both directions. In comparison, the porous medium is simulated in a $3L_x \times L_y$ box with the same boundary conditions.

In the RPF case, a sensitivity analysis with respect to the macroscale resolution \gls{dh} (average SPH particle spacing) was performed to determine the required number of micro-domains. As detailed in Section~\hyperref[SM]{SM6} in the \hyperref[SM]{Supplementary Material}, no significant differences were observed in the macroscopic polymer, solvent, or total stresses across the tested resolutions. Hence, a resolution of $\Delta h=0.1$ was adopted for the RPF case and all subsequent analyses.

\subsection{Reverse Poiseuille Flow (\gls{RPF})} \label{subsec:rpf}
The \gls{RPF} setup consists of two counter-flowing streams driven by a uniform external body force, avoiding the use of solid walls and no-slip boundary conditions \cite{Fedosov2010,Backer2005}. The simulation domain is divided into two halves along the $y$-direction. In the upper half, a constant force density $\mathbf{f} = (F,0,0)$ is applied in the $+x$ direction, while in the lower half the same magnitude is used in the opposite direction, $\mathbf{f} = (-F,0,0)$. Here, \gls{Fbody} denotes the scalar magnitude of the applied body force per unit mass. The theoretical shear stress profile derived from momentum balance is:
\begin{equation}
\tau_{xy}(y) = 
\begin{cases} 
F \rho \, \left( \tfrac{L_y}{4} - y \right), & 0 \leq y < L_y/2, \\
F \rho \, \left( y - \tfrac{3L_y}{4} \right), &  L_y/2 \leq y \leq L_y,
\end{cases}
\label{eq:RPF}
\end{equation}
where \gls{rho} represents the fluid density and \gls{Ly} denotes the box height. This stress field is exact, serving as a direct reference for validating the simulated polymeric and total stresses. Normalization of the stress is accomplished using the maximum theoretical shear stress derived from the RPF analytical solution (Eq.~\eqref{eq:RPF}), given by \gls{taumaxRPF}$= F \rho L_y / 4$. Consequently, the RPF case acts as a fundamental benchmark for evaluating the coupling dynamics and relaxation behavior of polymer melts within the \gls{LHMM} framework.

For these simulations, the coupling time step was set to \gls{delta_t}$= 0.32$, which corresponds to \gls{K}\(= 64\) DPD micro-steps. The SPH time step was \gls{Delta_t_SPH}\(= 0.04\), with a smoothing length of \gls{dh}$=0.1$, a fluid density of \gls{rho}$= 1.0$ and a solvent viscosity of \gls{eta_s}\(= 0.02\). The Weissenberg number (\gls{Wi}) and \gls{Re} were defined as \({\rm Wi} = \frac{2U_o\lambda}{L}\) and ${\rm Re}=\frac{U_oL}{2\eta_{tot}^0}$, respectively, where \gls{Uo} represents the characteristic average velocity, \gls{eta_tot}$=$\gls{eta0prime}$+$\gls{eta_s} denotes the total effective viscosity, and \(L = L_y = 4\) indicates the height of the box. 

\subsubsection{Transient response of different polymer chains}\label{subsub:RPF-N}
\begin{figure}[ht!]
\centering
\includegraphics[width=0.7\textwidth]{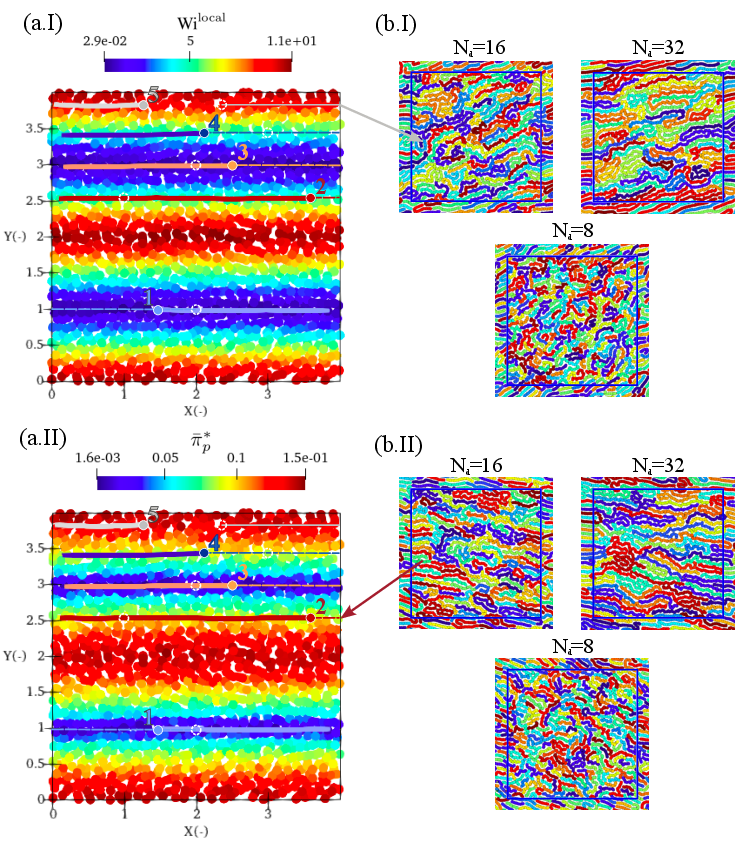}
\caption{Micro–macro coupling in \gls{RPF} for \gls{Fbody}$=1.0\cdot10^{-3}$ and \gls{alpha}$=1.0\cdot10^{-4}$ with \gls{Nb}$=16$ (\gls{beta}$=0.46$, \gls{Re}$=0.83$, and \gls{Wi}$=2.7$).  
(a) Spatial distribution of key variables at $t=\lambda(16)$. Sub-panel (a.I) shows the local Weissenberg number, \gls{Wilocal}$=(\underline{\underline{\dot{\boldsymbol{\gamma}}}}:\underline{\underline{\dot{\boldsymbol{\gamma}}}})^{1/2}\lambda$, and sub-panel (a.II) presents the dimensionless polymeric stress magnitude, \gls{piStarP}$=$\gls{alpha}$($\gls{pip}:\gls{pip}$)^{1/2}/\tau_{\rm max}^{\rm RPF}$. Streamlines of five representative particles are overlaid in both panels to illustrate their trajectories.  
s(b) Micro-domain (polymer melt \gls{Omega_core}) distributions at $t/\lambda(16)= 5.5$ for different chain lengths \gls{Nb}$=\{8,16,32\}$. Sub-panel (b.I) corresponds to the micro-domain linked to particle 5, while sub-panel (b.II) shows the one linked to particle 2. Both provide insight into the local polymeric structure, its dependence on chain length, and its deformation relative to the initial monomer distribution (Fig.~\ref{Fig:1}(b)). Videos showing the time evolution of the micro-domain deformation for each chain length are provided in the \hyperref[SM]{SUPPLEMENTARY VIDEOS}.}
\label{fig:Micro-Macro-RPF}  
\end{figure}

This section analyzes the transient response of the system for varying monomer numbers (\gls{Nb}) of each chain  in the \gls{RPF} problem. We utilize polymer microstructures characterized in Section \ref{subsec:steady_shear}. A detailed summary of the simulated conditions, including varying body forces (\gls{Fbody}), associated polymeric viscosity fractions (\gls{beta}), Reynolds numbers (Re), and Weissenberg numbers (\gls{Wi}), is provided in Tab.~\ref{tab:rpf_conditions}. Figure~\ref{fig:Micro-Macro-RPF} presents both the spatial distribution and time evolution of key microscopic and macroscopic variables for \gls{Fbody}$=1.0\cdot10^{-3}$ and \gls{alpha}$= 1.0\cdot10^{-4}$. Sub-panel (a.I) shows the particle distribution. The local viscosity is quantified through a local Weissenberg number, \gls{Wilocal}$=(\underline{\underline{\dot{\boldsymbol{\gamma}}}}:\underline{\underline{\dot{\boldsymbol{\gamma}}}})^{1/2}\lambda$, where \gls{gammadotTensor}\(= \nabla \boldsymbol{v} + \nabla \boldsymbol{v}^{\mathsf{T}}\), representing the macroscopic velocity gradient ($\nabla \boldsymbol{v}_I$) on each SPH particle $I$.  Sub-panel (a.I) shows the particle distribution and Sub-panel (a.II) displays the corresponding dimensionless polymeric stress magnitude, \gls{piStarP}$=$\gls{alpha}$($\gls{pip}:\gls{pip}$)^{1/2}/$\gls{taumaxRPF}, which is passed from the micro-domain to its associated macro-particle. These sub-panels explicitly demonstrate the micro–macro coupling: how the local shear rate from the macroscopic flow drives the evolution of polymeric stress at the microscopic level, and how the microstructure responds by aligning under the imposed deformation. Streamlines for five representative particles further illustrate the trajectories and the coherent interaction between the micro and macro scales at $t=\lambda(N_{\rm d}=16)$.

\begin{table}[htbp!]
    \centering
    \caption{Set-up for various polymer chain lengths (\gls{Nb}) and body force (\gls{Fbody}) values in Reverse Poiseuille Flow, and the response of Weissenberg (\gls{Wi}) and \gls{Re}.} % Título de la tabla
    \label{tab:rpf_conditions}
    \begin{tabular}{l *{3}{c c c }} % 'l' para Nd, y luego 3 bloques de 'c c c c' para Fbody, beta, Re, Wi
        \toprule
        \multirow{2}{*}{\gls{Nb}} & \multicolumn{3}{c}{(a) \gls{Fbody}$=1.8\cdot10^{-4}$} & \multicolumn{3}{c}{(b) \gls{Fbody}$=1.0\cdot10^{-3}$} & \multicolumn{3}{c}{(c) \gls{Fbody}$=1.0\cdot10^{-2}$} \\
        \cmidrule(lr){2-4} \cmidrule(lr){5-7} \cmidrule(lr){8-10}
       & \gls{beta} & Re & \gls{Wi} & \gls{beta} & Re & \gls{Wi} & \gls{beta} & Re & \gls{Wi} \\
        \midrule
        8   & 0.15 & 0.24 & 0.13  & 0.15  & 1.3  & 0.7  & 0.77 & 1.5 & 4.15 \\
        16  & 0.46 & 0.14 & 0.45  & 0.46  & 0.83  & 2.7  & 0.94 & 0.48  & 15.35 \\
        32  & 0.55 & 0.10  & 1.52  & 0.55  & 0.69   & 9.9  & 0.96 & 0.2  & 53.43 \\
        \bottomrule
    \end{tabular}
\end{table}

Figure~\ref{fig:Micro-Macro-RPF}(b) illustrates the conformational state of the micro-domain (polymer melt) at a dimensionless time of $t/\lambda(16)= 5.5$ for various polymer chain lengths (\gls{Nb}=8,16,32). Sub-panels (b.I) and (b.II) depict the microstructural configuration within the core region (\gls{Omega_core}) corresponding to two distinct material points, particles 5 and 2, respectively. These snapshots reveal the dependence of the polymer conformation on chain length, quantifying the degree of deformation and alignment relative to the initial, isotropic monomer distribution shown in Fig.~\ref{Fig:1}(b). Furthermore, the \hyperref[SM]{SUPPLEMENTARY VIDEOS} material visualizes the transient evolution of the micro-domain, capturing the chain reorganization driven by the macroscopic flow field. Collectively, these results offer a direct representation of the dynamic micro-macro coupling, wherein the local polymer structure continuously adapts to the imposed macroscopic deformation.

\begin{figure}[htbp!]
\centering
\includegraphics[width=0.9\textwidth]{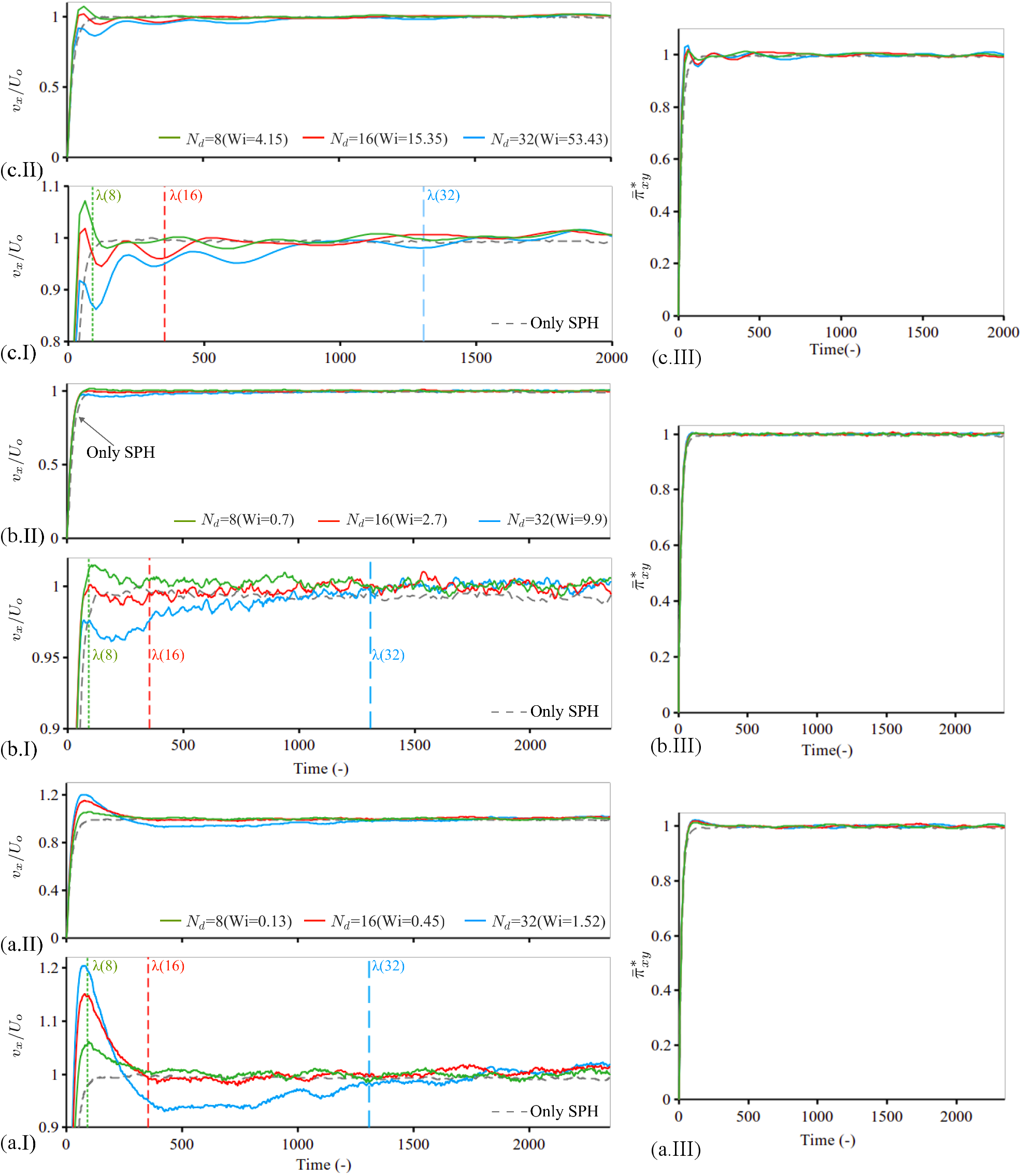}
\caption{Variation of \gls{Nb}$=\{8, 16, 32\}$ and their effects on the time evolution of the average velocity in the $x$-direction ($v_{x}$) and the shear stress of \gls{pibar}, compared with Newtonian fluid reference simulations (\gls{SPH}-only). (a) and (b) with $\alpha = 1.0\cdot10^{-4}$ and $\beta=\{0.15, 0.46, 0.55\}$ for $N_{\rm d}=\{8,16,32\}$, and (c) with $\alpha = 2.0\cdot10^{-3}$ and $\beta=\{0.77, 0.94, 0.96\}$ for $N_{\rm d}=\{8,16,32\}$. Results are shown for: (a) $F=1.8\cdot10^{-4}$ with ${\rm Wi}=\{0.13, 0.45, 1.52\}$. (b) $F=1.0\cdot10^{-3}$ with ${\rm Wi}=\{0.7, 2.7, 9.9\}$. (c) $F=1.0\cdot10^{-2}$ with ${\rm Wi}=\{4.15, 15.35, 53.43\}$. (I) Zoomed-in view of the velocity plateau, with vertical lines marking $\lambda(N_{\rm d})$. (II) Velocity response over time for different $N_{\rm d}$. (III) Shear stress response over time, showing the dimensionless total shear stress \gls{piStarXY}$=\bar{\pi}_{xy}/\tau_{\rm max}^{\rm RPF}$.}
\label{fig:VelocityVsTime-RPF}
\end{figure}

Figure~\ref{fig:VelocityVsTime-RPF} illustrates the dynamic response of the system, showing the time evolution of the dimensionless velocity $v_x/U_o$ and the dimensionless total shear stress \gls{piStarXY}$=\bar{\pi}_{xy}/\tau_{\rm max}^{\rm RPF}$, where \gls{piXY} denotes the $xy$-component of the total stress tensor \gls{pibar} (Eq.~\eqref{eq:Hydro_hybrid}). Results are reported for different body forces (\gls{Fbody}), weighting parameters (\gls{alpha}), and chain lengths ($N_{\rm d}$), leading to differents \gls{Re}, \gls{Wi}, and \gls{beta} numbers. Panels (a)–(c) correspond to distinct $(F,\alpha)$ combinations. Sub-panels (I) provide a magnified view of the velocity plateau, with vertical lines marking the relaxation times $\lambda(N_{\rm d})$. Sub-panels (II) and (III) show the temporal evolution of velocity and shear stress, respectively, for different $N_{\rm d}$. Overall, these results highlight the role of chain length and micro–macro weighting in governing the relaxation dynamics and transient response.

The transient behavior is analyzed across three regimes. In Figs.~\ref{fig:VelocityVsTime-RPF}(a)–(b), the weighting parameter $\alpha$ is fixed while the forcing $F$ is increased, probing a range of \gls{beta} and \gls{Wi} as detailed in Tab.~\ref{tab:rpf_conditions}. A key feature in these cases is the presence of overshoots in both velocity and stress before the system reaches a steady state. In the low-\gls{Wi} regime (panel a), the overshoots are broad and subtle: the velocity peak (sub-panel II) reflects the initial resistance of the polymer melt, followed by an acceleration beyond the equilibrium point. The corresponding stress peak (sub-panel III) is similarly modest. At larger \gls{Wi} (panel b), the response is faster: the overshoots become sharper but less pronounced relative to the steady-state values.

Figure~\ref{fig:VelocityVsTime-RPF}(c) examines a high-forcing scenario (${\rm Wi}$ reaching up to 53.43), where the polymeric contribution becomes predominant ($\beta > 0.94$). In this case, the system markedly deviates from the Newtonian reference: both velocity and stress exhibit sharp transient peaks followed by damped overshoots. This indicates that a substantial amount of elastic energy is stored in the polymer microstructure and subsequently released, temporarily overshooting the steady state. Despite these strong transients, the LHMM-based coupled solver maintains stability and successfully reaches equilibrium. This emphasizes the feasibility of operating under polymer-dominated conditions within the micro-macro coupling. Furthermore, the time required to reach steady state scales with the relaxation time of the Carreau–Yasuda fit for each chain length \gls{Nb}, as shown in panels (I). Specifically, longer chains yield longer relaxation times, with the approach to equilibrium consistently following the magnitude of $\lambda(N_{\rm d})$.

\subsubsection{Steady-state response for different polymer chains}\label{subsub:RPF-N}
Following the analysis of the transient response, we now examine the steady-state profiles obtained from the coupled \gls{LHMM} simulations. Steady state is identified by monitoring both velocity and shear stress profiles, and is considered reached once they display a constant plateau, as shown in Fig.~\ref{fig:VelocityVsTime-RPF}. 

\begin{figure}[htbp!]
\centering
\includegraphics[width=0.8\textwidth]{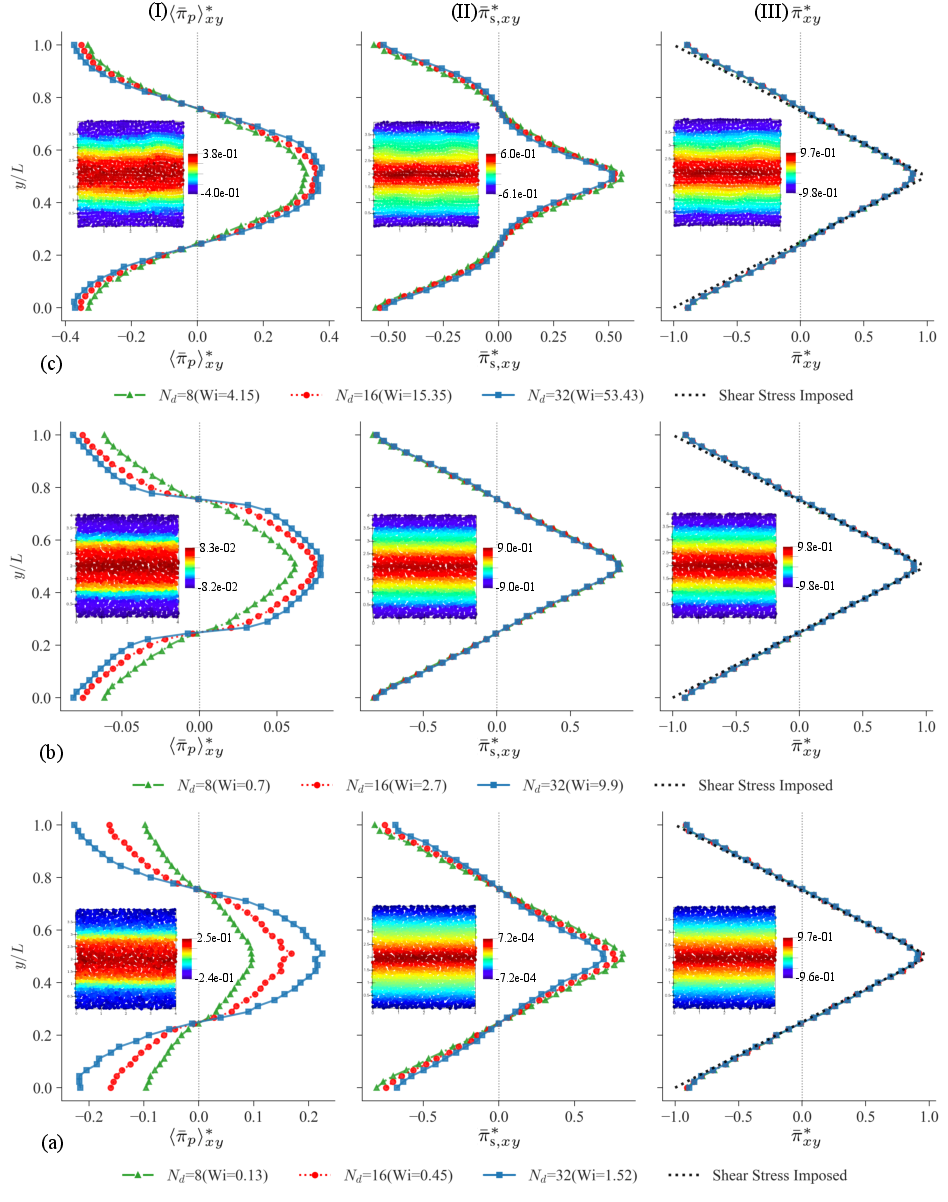}
\caption{Average steady-state shear stress profiles for varying $N_{\rm d}=\{8, 16, 32\}$. The figure shows the average dimensionless stress components profile at $y$ along the line $x=2.0$. Contributions are shown for: (I) The dimensionless shear stress of the microscopic polymeric contribution ($\langle\bar{\pi}_p\rangle_{xy}^*$), (II) the dimensionless shear stress of the macroscopic solvent ($\bar{\pi}^{*}_{\text{s},xy}$), and (III) total shear stress ($\bar{\pi}^{*}_{xy}$) compared to the shear stress imposed ($\tau_{xy}(y)$/$\tau_{\rm max}^{\rm RPF}$) with Eq.~\eqref{eq:RPF}. (a) and (b) with $\alpha = 1.0\cdot10^{-4}$ and $\beta=\{0.15, 0.46, 0.55\}$ for $N_{\rm d}=\{8,16,32\}$, and (c) with $\alpha = 2.0\cdot10^{-3}$ and $\beta=\{0.77, 0.95, 0.97\}$ for $N_{\rm d}=\{8,16,32\}$. Results are shown for: (a) $F=1.8\cdot10^{-4}$ with ${\rm Wi}=\{0.13, 0.45, 1.52\}$. (b) $F=1.0\cdot10^{-3}$ with ${\rm Wi}=\{0.7, 2.7, 9.9\}$. (c) $F=1.0\cdot10^{-2}$ with ${\rm Wi}=\{4.15, 15.35, 53.43\}$. Futhermore, each plot also includes an instantaneous snapshot of the particle distribution at $t=2000$, illustrating the stress field for the corresponding component and $N_{\rm d}$ (in panel (a), $N_{\rm d}=8$; (b), $N_{\rm d}=16$; (c), $N_{\rm d}=32$).}
\label{fig:ShearStressN}
\end{figure}

Figure~\ref{fig:ShearStressN} presents the average steady-state shear stress profiles along the domain centerline ($x=L_{x}/2$) for different body forces (\gls{Fbody}), weighting parameters (\gls{alpha}), and monomer numbers ($N_{\rm d}=\{8,16,32\}$) effecting directly \gls{beta} and \gls{Wi}, following the configurations of Fig.~\ref{fig:VelocityVsTime-RPF} and Tab.~\ref{tab:rpf_conditions}. The decomposition is shown as: (I) the microscopic polymeric contribution, \gls{piStarPxy}=\gls{alpha}\gls{pip_xy}/\gls{taumaxRPF}; (II) the macroscopic solvent contribution, \gls{piStarSXY}=\gls{pis_xy}/\gls{taumaxRPF} (where \gls{pis_xy} is the $xy$-component of the solvent stress tensor \gls{pis}); and (III) the total shear stress, \gls{piStarXY}. In all cases, the simulated total stress accurately reproduces the linear analytical profile $\tau_{xy}(y)$ from Eq.~\eqref{eq:RPF}, in agreement with previous works~\cite{Fedosov2010,NietoSimavilla2022,Backer2005}. This consistency check ensures that the \gls{LHMM} framework correctly conserves momentum.

The balance between polymeric and solvent contributions is clearly illustrated in panels (I) and (II). Their distributions are directly influenced by \gls{beta} and the chain length \gls{Nb}. Distortions in the individual \gls{piStarPxy} and \gls{piStarSXY} profiles are most evident in case (a), due to lower forcing and ${\rm Wi}<1.6$, and in case (c), where higher forcing, polymer viscosity fraction \gls{beta}, and \gls{Wi} are presented. In case (a), the polymeric stress contribution is minimal, leaving the solvent stress to dominate and closely match the total stress profile. In contrast, in case (c), the polymeric contribution becomes dominant, causing the solvent stress profile to deform into a complex shape that maintains the overall linear profile of the resulting total stress (Panel III). These findings confirm that the coupling scheme effectively captures the redistribution of stress between the solvent and polymers. Additionally, Panel (I) illustrates that the polymeric contribution increases with chain length \gls{Nb}. In cases (a) and (b), distinct values of \gls{beta} result in clearly separated stress curves for each \gls{Nb}. However, in case (c), where all \gls{beta} values approach unity, the polymeric profiles nearly converge into a single curve, suggesting similar contributions across different \gls{Nb}.

\begin{figure}[htbp!]
\centering
\includegraphics[width=0.8\textwidth]{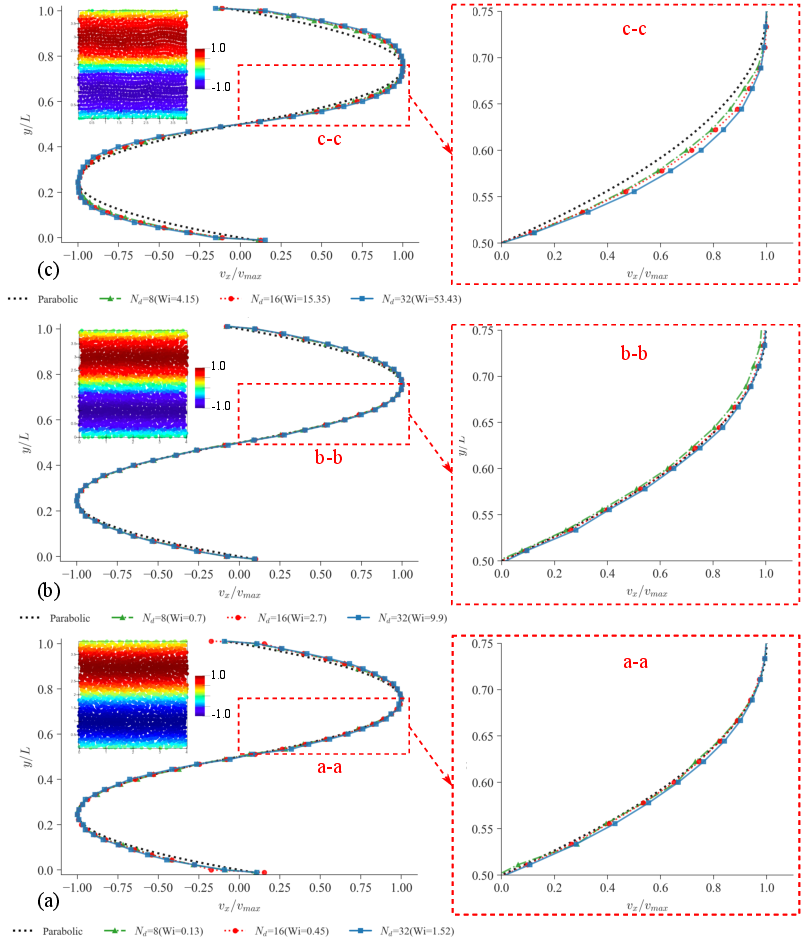}
\caption{Average steady-state velocity profiles for varying $N_{\rm d}=\{8, 16, 32\}$. The figure shows the average dimensionless velocity in the $x$-axis ($v_{x}/v_{\rm max}$) profile at $y$ along the line $x=2.0$, and its comparison with the Newtonian parabolic result, where $v_{\rm max}$ is the maximum horizontal velocity in the average velocity profile and is compared with the parabolic velocity profile defined in Eq.~\eqref{eq:velocity}. (a) and (b) with $\alpha = 1.0\cdot10^{-4}$ and $\beta=\{0.15, 0.46, 0.55\}$ for $N_{\rm d}=\{8,16,32\}$, and (c) with $\alpha = 2.0\cdot10^{-3}$ and $\beta=\{0.77, 0.95, 0.97\}$ for $N_{\rm d}=\{8,16,32\}$. Results are shown for: (a) $F=1.8\cdot10^{-4}$ with ${\rm Wi}=\{0.13, 0.45, 1.52\}$. (b) $F=1.0\cdot10^{-3}$ with ${\rm Wi}=\{0.7, 2.7, 9.9\}$. (c) $F=1.0\cdot10^{-2}$ with ${\rm Wi}=\{4.15, 15.35, 53.43\}$. Each plot also includes an instantaneous snapshot of the velocity particle distribution at $t=2000$ (in panel (a), $N_{\rm d}=8$; (b), $N_{\rm d}=16$; (c), $N_{\rm d}=32$).}
\label{fig:VelosityN}
\end{figure}

The instantaneous snapshots included in each subpanel of Fig.~\ref{fig:ShearStressN} provide additional spatial insight into the steady-state stress field distributions at the particle level for different monomer lengths (\gls{Nb}) at $t=2000$ ($t/\lambda\left(N_{\rm d}=16\right)=5.7$). These visualizations confirm that the differences in \gls{piStarPxy} and \gls{piStarSXY} intensity are primarily driven by the \gls{Fbody} and the (\gls{Nb},\gls{beta}), as discussed above. Nonetheless, despite local deviations, the particle-level total stress fields align well with the imposed averaged stress profiles, reinforcing the consistency of the micro-to-macro coupling in the LHMM framework.
\\
\\
Having established the redistribution of stresses between the contributions of the solvent and the polymer, we now turn our attention to how these characteristics are reflected in the corresponding velocity profiles. Figure~\ref{fig:VelosityN} illustrates the average steady-state velocity profiles for \gls{Nb} values of 8, 16, and 32, all under the same \gls{Fbody} and (\gls{alpha}, \gls{beta}) configurations as presented in Fig.~\ref{fig:ShearStressN}. For reference, the simulated velocity profiles are compared against the Newtonian parabolic solution for a pure solvent (see \ref{app:velocity}) \cite{Fedosov2010,Backer2005}.

Figures~\ref{fig:VelosityN}(a) and (b) demonstrate that the simulated velocity aligns closely with the Newtonian parabolic reference (Eq.~\eqref{eq:velocity}), confirming that the coupled \gls{LHMM} system accurately reproduces the expected Newtonian-like flow under conditions of weak to moderate polymeric effects. In contrast, Figure~\ref{fig:VelosityN}(c) reveals a noticeable flattening of the velocity profile near the center of the channel (in low-shear regions), which is consistent with the shear-thinning behavior observed in previous SDPD and DPD studies~\cite{Litvinov2016,Litvinov2014,NietoSimavilla2022}. This deviation becomes increasingly pronounced with higher \gls{Nb}, since the shear-thinning becomes stronger (Fig.~\ref{fig:VelosityN}(a)).

The zoomed-in views (detailed in $y=[0.5;0.75]$) in Fig.~\ref{fig:VelosityN} further highlight the alignment with the Newtonian parabolic shape profile observed in cases (a) and (b), while also illustrating the systematic deviation noted in case (c). Additionally, the instantaneous snapshots presented in each subpanel of Fig.~\ref{fig:VelosityN} showcase the $v_x$ field distributions at the particle level for varying monomer lengths ($N_{\rm d}$) at $t=2000$ ($t/\lambda\left(N_{\rm d}=16\right)=5.7$). These snapshots reinforce the finding that the alignment of particle-level field distributions continues to mirror the averaged velocity profiles $v_x$ across different configurations, thereby enhancing the robustness of the \gls{LHMM} coupling and supporting the conclusions derived from the stress field analysis (Fig.~\ref{fig:ShearStressN}).

In this section, we briefly evaluate the impact of the weighting parameters over polymeric viscosity fraction ($\alpha\longrightarrow$\gls{beta}), which regulate the relative contributions of polymeric and solvent stresses within the coupled momentum balance. To explore their influence under moderate and high forcing, we conducted a series of additional simulations at $N_{\rm d}=16$, covering a range of Weissenberg numbers ($\text{Wi} \sim \mathcal{O}(1-10)$). The findings indicate that increasing $\alpha$ (and correspondingly increase $\beta$) amplifies the polymer contribution, resulting in prolonged stress overshoots and a slower relaxation process during the transient phase. In the steady state, the velocity profiles exhibit a noticeable flattening, which aligns with shear-thinning behavior. Notably, the balance between polymeric and solvent stresses remains stable across the various Weissenberg numbers investigated, demonstrating the robustness of the coupling scheme against variations in $(\alpha,\beta)$. A comprehensive examination of these effects, including insights into transient dynamics and systematic comparisons under different forcing conditions, can be found in Section~\hyperref[SM]{SM7} in the \hyperref[SM]{Supplementary Material}.

\subsection{Complex flow through a Periodic Array of Cylinders (PAC)} \label{complexflow} 
We now analyze flow through a \gls{PAC}, which is designed to investigate fluid–structure interactions and complex flow with curved streamlines. For systems with finite memory, the loss of space invariance renders the actual Lagrangian dynamics transient, even when the fluid appears steady in an Eulerian frame. This setup introduces a stationary obstacle into the flow path, creating heterogeneous flow fields relevant in confined and porous media~\cite{Morii2021,King2021,Datta2022}. The configuration features a single cylinder with a radius of \gls{Rcy}$=0.56$, positioned at the center of a box defined by $[L_x/2, L_y/2]$ (\gls{Lx}=\gls{Ly}=4), with periodic boundary conditions applied in both directions. An external body force \gls{f} is defined as \gls{f}$=[F,0]$. For the \gls{LHMM} simulations, the exchange time step is set at \gls{delta_t}=$0.125$, with a microscale exchange interval of \gls{K}$=25$. The polymer chain length is fixed to \gls{Nb}$=16$, and the solvent viscosity is \gls{eta_s}$=0.08$. This arrangement enables us to investigate the impact of the cylindrical obstacle and the resulting complex flow patterns (e.g., extensional flow around the cylinder, shear flow in the gaps) on the polymeric chains.

We examine how the polymer's rheological response affects overall flow resistance and velocity profiles. Key simulation parameters for the PAC flow include the Weissenberg number (${\rm Wi} = U_{o}\lambda/R_{cl}$), the Reynolds number (${\rm Re} = R_{cl}U_{o}\rho/$\gls{eta_tot}), and the coupling parameters (\gls{beta}).

\begin{figure}[ht!]
\centering
\includegraphics[width=0.98\textwidth]{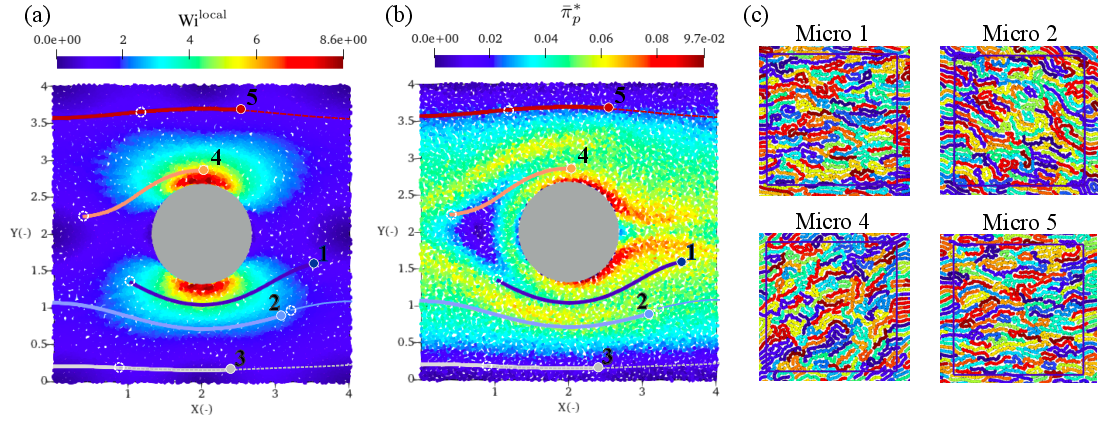}
\caption{Micro–macro coupling in PAC flow for $dh=1/25$, $\Delta t_{SPH}=1.25\cdot10^{-3}$, $F=1.0\cdot10^{-3}$, $\beta=0.46$ (corresponding to $\alpha = 4.0\cdot10^{-4}$ and $N_{\rm d}=16$), with resulting ${\rm Wi}=5.2$ and $Re=0.032$. Spatial distribution of micro–macro variables at an early transient state ($t=\lambda/4$). Sub-panel (a) shows the local Weissenberg number, \gls{Wilocal}$=($\gls{gammadotTensor}$:$\gls{gammadotTensor}$)^{1/2}\lambda$, and (b) the dimensionless magnitude of the polymeric stress, \gls{piStarP}$=$\gls{alpha}$($\gls{pip}:\gls{pip}$)^{1/2}/$\gls{tauStarPAC}. In both, streamlines for five representative particles are delineated to illustrate their local trajectories. (c) Microdomain (polymer melt) distributions at $t=\lambda/4$ showing the spatial organization of polymer chains (colored by each chain's beads) for selected micro-domains. These panels illustrate the micro-domains linked to particles 1, 2, 4, and 5, providing further insight into the local polymeric structure and its response within different flow regions around the cylinder. Videos showing the time evolution of the micro-domain deformation for each chain length are provided in the \hyperref[SM]{SUPPLEMENTARY VIDEOS} (See panels (c)).}
\label{fig:Micro-Macro-PAC}
\end{figure}

To illustrate the complex interaction between microscopic polymer dynamics and macroscopic flow behavior in a PAC configuration, Figure~\ref{fig:Micro-Macro-PAC} provides a comprehensive view of the micro–macro coupling. It depicts the spatial distribution of particles and the temporal evolution of critical microscopic and macroscopic variables associated with flow dynamics, as well as the resulting macroscopic response. The figure also examines how the conformation of polymer micro-domains corresponds to the microstructure's response to stress. Results are shown for \gls{dh}$=1/25$, \gls{Delta_t_SPH}$=1.25\cdot10^{-3}$, \gls{Fbody}$=1.0\cdot10^{-3}$, $\beta=0.46$ (corresponding to $\alpha = 4.0\cdot10^{-4}$), with resulting average steady ${\rm Wi}=5.2$ and \gls{Re}$=0.032$.

Sub-panels (a.I) and (a.II) in Figure~\ref{fig:Micro-Macro-PAC} display the particle distributions alongside the local Weissenberg number, defined as \gls{Wilocal}$=($\gls{gammadotTensor}$:$\gls{gammadotTensor}$)^{1/2}\lambda$, and the dimensionless magnitude of the polymeric stress, \gls{piStarP}$=$\gls{alpha}$($\gls{pip}:\gls{pip}$)^{1/2}/$\gls{tauStarPAC} (where \gls{tauStarPAC}$=$\gls{Uo}\gls{eta_tot}$/$\gls{Rcy} and \gls{Uo} represents the average steady velocity), respectively. Both sub-panels illustrate the interplay at $t=\lambda/4$ between the shear rate and the polymeric stress. Streamlines for five representative particles highlight their nearly straight trajectories in this PAC case, indicating a transient state before significant deformation near the cylinder.

Figure~\ref{fig:Micro-Macro-PAC}(b) shows the temporal evolution of micro–macro variables for the five representative particles highlighted in panel (a). Sub-panels (b.I) and (b.II) display the local shear rate and the dimensionless polymeric stress magnitude, respectively, over one characteristic period $\lambda$. These results demonstrate the strong correlation between stress and shear rate along particle trajectories, which, unlike the nearly straight paths in the \gls{RPF} case, are highly sensitive to the initial particle position in the \gls{PAC} flow due to the presence of the obstacle.

The corresponding micro-domains at $t=\lambda/4$ for particles 1, 2, 4, and 5 are shown in Fig.~\ref{fig:Micro-Macro-PAC}(c), with detailed views of the core region (\gls{Omega_core}) for each micro-domain associated with its respective particle. These snapshots illustrate how local chain stretching, alignment, and deformation vary with the particle trajectory, establishing a clear connection between microstructural evolution and the heterogeneous flow field traced by the streamlines. The \hyperref[SM]{SUPPLEMENTARY VIDEOS} material provides a direct visualization of these transient Lagrangian dynamics, documenting the continuous evolution of the micro-domain conformation as material points traverse the heterogeneous flow field. This visualization reinforces the concept that even in a macroscopically steady flow (Eulerian frame), the polymer microstructure undergoes a history-dependent, non-stationary deformation process (Lagrangian frame). These videos thus animate and extend the static microstructural snapshots presented in Fig.~\ref{fig:Micro-Macro-PAC}(c).

\subsubsection{Resolution Analysis}~\label{subsub:Reso}
In the PAC case involving a single cylinder, we evaluate the sensitivity of the average macro–particle spacing (\gls{dh}) with respect to the total number of particles in the domain, given by (\gls{Lx}/\gls{dh})$\times$(\gls{Ly}/\gls{dh}). The body force is fixed at \gls{Fbody}$=1.0\cdot10^{-3}$, and two coupling-viscosity ratios are considered: $\beta=0.46$ and $\beta=0.65$. We investigate three spatial resolutions: $\Delta h=1/16$ (\(64\times64\) particles), $\Delta h=1/20$ (\(80\times80\) particles), and $\Delta h=1/25$ (\(100\times100\) particles). The corresponding exchange intervals for the macroscale solver (\gls{M}) are set to $50$, $100$, and $200$ to determine \gls{Delta_t_SPH}.

\begin{figure}[ht!]
\centering
\includegraphics[width=0.95\textwidth]{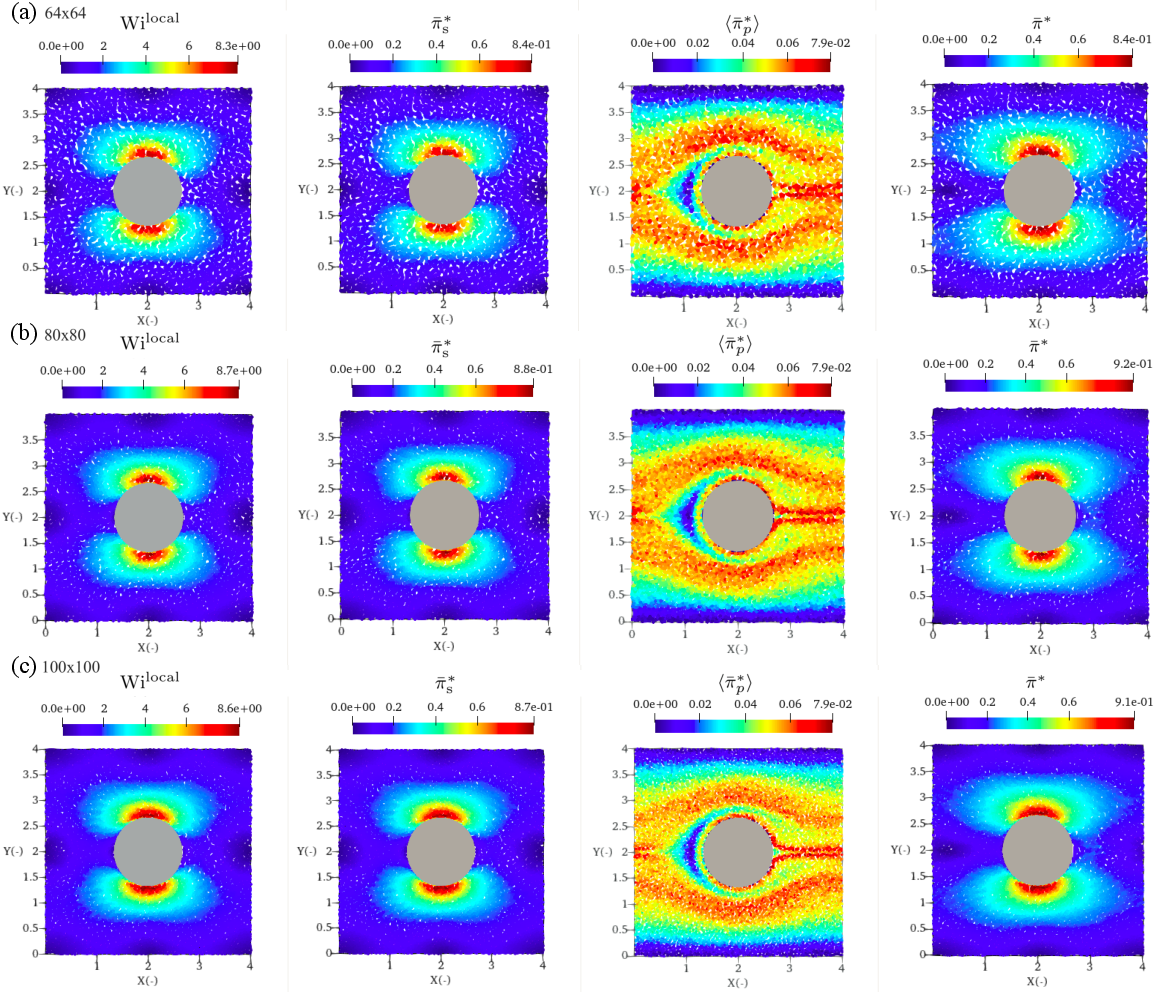}
\caption{Flow/stress field during initial transient at \( t = \lambda/2 \) for $\beta = 0.46$ with resulting ${\rm Wi}=5.2$ and $Re=0.032$, showing particle distribution of local Wi number, \gls{Wilocal} (leftmost panel), the dimensionless stress magnitude of the polymeric stress, $\bar{\pi}^*_{p}$, the solvent, $\bar{\pi}^*_{\rm s}$, and the total, $\bar{\pi}^*$ (rightmost Panel). Panels correspond to different spatial resolutions: (a) $\Delta h=1/16$ (\texttt{64x64} particles), (b) $\Delta h=1/20$ (\texttt{80x80} particles), and (c) $\Delta h=1/25$ (\texttt{100x100} particles).}
\label{fig:B46IK}
\end{figure}

\begin{figure}[ht!]
\centering
\includegraphics[width=0.95\textwidth]{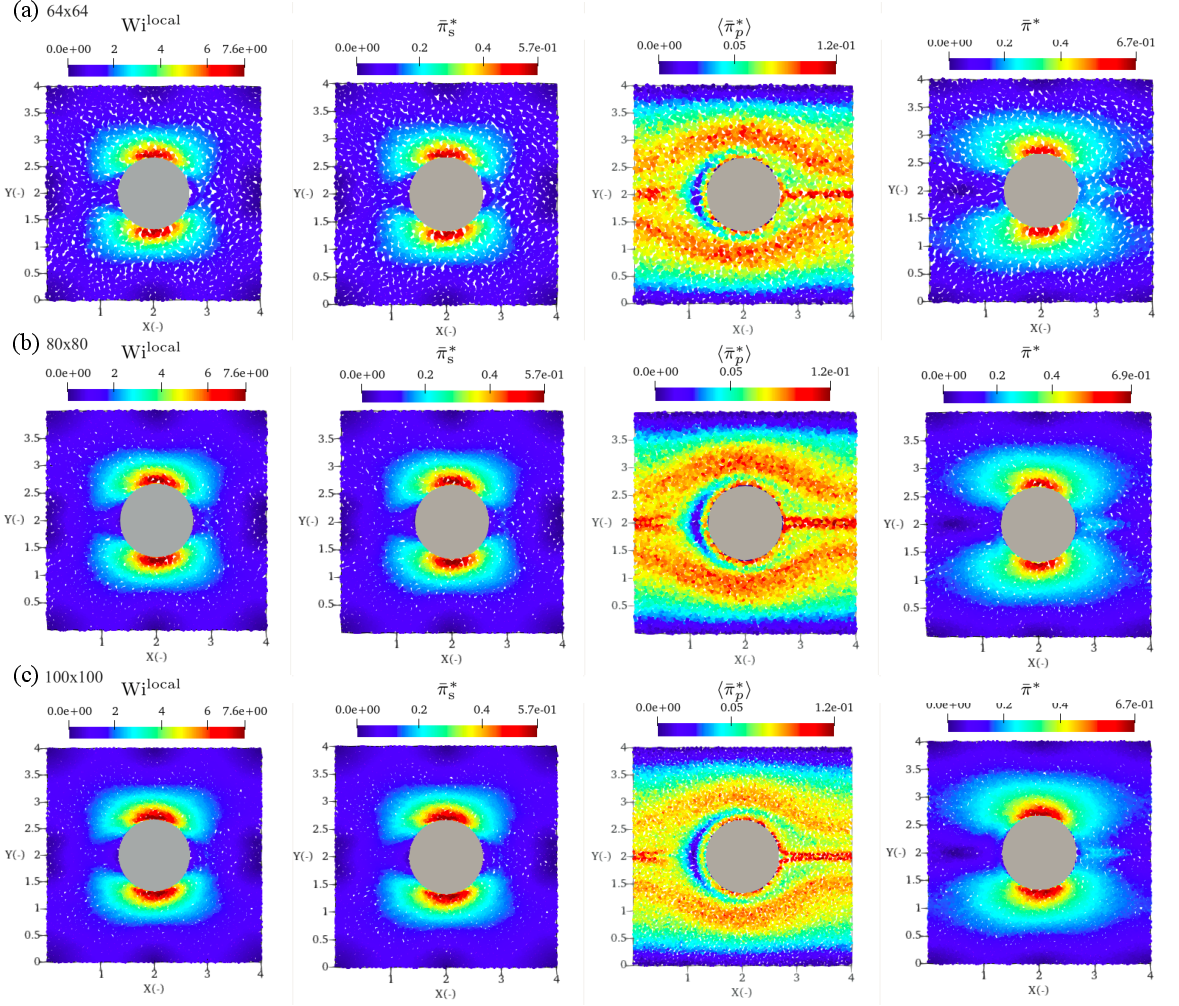}
\caption{Flow/stress field during initial transient at \( t = \lambda/2 \) for $\beta = 0.65$ with resulting ${\rm Wi}=4.5$ and $Re=0.018$, showing particle distribution of local Wi number, ${\rm Wi}^{\rm local}$ (leftmost panel), the dimensionless stress magnitude of the polymeric stress, $\langle\bar{\pi}^*_{p}\rangle$, the solvent, $\bar{\pi}^*_{\rm s}$, and the total, $\bar{\pi}^*$ (rightmost Panel). Panels correspond to different spatial resolutions: (a) $\Delta h=1/16$ (\texttt{64x64} particles), (b) $\Delta h=1/20$ (\texttt{80x80} particles), and (c) $\Delta h=1/25$ (\texttt{100x100} particles).}
\label{fig:B65IK}
\end{figure}

Building on the transient analysis shown in Figure~\ref{fig:Micro-Macro-PAC}, Figures~\ref{fig:B46IK} and \ref{fig:B65IK} illustrate the particle-level distributions during the initial transient $t = \lambda/2$ for $\beta=0.46$ and $\beta=0.65$, respectively. Each figure displays four dimensionless quantities: the local Weissenberg number, \gls{Wilocal}$=($\gls{gammadotTensor}$:$\gls{gammadotTensor}$)^{1/2}/$\gls{lambda}; the normalized polymeric stress, \gls{piStarP}$=$\gls{alpha}$($\gls{pip}$:$\gls{pip}$)^{1/2}/$\gls{tauStarPAC}; the solvent stress, \gls{piStarS}$=($\gls{pis}$:$\gls{pis}$)^{1/2}/$\gls{tauStarPAC}; and the total stress, \gls{piStar}$=($\gls{pibar}$:$\gls{pibar}$)^{1/2}/$\gls{tauStarPAC}. As expected, increasing the resolution (down to $\Delta h=1/25$) yields more clearly resolved flow structures, especially in the polymeric stress field. Nonetheless, the three resolutions exhibit consistent results in both the order of magnitude and the spatial distribution of shear rate and stress components. This consistency indicates that the LHMM coupling provides adequate accuracy across different discretizations. The findings hold for both parameter sets: ${\rm Wi}=5.2$ with $Re=0.032$ and $\beta=0.46$ (Fig.~\ref{fig:B46IK}), and ${\rm Wi}=4.5$ with $Re=0.018$ and $\beta=0.65$ (Fig.~\ref{fig:B65IK}).

\begin{figure}[htbp!]
\centering
\includegraphics[width=1.0\textwidth]{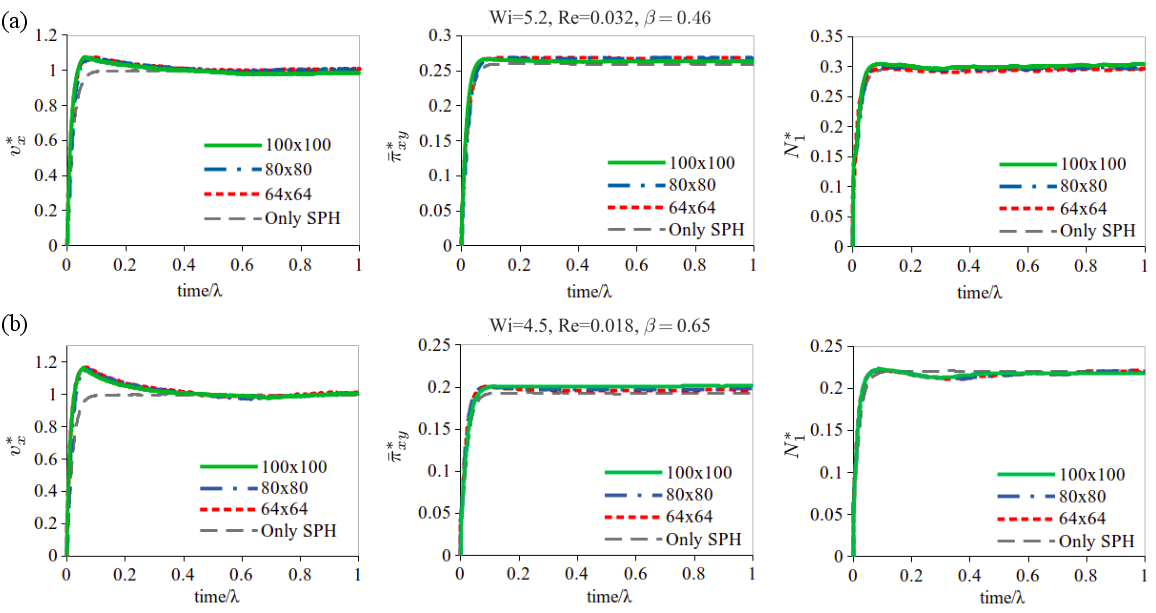}
\caption{The effect of the resolution \(\Delta h\) and \(\beta\) on the temporal evolution of the following parameters is examined: the dimensionless magnitude of the x-component of velocity \(\left(v^*_x={\|v_x\|}/{U_o}\right)\) (leftmost Panel), the total shear stress \(\left(\bar{\pi}^*_{xy}={\|\bar{\pi}_{xy}\|}/{\tau_{\text{PAC}}^{*}}\right)\), and the total normal stress \(\left(N^*_{1}={\|N_{1}\|}/{\tau_{\text{PAC}}^{*}}\right)\) (rightmost panel).  Panel (a) shows results for $\beta = 0.46$, \gls{Wi} = 5.2, and Re = 0.032, while panel (b) displays results for \(\beta = 0.65\), \(\text{Wi} = 4.5\), and \(\text{Re} = 0.018\). Here, \gls{Uo} is the magnitude of the total average velocity, and \gls{tauStarPAC}$=$\gls{Uo}\gls{eta_tot}$/$\gls{Rcy} is the dimensionless Stress parameter.}
\label{fig:1Reso}
\end{figure}

To assess the impact of numerical parameters on transient flow dynamics, Fig.~\ref{fig:1Reso} illustrates the temporal evolution of average velocity along the x-axis ($v_x$) and stress magnitudes (\gls{piXY} and \gls{N1}) for three different spatial resolutions ($\Delta h$) at two different $\beta$ values: $0.46$ (Panel (a)) and $0.65$ (Panel (b)). This analysis focuses on the shear stress component \gls{piXY} and the first normal stress difference, defined as \gls{N1} = \gls{piXX}$-$\gls{piYY} (where \gls{piXX}and \gls{piYY} are the components $-xx$ and $-yy$ of the tensor \gls{pibar}, respectively), both of which are presented in dimensionless form in Figure~\ref{fig:1Reso}.

In Panel (a) ($\beta=0.46$) in Fig.~\ref{fig:1Reso}, we observe behavior consistent with long-term relaxation effects around one characteristic period of $t=\lambda$ compared with the case with Newtonian \gls{SPH}. This is marked by a small peak in $\left\|v_x\right\|$ followed by rapid damping to an apparent equilibrium steady state, where both stress and velocity evolve constantly. When compared to Panel (b) ($\beta=0.65$), the velocity peaks are less pronounced in Panel (a), whereas the peaks in Panel (b) are much clearer, and the damping recovery to reach steady state takes more time. A similar trend is observed for the first normal stress difference, \(N_1\), while for \(\bar{\pi}_{xy}\), a constant and steady state emerges abruptly. Notably, the various cases of \(\Delta h\) produce remarkably similar results, with minimal variation evident in Fig.~\ref{fig:1Reso}. This indicates that the resolution has a negligible impact as the simulated transient dynamics within this range of \(\Delta h\) and \(\beta\).

\begin{figure}[ht!]
\centering
\includegraphics[width=0.9\textwidth]{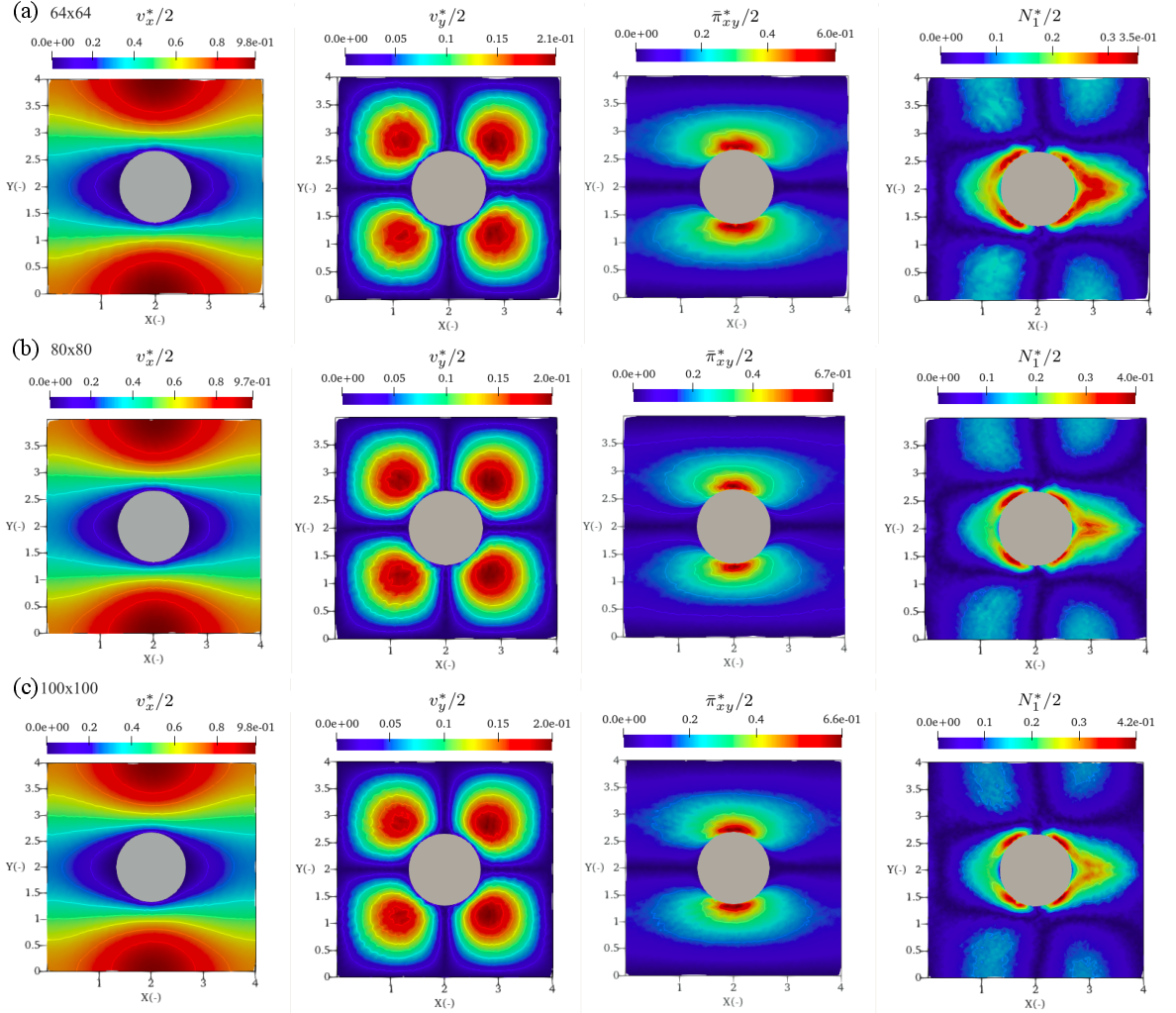}
\caption{Spatial distribution of the fields for different resolutions at steady-state for \gls{Wi} = 5.2, Re = 0.032, and $\beta = 0.46$. The panels display the dimensionless magnitude of the x-component of velocity (\gls{vxStar}$={\|v_x\|}/{U_o}$, leftmost panel), the y-component of velocity (\gls{vyStar}$={\|v_y\|}/{U_o}$), the total shear stress (\gls{piStarXY}$={\|\bar{\pi}_{xy}\|}/{\tau_{\text{PAC}}^{*}}$), and the total normal stress (\gls{N1Star}$={\|N_{1}\|}/{\tau_{\text{PAC}}^{*}}$, rightmost panel). Results are computed with different spatial resolutions: (a) \texttt{64x64} particles (\gls{dh}$=1/16$), (b) \texttt{80x80} particles (\gls{dh}$=1/20$), and (c) \texttt{100x100} particles (\gls{dh}$=1/25$). Here, \gls{Uo} is the magnitude of the total average velocity, and \gls{tauStarPAC}$=$\gls{Uo}\gls{eta_tot}$/$\gls{Rcy} is the dimensionless Stress parameter.}
\label{fig:B46SS}
\end{figure}

\begin{figure}[ht!]
\centering
\includegraphics[width=0.9\textwidth]{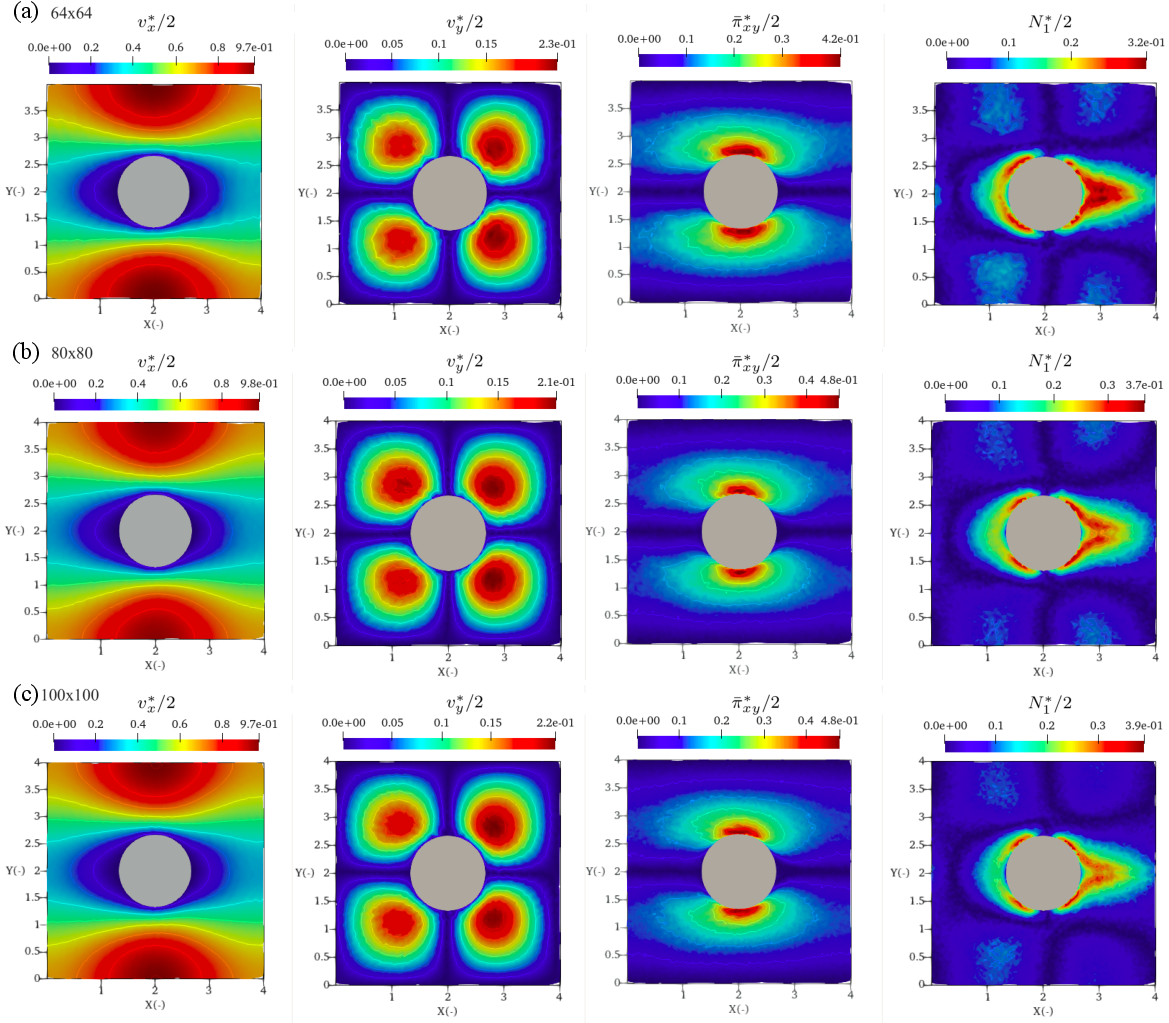}
\caption{Spatial distribution of the fields for different resolutions at steady-state for \gls{Wi} = 4.5, Re = 0.032, and $\beta = 0.65$. The panels display the dimensionless magnitude of the x-component of velocity (\gls{vxStar}$={\|v_x\|}/{U_o}$, leftmost panel), the y-component of velocity (\gls{vyStar}$={\|v_y\|}/{U_o}$), the total shear stress (\gls{piStarXY}$={\|\bar{\pi}_{xy}\|}/{\tau_{\text{PAC}}^{*}}$), and the total normal stress (\gls{N1Star}$={\|N_{1}\|}/{\tau_{\text{PAC}}^{*}}$, rightmost panel). Results are computed with different spatial resolutions: (a) \texttt{64x64} particles (\gls{dh}$=1/16$), (b) \texttt{80x80} particles (\gls{dh}$=1/20$), and (c) \texttt{100x100} particles (\gls{dh}$=1/25$). Here, \gls{Uo} is the magnitude of the total average velocity, and \gls{tauStarPAC}$=$\gls{Uo}\gls{eta_tot}$/$\gls{Rcy} is the dimensionless Stress parameter.}
\label{fig:B65SS}
\end{figure}

In addition to the transient analysis, Figures~\ref{fig:B46SS} and \ref{fig:B65SS} depict the steady-state spatial distributions of these magnitudes at various resolutions. Specifically, we present the spatial distributions of the dimensionless magnitudes of \gls{vxStar}, \gls{vyStar}, \gls{piStarXY}, and \gls{N1Star} for the three \gls{dh} cases within the PAC problem.

In line with the transient analyses, the steady-state distributions for both $\beta$ values (Figures~\ref{fig:B46SS} and \ref{fig:B65SS}) reveal consistent trends across resolutions. The profiles of \gls{vxStar}, \gls{vyStar}, \gls{piStarXY}, and \gls{N1Star} exhibit comparable magnitudes and spatial patterns, with only minor differences in sharpness between $\Delta h$ cases. A notable feature is the asymmetric peak in $N_1$ downstream of the cylinder, coinciding with the thin sheet of polymeric stress (\gls{piStarP}) wrapping around the cylinder and advected along the channel centerline (see Figures~\ref{fig:B46IK} and \ref{fig:B65IK}), a marker of steady viscoelastic behavior and the aasociated fluid memory. This peak appears smoother at the finest resolution ($\Delta h=1/25$), while the coarser case ($\Delta h=1/16$) produces a more pronounced intensity.

\begin{figure}[ht!]
\centering
\includegraphics[width=0.85\textwidth]{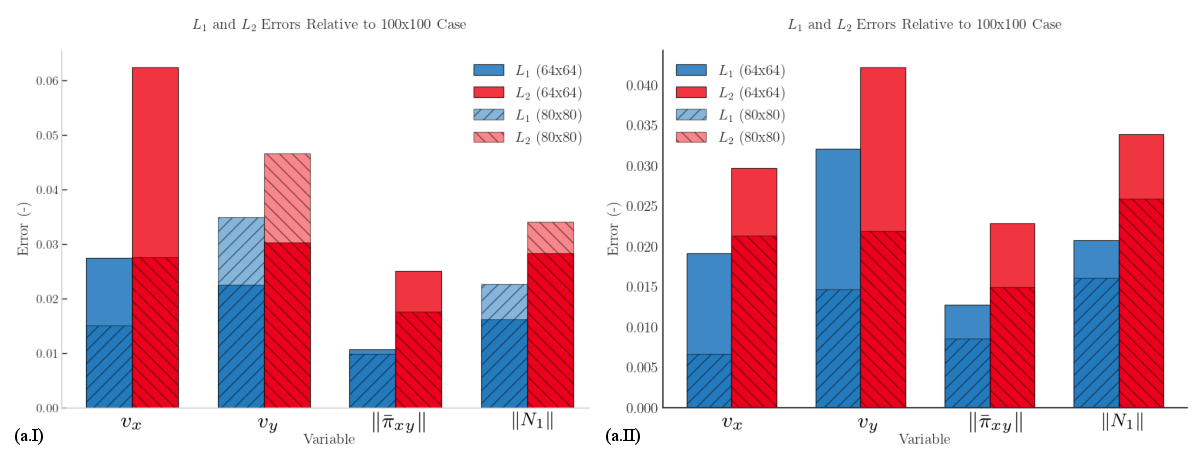}
\caption{L1 and L2 errors for \( v_x \), \( v_y \), \( \|\bar{\pi}_{xy}\| \), and \( \|N_1\| \) relative to the \texttt{100x100} particles ($\Delta h=1/25$) reference case with (a.I) $\beta = 0.46$, and (a.II) $\beta = 0.65$.}
\label{fig:errors}
\end{figure}

Figure~\ref{fig:errors} provides additional insights into numerical convergence. Panels (a.I) and (a.II) show the normalized L1 and L2 errors of  \gls{vxStar}, \gls{vyStar}, $\|$\gls{piStarXY}$\|$, and \gls{N1Star} relative to the \texttt{100x100} reference for $\beta=0.46$ and $\beta=0.65$, respectively. In both cases, the L1 and L2 errors are closely correlated and consistently decrease with finer resolution, particularly when comparing the \texttt{64x64} and \texttt{80x80} cases against the reference. The precise definitions and tabulated values of these error metrics are provided in Section~\hyperref[SM]{SM8} in the \hyperref[SM]{Supplementary Material}.

\subsubsection{Different Wi number regimes}
\begin{figure}[ht!]
\centering
\includegraphics[width=1.0\textwidth]{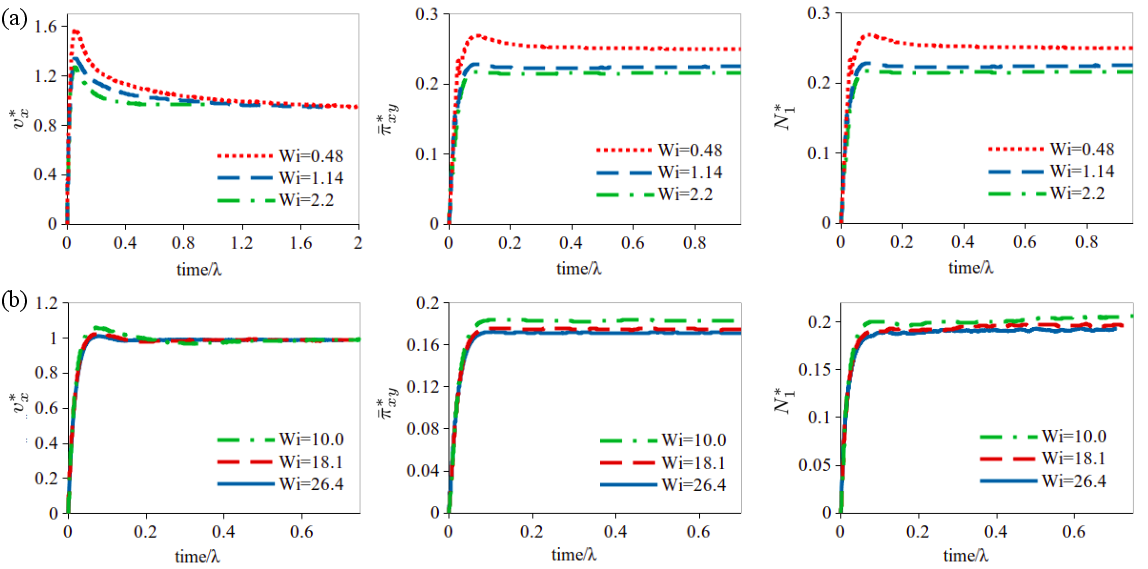}
\caption{Temporal evolution of fields for varying \gls{Wi}, achieved by adjusting the body force $F$. For each Panel (a) and (b), the results display, from left to right: The panels display the dimensionless magnitude of the x-component of velocity (\gls{vxStar}$={\|v_x\|}/{U_o}$, leftmost panel), the total shear stress (\gls{piStarXY}$={\|\bar{\pi}_{xy}\|}/{\tau_{\text{PAC}}^{*}}$), and the total normal stress (\gls{N1Star}$={\|N_{1}\|}/{\tau_{\text{PAC}}^{*}}$, rightmost panel). Panel (a) shows results for low \gls{Wi} numbers: \gls{Wi}=0.48 (Re=0.002, $F=1.25\cdot10^{-4}$), \gls{Wi}=1.14 (Re=0.004, $F=2.5\cdot10^{-4}$), and \gls{Wi}=2.2 (Re=0.009, $F=3.5\cdot10^{-4}$). Panel (b) shows results for moderate \gls{Wi} numbers: \gls{Wi}=10.0 (Re=0.04, $F=2.0\cdot10^{-3}$), \gls{Wi}=18.1 (Re=0.07, $F=3.5\cdot10^{-3}$), and \gls{Wi}=26.4 (Re=0.1, $F=5.0\cdot10^{-2}$). Here, \gls{Uo} is the magnitude of the total average velocity, and \gls{tauStarPAC}$=$\gls{Uo}\gls{eta_tot}$/$\gls{Rcy} is the dimensionless Stress parameter.}
\label{fig:1Low-Mod}
\end{figure}

To investigate a wider range of \gls{Wi} numbers, we systematically varied the body force \gls{Fbody} with values of \(\left[1.25 \cdot 10^{-4}; 2.5 \cdot 10^{-4}; 3.5 \cdot 10^{-4}; 2.0 \cdot 10^{-3}; 3.5 \cdot 10^{-3}; 5.0 \cdot 10^{-2}\right]\), while maintaining a fixed polymer viscosity ratio of \gls{beta} = 0.65, a resolution of \gls{dh} = 1/16 (using \texttt{64x64} particles), and a parameter \gls{M} = 50. This approach resulted in \gls{Wi} numbers ranging from \([0.48; 1.14; 2.2; 10.0; 18.1; 26.4]\), allowing us to examine a broad spectrum of regimes and their impact on transient flow dynamics. Figure~\ref{fig:1Low-Mod} illustrates the temporal evolution of the dimensionless velocity component (\gls{vxStar}), the total shear stress (\gls{piStarXY}), and the total first normal stress difference (\gls{N1Star}).

In the low-\gls{Wi} regime panel (a) (Fig.~\ref{fig:1Low-Mod}), the system exhibits long relaxation times (\(\sim 2\lambda\)), visible as an overshoot in \(\|v_x\|\) followed by gradual damping towards steady state. This trend is more pronounced and slower at \gls{Wi}=0.48, where both \(\bar{\pi}_{xy}\) and \(N_1\) display smoother but delayed relaxation compared with higher-\gls{Wi} cases.  

By contrast, at moderate \gls{Wi} panles (b)(Figure~\ref{fig:1Low-Mod}), the overshoot peaks are sharper and the relaxation to equilibrium requires longer times. Here, both \(N_1\) and \(\bar{\pi}_{xy}\) stabilize into well-defined steady values, reflecting a reduced impact of transient microstructural dynamics. These differences emphasize how the imposed body force (and resulting \gls{Wi}) strongly modulates the transient response within the LHMM framework.  

\begin{figure}[ht!]
\centering
\includegraphics[width=0.9\textwidth]{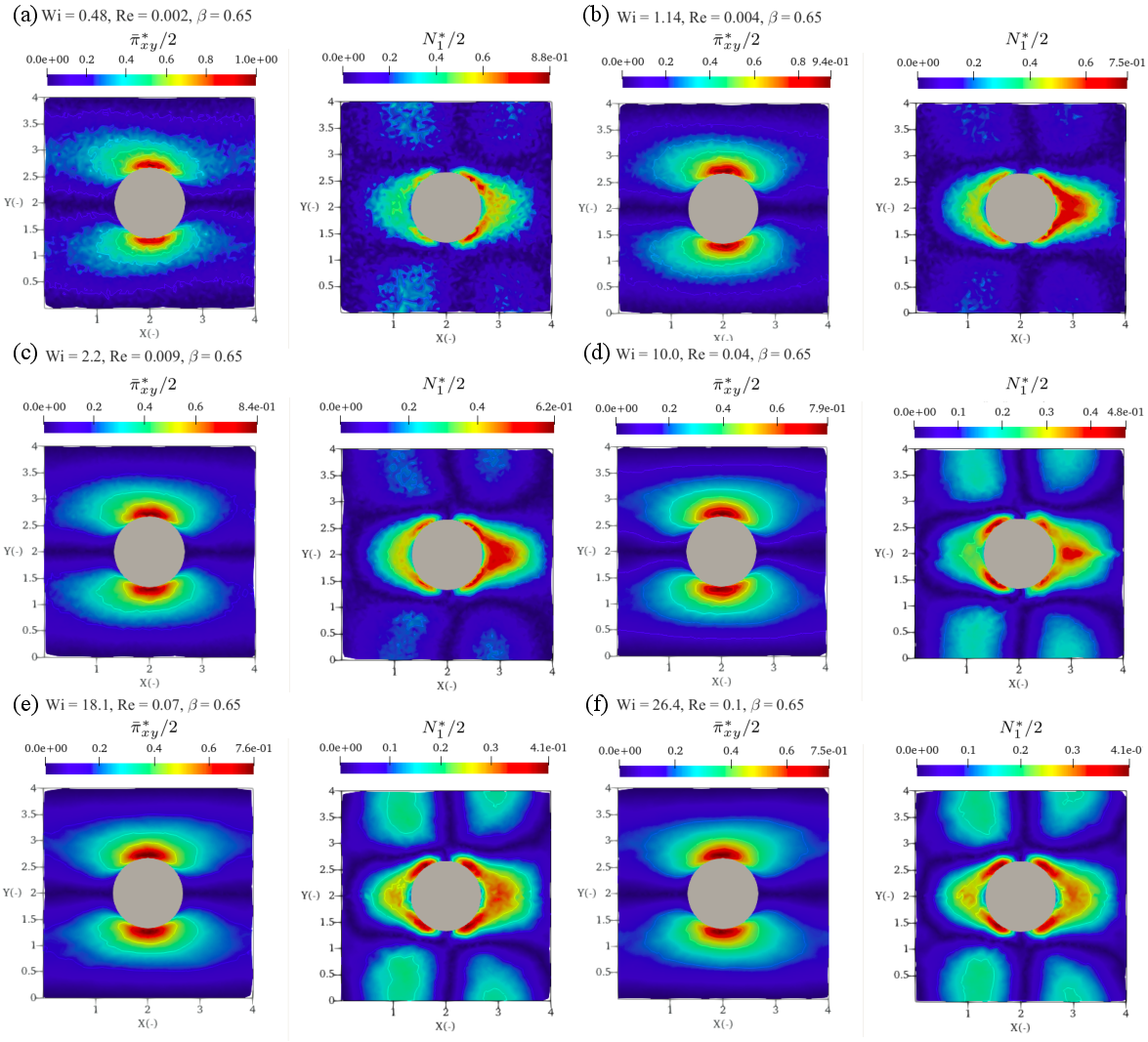}
\caption{Spatial distribution of fields at steady-state for three low \gls{Wi}, obtained by varying the body force \gls{Fbody}. The panels display, from left to right: the total shear stress (\gls{piStarXY}$={\|\bar{\pi}_{xy}\|}/{\tau_{\text{PAC}}^{*}}$), and the total normal stress (\gls{N1Star}$={\|N_{1}\|}/{\tau_{\text{PAC}}^{*}}$. Panels correspond to: (a) \gls{Wi}=0.48 (Re=0.002, $F=1.25\cdot10^{-4}$), (b) \gls{Wi}=1.14 (Re=0.004, $F=2.5\cdot10^{-4}$), (c) \gls{Wi}=2.2 (Re=0.009, $F=3.5\cdot10^{-4}$), (d) \gls{Wi}=10.0 (Re=0.04, $F=2.0\cdot10^{-3}$), (e) \gls{Wi}=18.1 (Re=0.07, $F=3.5\cdot10^{-3}$), and (f) \gls{Wi}=26.4 (Re=0.1, $F=5.0\cdot10^{-2}$). Here, \gls{tauStarPAC}$=$\gls{Uo}\gls{eta_tot}$/$\gls{Rcy} is the dimensionless Stress parameter.}
\label{fig:LowSS}
\end{figure}

We next examine the corresponding steady-state spatial distributions, shown in Figure~\ref{fig:LowSS}. Across the tested \gls{Wi}, the shear stress fields remain consistent in shape and magnitude. At the same time, \(N_1\) displays a peak downstream of the cylinder, forming a sheet-like region of elevated polymeric stress advected along the centerline. This feature, associated with extensional-rate, becomes sharper with increasing \gls{Wi}. At very low \gls{Wi} (e.g., 0.48), the peak is broader and less intense, whereas moderate-\gls{Wi} cases yield more localized and intense structures.  

Consequently, we observe that fluctuations are more prominent in the low-\gls{Wi} fields shown in Panels (a)-(c), particularly in areas distant from the cylinder where shear rates are low. In contrast, their impact diminishes in the higher \gls{Wi} panels (d)-(f). Despite these local variations, the overall similarity of profiles across different regimes underscores the consistency of the numerical scheme in accurately capturing steady-state flow characteristics. In summary, both the low- and moderate-\gls{Wi} regimes display consistent steady-state stress and velocity distributions, with variations primarily in intensity (micro and macro coupling) and relaxation time dynamics. This evidence supports the notion that the chosen resolution and parameter configuration are sufficient for capturing essential flow features across a wide range of \gls{Wi}.

\subsection{Random porous media (\gls{RPM})} \label{subsec:porous_media}
\begin{figure}[ht!]
\centering
\includegraphics[width=0.70\textwidth]{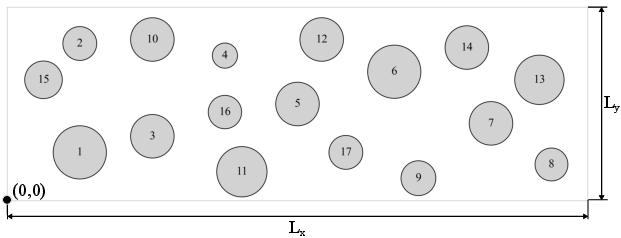}
\caption{The schematic setup of the 17 cylindrical solid obstacles that define the porous medium geometry (positions and radii), as detailed in Table~\hyperref[SM]{S3}. The rectangular domain is defined by dimensions $L_x=3L_y$ and $L_y=4$, with the origin $(0,0)$ located at the bottom-left corner of the cross-section.} 
\label{fig:Porous_Setup}
\end{figure}

Simulating fluid flow through porous media serves as a rigorous test for our coupled micro-macro framework, as it integrates geometric complexity with significant viscoelastic effects. In our model of a porous medium, we utilize a rectangular domain with a length in the x-direction that is three times that of the y-direction ($L_x = 3L_y$). This domain contains 17 non-uniformly distributed cylindrical solid obstacles, yielding a porosity of $\phi = 0.792$ (see Fig.~\ref{fig:Porous_Setup}). For detailed information on the coordinates and radii of the obstacles, please refer to Table~\hyperref[SM]{S3}, which is discussed in Section~\hyperref[SM]{SM9} in the \hyperref[SM]{Supplementary Material}. The Weissenberg number is defined as \gls{Wi} = \gls{Uo}\gls{lambda}$/$\gls{Rav}, where \gls{Uo} represents the time-averaged velocity and \gls{Rav} denotes the mean solid obstacle radius (0.425). The system is characterized by a chain length of \gls{Nb} = 16, a spatial resolution of \gls{dh} = 1/16, and coupling parameters \gls{delta_t} = 0.125 and \gls{M} = 50(\gls{Delta_t_SPH}$=2.5\cdot10^{-3}$).

From a computational standpoint, the domain consists of 12,288 \gls{SPH} particles, each associated with a micro-domain comprising 23,104 \gls{DPD} particles (\gls{Nb}$=16$), leading to a total of approximately $2.83 \times 10^8$ \gls{DPD} particles. To advance to $t=\lambda$ (approximately $1.4 \times 10^4$ steps), the simulation required 65.7 hours ($\approx 1.68$ s per step) on the BCAM workstation (see Table~\hyperref[SM]{S4}). For further details, please refer to Section~\hyperref[SM]{SM10} in the \hyperref[SM]{Supplementary Material}.  

\begin{figure}[ht!]
\centering
\includegraphics[width=0.95\textwidth]{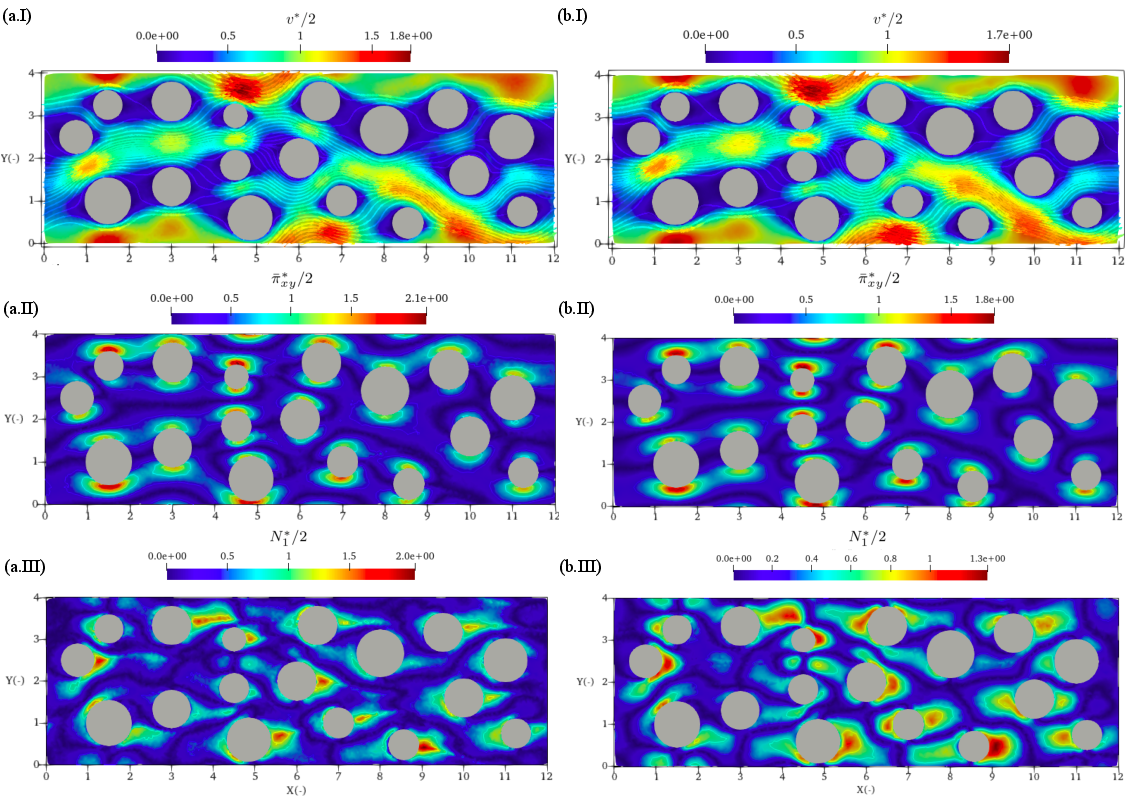}
\caption{Steady-state spatial distribution of fields in the porous medium case for (a) \gls{Wi}=0.52 ($F=1.0\cdot10^{-3}$) and (b) \gls{Wi}=6.5 ($F=5.0\cdot10^{-2}$). For each panel (a) and (b), the figure displays: (I) the dimensionless magnitude of the velocity (\gls{vStar}=$\|v\|/U_o$) and its streamlines; (II) the dimensionless magnitude of the total shear stress \gls{piStarXY}$=$\gls{piXY}$/$\gls{tauStarPore}; and (III) the dimensionless magnitude of the total normal stress \gls{N1Star}$=$\gls{N1}$/$\gls{tauStarPore}. Here, \gls{Uo} represents the magnitude of the total average velocity, and \gls{tauStarPore}$=$\gls{Uo}\gls{eta_tot}$/$\gls{Rav}.}
\label{fig:PorousSS}
\end{figure}

The steady-state flow fields, obtained after long integration times, are presented in Figure~\ref{fig:PorousSS} for two distinct Weissenberg numbers: a low-\gls{Wi}$=0.52$ ($F=1.0\cdot10^{-3}$) and a high-\gls{Wi}$=6.5$ ($F=5.0\cdot10^{-2}$). The velocity fields (Panel I) reveal complex streamlines wrapping around the obstacles, with high-velocity channels confined to the narrowest constrictions between solid obstacles. The total shear stress, \gls{piStarXY}, concentrates along the cylinder surfaces and within these same constricted passages (Panel II).

In contrast, the first normal stress difference, \gls{N1Star}, develops sharp, localized peaks in the wakes downstream of the obstacles, a characteristic signature of the intense extensional deformation in these regions (Fig.~\ref{fig:PorousSS} Panel II)\cite{Kawale2017a,Kumar2023a}. The magnitude and localization of both stress components are markedly more pronounced at \gls{Wi}$=6.5$. This highlights the significant role of the polymer's response in altering the flow field~\cite{Kumar2023a,Varshney2017,Browne2020a}.

This benchmark case demonstrates the scalability and robustness of the \gls{LHMM} framework in handling geometrically complex flows. The model successfully captures both the global flow organization and the highly localized stress concentrations that are critical for understanding viscoelastic flow conditions. For a detailed analysis of the initial transient dynamics of this case, please refer to Section~\hyperref[SM]{SM9.2} in the \hyperref[SM]{Supplementary Material}.

\section{Conclusions and Future Work} \label{sec:conclusions}
This study successfully developed and applied a Lagrangian Heterogeneous Multiscale Method (\gls{LHMM}), uniquely integrating Dissipative Particle Dynamics (\gls{DPD}) at the microscale with a thermodynamics structure-preserving Particle Hydrodynamics (\gls{SPH}) formulation at the macroscale.

Our approach rigorously characterized microscopic rheological properties by employing \gls{DPD} as a virtual rheometer. This provided crucial inputs, such as shear-thinning viscosity and relaxation times, necessary for parameterizing the \gls{LHMM} across various \gls{Wi} regimes. The effectiveness of the \gls{LHMM} was validated through benchmark tests, including Reverse Poiseuille Flow (RPF) and flow through a Periodic Array of Cylinders (PAC). These assessments consistently demonstrated the framework’s capability to capture both transient and steady-state viscoelastic behaviors.

Furthermore, the \gls{LHMM} exhibited scalability and computational efficiency, enabling its application to highly complex geometries, as illustrated by detailed simulations of flow through a \gls{RPM}. This case underscored the framework’s ability to resolve intricate flow patterns, localized stress concentrations, and microstructural responses in heterogeneous environments, with particle counts exceeding $10^8$. Collectively, these findings establish the \gls{LHMM} as a predictive and versatile tool for modeling polymer melt rheology in complex \gls{2D} flows, effectively bridging microscopic structure with macroscopic transport. Beyond its methodological significance, this work opens new avenues for applications in polymer processing, porous media transport, and the design of advanced materials.

\begin{itemize}
\item Advanced viscoelastic instability analysis: deeper investigation into the onset and evolution of viscoelastic instabilities in complex flow geometries.
\item Extension to three-dimensional flows: expanding the \gls{LHMM} framework to simulate 3D configurations for broader real-world applicability.
\item Possible extension to other complex system for which a LAMMPS-based implementation is available, for example coupling \gls{DEM} and \gls{SPH} for multiscale modeling of dense particle suspensions. 
\end{itemize}

\section*{Acknowledgement}
The authors acknowledge the funding provided by IKUR–HPC\&AI – (HPCAI10: MOLD-POLY) and the IKUR Strategy funded by  Basque Government and the European Union NextGenerationEU/PRTR. The research is also partially funded by the Spanish State Research Agency through BCAM Severo Ochoa excellence accreditation CEX2021-0011 42-S/MICIN/AEI/10.13039/501100011033, and through the project PID2024-158994OB-C42 (‘Multiscale Modeling of Friction, Lubrication, and Viscoelasticity in Particle Suspensions’ and acronym ‘MMFLVPS’) funded by MICIU/AEI/10.13039/501100011033 and cofunded by the European Union. The authors thankfully acknowledges the computer resources at MareNostrum and the technical support provided by Barcelona Supercomputing Center (IM-2024-3-0013 and IM-2025-2-0042).

\section*{Supplementary Information}\label{SM}
Supplementary data associated with this article can be found in the online version. The detailed Supplementary Material, which expands upon the research presented here, including the Supplementary Document and Video, is available temporarily via Mendeley's Dataset repository at: {\color{blue}\url{https://data.mendeley.com/preview/vxtbhpfzyt?a=8afc905e-ca15-4325-9dba-c5b8f1edfef3}}. Additionally, the complete coding repository for the LHMM framework with DPD-SPH are publicly accessible at: {\color{blue}\url{https://github.com/BCAM-CFD/LHMM/tree/main/LHMM-PolymerDPD}}.

\printglossary[type=acronym]
\printglossary[type=nomencl] % Si tienes otra lista, también va aquí

\appendix

\section{Kernel function approximation}\label{AppSPH}
The Lucy kernel for \gls{2D} \gls{SPH} simulations employs a smoothing length  $h = 4\text{dx}$, where $\text{dx}$ is the initial particle spacing. The kernel function is defined as:
\begin{equation} \label{eq:lucy_kernel}
    W(r_{IJ}) = 
    \begin{cases}
    \dfrac{5}{\pi h^{2}} \left(1 + \dfrac{3r_{IJ}}{h}\right)\left(1 - \dfrac{r_{IJ}}{h}\right)^3, & \dfrac{r_{IJ}}{h} < 1 \\ \\
    0, & \dfrac{r_{IJ}}{h} \geq 1.
    \end{cases}
\end{equation}

This formulation provides third-order accuracy while maintaining compact support within the smoothing domain.

\section{Artificial sound speed}\label{SoundAn}
  The numerical speed $c_s$ is calculated to enforce near-incompressible flow conditions ($Ma<0.1$) through:

\begin{equation} \label{eq:sound_speed}
    c_s = \max \left(
    \dfrac{|\mathbf{v}|{\text{max}}}{\delta},
    \sqrt{\dfrac{\eta_{\text{total}} |\mathbf{v}|_{\text{max}}}{\delta L}},
    \sqrt{\dfrac{|\mathbf{f}^{\text{ext}}|L}{\delta}}
    \right)
\end{equation}

where $\delta=0.1$ serves as a compressibility parameter and $L$ represents the characteristic domain length scale.
  
\section{Time-Step Calculation for SPH (Macroscopic scale)} \label{sec:time_stepping} 
The time-step $\Delta t$ is determined by the CFL condition:

\begin{equation} \label{eq:cfl_condition}
\Delta t_{\rm SPH} = 0.25 \cdotp\min \left\{
\frac{h}{\|\mathbf{v}\|_{\text{max}} + c_s}, 
\left(\frac{h}{\|\mathbf{f}\|}\right)^{1/2}, 
\frac{h^2}{2\nu}
\right\}
\end{equation}

where $\nu = \eta_{\text{tot}}/\rho$ is the polymer solution's kinematic viscosity. The safety factor 0.25 was found optimal for maintaining stability in viscoelastic flow simulations.

\section{LHMM flowchart}\label{sec:Flowchart}
\begin{figure}[ht!]
    \centering
    \includegraphics[width=0.9\textwidth]{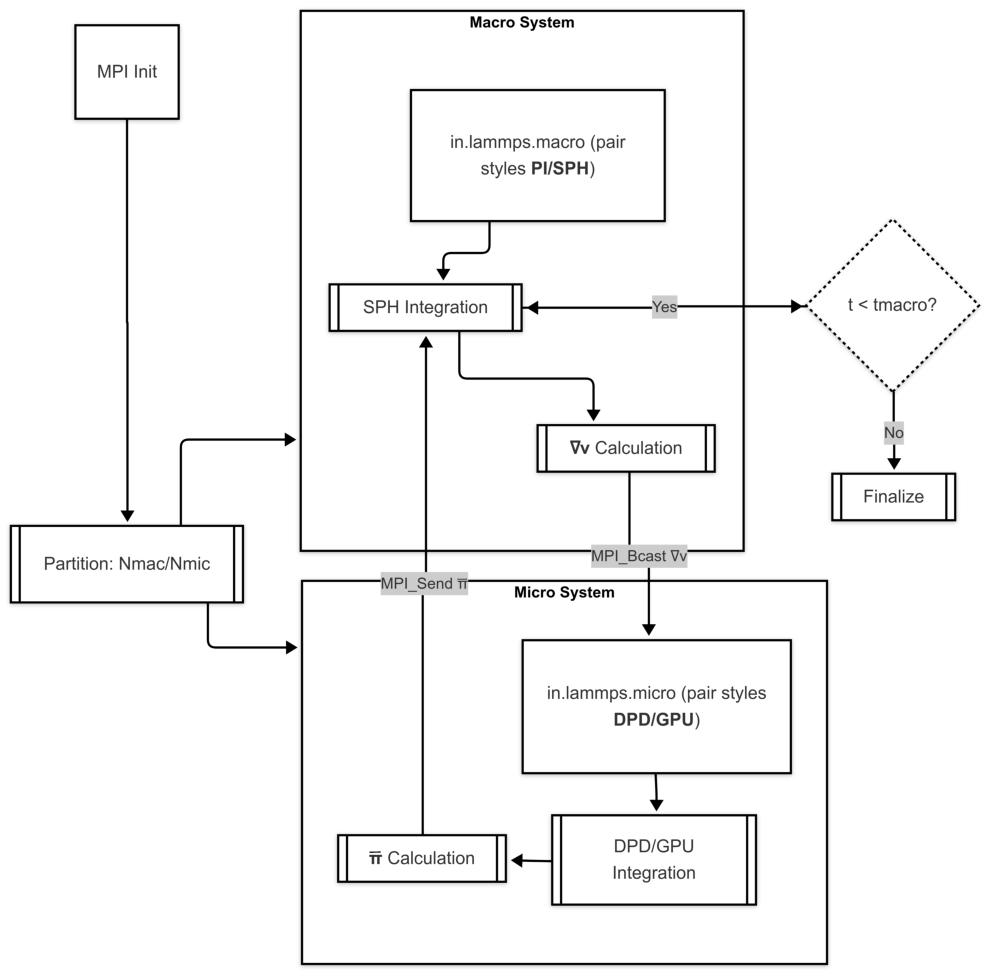}
    \caption{\gls{LHMM} multiscale coupling workflow. The \textbf{Macro System} (top) executes \gls{SPH} simulations using \texttt{PI/SPH} pair styles, computing velocity gradients ($\nabla\mathbf{v}$) that are broadcast to the \textbf{Micro System} (bottom) via \texttt{MPI\_Bcast}. The micro system runs GPU-accelerated \gls{DPD} simulations (\texttt{DPD/GPU} pair styles) and returns stress tensors (\gls{pibar}) via \texttt{MPI\_Send}. The loop continues until completing the total macro steps. Box styles indicate active computation (bold borders) and the termination check (dashed border).}
    \label{fig:LHMM_Flowchart}
\end{figure}

This appendix provides a detailed visual representation of the \gls{LHMM} multiscale coupling scheme's workflow, supplementing the main text for readers seeking a comprehensive overview of the implementation. The diagram outlines the two-way information exchange between the macro and micro systems in LHMM, including the specific MPI communication commands and key simulation steps. The \gls{LHMM} multiscale coupling scheme employs a C++/MPI architecture that coordinates LAMMPS simulations \cite{Xu2024,Hue2025} across $N_{\text{mac}}$ (macro) and $N_{\text{mic}}$ (micro) MPI ranks, where $N_{\text{mac}} + N_{\text{mic}} = P$. As detailed in Figure~\ref{fig:LHMM_Flowchart}, the macro system executes \texttt{in.lammps.macro} with \texttt{pair\_style PI/SPH}, computing macroscopic velocity gradients ($\nabla\mathbf{v}$) after \gls{M} \gls{SPH} steps (defining the coupling time step $\delta t$) before broadcasting via \texttt{MPI\_Bcast}. The micro system runs \texttt{in.lammps.micro} with \texttt{pair\_style \gls{DPD}/GPU}, processing the received $\nabla\mathbf{v}$ for \gls{K} \gls{DPD} steps (also defining $\delta t$) and returning microscopic stress tensors (\gls{pibar}) via \texttt{MPI\_Send}. This cycle repeats for the total simulation duration ($t_{\text{macro}}$ steps), with dynamic load balancing across the partitioned communicators. The detailed step-by-step procedure governing this iterative exchange is provided in \ref{subsec:coupling_protocol}.

\section{Detailed Coupling Protocol} \label{subsec:coupling_protocol}

The macro and micro systems exchange information iteratively through a synchronized protocol at each exchange time step $\delta t$. This coupling time step is defined by the relation $\delta t = \bar{M} \cdot \Delta t_{\rm SPH} = k' \cdot \Delta t_{\rm DPD}$, where \gls{M} and \gls{K} are integers representing the number of \gls{SPH} and \gls{DPD} steps within one coupling interval, respectively.

\begin{enumerate}[label=\textbf{Step \arabic*:}, leftmargin=*, align=left, start=0]
    \item \textit{Initialization}: \\Initialization of the particle distribution for both macroscale and micro-domains (linked to macroparticles). This step includes defining boundary and initial conditions, the sound speed in the macroscale (Eq.~\eqref{eq:sound_speed}), initial time steps for the macroscale (\gls{Delta_t_SPH}, determined by Eq.~\eqref{eq:cfl_condition}) and microscale (\gls{Delta_t_DPD}, defined in Section~\ref{sec:micro_results} from viscometric characterization). Additionally, parameters \gls{alpha} and \gls{beta} for Eqs.~\eqref{eq:macro} and \eqref{eq:beta} are set.
    \item[]\textit{Coupling Cycle Begins:}
    \item \textit{Macroscopic Integration (over \gls{M} SPH steps, defining $\delta t$):}
        \begin{itemize}
            \item The SPH simulation runs for \gls{M} time steps. During this interval, particle pressure is computed using the equation of state (Eq.~\eqref{eq:tait}).
            \item The \gls{SPH} approximation for the continuity and momentum conservation equations (Eqs.~\eqref{eq:sph_continuity}) is applied.
            \item A Velocity-Verlet scheme updates particles' physical properties such as position, velocity, and forces.
        \end{itemize}
    \item \textit{Top-down (Macro $\rightarrow$ Micro, at the end of $\delta t$):}
        \begin{itemize}
            \item The macroscopic velocity gradient \gls{grad_vI} is computed from \gls{SPH} (via Eq.~\eqref{eq:macro_micro}) and transferred to each associated \gls{DPD} simulation in its micro-domain.
            \item Boundary conditions are imposed in \gls{Omega_bc} using Eq.~\eqref{eq:micro_bc}. This local flow information deforms the polymer microstructure by applying a linear background velocity field or through explicit boundary-driven shearing in the \gls{DPD} microscale, affecting the ensemble of $i$ particles and bead-chains.
        \end{itemize}
    \item \textit{Microscopic Integration (over \gls{K} DPD steps, defining $\delta t$):}
        \begin{itemize}
            \item The \gls{DPD} simulations evolve for \gls{K} time steps, receiving the imposed \gls{vi_prime} in \gls{Omega_bc} from the previous macro-step.
            \item Forces are computed according to \gls{DPD} interaction forces (Eqs.~\eqref{eq:conservative}-\eqref{eq:random}) and the \gls{FENE} bonding forces (Eq.~\eqref{eq:fene_force}) for each bead within the same polymer chain.
            \item A Velocity-Verlet scheme updates particles' physical properties (position, velocity, and forces) within the microscale from the Eq.~\eqref{eq:motion}. 
        \end{itemize}
    \item \textit{Bottom-up (Micro $\rightarrow$ Macro, at the end of $\delta t$):}
        \begin{itemize}
            \item After the microstructure has evolved, the polymeric stress tensor \gls{pip} is calculated via \gls{IK} averaging over the \gls{DPD} ensemble, employing Eqs.~\eqref{eq:IKtotal}-\eqref{eq:potential_stress} within the core region (\gls{Omega_core}). 
            \item This stress is then passed to the corresponding macro-particles and incorporated into the macroscopic momentum equations (specifically, the \gls{SPH} approximation for the continuity of mass and momentum conservation equations in Eq.~\eqref{eq:sph_continuity}) to update the macroscopic flow at time $t+\delta t$.
        \end{itemize}
    \item Return to Step~1 for the next coupling interval.
\end{enumerate}

\section{Carreau-Yasuda Model Details} \label{app:CY_details}
This appendix provides further details on the Carreau-Yasuda (C-Y) model used for fitting the DPD simulation data. The C-Y equation is given by:
\begin{equation} 
\eta(\dot{\gamma}_{xy}) = \eta_{\text{cy}}^\infty + (\eta_{\text{cy}}^0 - \eta_{\text{cy}}^\infty) \left[1 + (\lambda \dot{\gamma}_{xy})^{a_{\text{cy}}} \right]^{(n_{\text{cy}}-1)/a_{\text{cy}}},
\label{eq:Yasuda} % Cambia la etiqueta si la mantienes en main también, o elimínala si solo va en apéndice
\end{equation}
where \gls{eta0cy} is the zero-shear viscosity, \gls{etainf} is the infinite-shear viscosity, $\lambda$ is the mean relaxation time, \gls{acy} is the transition sharpness parameter, and \gls{ncy} is the power-law index. These parameters are optimized during the fitting process for each polymer chain length.

\section{Normal Stress Coefficients and Relaxation Time Definitions} \label{app:NormalStressRelaxation}
This appendix provides an overview of the definitions and calculations for normal stress coefficients and relaxation times utilized in our rheological analysis. The first normal stress coefficient (\gls{psi1}), crucial for characterizing the elastic properties of viscoelastic fluids, is defined as the difference between the normal components of the polymer stress tensor, normalized by the square of the applied shear rate.
\begin{equation}
 \psi_1 = \frac{\bar{N}1}{\dot{\gamma}_{xy}^2},
 \label{eq:psi1_appendix}
 \end{equation}
where $\bar{N}_1 = \langle \bar{\pi}_p \rangle_{xx} - \langle \bar{\pi}_p \rangle_{yy}$ represents the first normal stress difference within the \gls{Omega_core}, computed from the polymer stress response \gls{pip}.
The mean relaxation time (\gls{lambda})~\cite{Fedosov2010} is defined as a function of the shear rate:
\begin{equation}
 \lambda(\dot{\gamma}_{xy}) = 0.5 \frac{\psi_1}{\langle \bar{\eta}_{p} \rangle_s},
 \label{eq:tau0_def}
 \end{equation}
and its zero-shear value is obtained from the asymptotic limits as $\dot{\gamma}_{xy} \to 0$:
\begin{equation}
 \lambda = 0.5 \frac{\psi_1(\dot{\gamma}_{xy} \to 0)}{\eta_p^0}.
 \label{eq:tau0_def_appendix}
 \end{equation}
\section{Analytical velocity profile for pure solvent}\label{app:velocity}

For reference, the steady-state velocity profile of a Newtonian solvent in planar Poiseuille flow under a constant body force \gls{Fbody} can be expressed analytically as~\cite{Fedosov2010,NietoSimavilla2022}:

\begin{equation}
v_x(y) = v_{\text{max}} \left[ 1 - \left(\frac{y}{L_y/4}\right)^2 \right],
\label{eq:velocity}
\end{equation}

Where \gls{Ly} is the channel width and $v_{\text{max}}$ is the peak velocity at the channel centerline, given by

\begin{equation}
v_{\text{max}} = \frac{F n L_y^2}{32 \eta},
\end{equation}

with $n$ the particle number density and $\eta$ the Newtonian viscosity. This parabolic profile is used as a benchmark in Section~\ref{subsub:RPF-N} to assess the velocity response of the coupled \gls{LHMM} system.

\bibliography{library3}

\end{document}